\documentclass[iop,revtex4]{emulateapj}

\usepackage{natbib}
\usepackage{wasysym}
\usepackage{lscape}

\shorttitle{Multiplicity in Taurus}
\shortauthors{Daemgen et al.}

\begin{document}

\title{Sub-stellar Companions and Stellar Multiplicity in the Taurus Star-Forming Region}

\author{Sebastian Daemgen}
\affil{Department of Astronomy \& Astrophysics, University of Toronto, 50 St. George Street, Toronto, ON, Canada M5H 3H4}
\email{daemgen@astro.utoronto.ca}
\and
\author{Mariangela Bonavita}
\affil{The University of Edinburgh, Royal Observatory, Blackford Hill, Edinburgh EH9 3HJ, U.K.}
\and
\author{Ray Jayawardhana}
\affil{Physics \& Astronomy, York University, Toronto, Ontario L3T 3R1, Canada}
\and
\author{David Lafreni\`ere}
\affil{Department of Physics, University of Montr\'eal, Montr\'eal, QC, Canada}
\and
\author{Markus Janson}
\affil{Department of Astronomy, Stockholm University, Stockholm, Sweden}

\begin{abstract}
We present results from a large, high-spatial-resolution near-infrared imaging search for stellar and sub-stellar companions in the Taurus-Auriga star-forming region. The sample covers 64 stars with masses between those of the most massive Taurus members at $\sim$3\,$M_\odot$ and low-mass stars at $\sim$0.2\,$M_\odot$. We detected 74 companion candidates, 34 of these reported for the first time. Twenty-five companions are likely physically bound, partly confirmed by follow-up observations. Four candidate companions are likely unrelated field stars. Assuming physical association with their host star, estimated companion masses are as low as $\sim$2\,M$_\mathrm{Jup}$. The inferred multiplicity frequency within our sensitivity limits between $\sim$10--1500\,AU is $26.3^{+6.6}_{-4.9}\%$. {Applying a completeness correction, 62\%$\pm$14\% of all Taurus stars between 0.7 and 1.4\,M$_\odot$ appear to be multiple. Higher order multiples were found in $1.8^{+4.2}_{-1.5}\%$ of the cases, in agreement with previous observations of the field. We estimate a sub-stellar companion frequency of $\sim$3.5--8.8\% within our sensitivity limits from the discovery of two likely bound and three other tentative very low-mass companions. This frequency appears to be in agreement with what is expected from the tail of the stellar companion mass ratio distribution, suggesting that stellar and brown dwarf companions share the same dominant formation mechanism. Further, we find evidence for possible evolution of binary parameters between two identified sub-populations in Taurus with ages of $\sim$2\,Myr and $\sim$20\,Myr, respectively.}
\end{abstract}


\section{Introduction}
A large fraction of stars in our galaxy are members of binary or multiple systems \citep{duq91}, suggesting that multiple star formation plays a major role in processes such as cluster formation, protoplanetary disk evolution, and planet formation. As a consequence, more and more attention has been drawn to investigating the occurrence and properties of binary and multiple systems, resulting in an increasing number of multiplicity studies being performed in the last decade \citep[see, e.g.,][]{duc13}.

Significant improvements in high-contrast instruments and techniques have allowed us to reach a high level of completeness down to very low primary and companion masses, especially for favorable targets such as young stars in nearby star-forming regions. As a result, recent searches have revealed a new population of wide sub-stellar companions \citep[see, e.g.,][]{cha05,luh06,mar08,mar10,lag10}. These findings imply that the so-called {\it brown dwarf desert}, invoked due to a paucity of brown dwarf companions up to a few AU seen in radial velocity surveys \citep[e.g.,][]{lin03}, might not extend farther out. The existence of a seemingly continuous population, like that shown by the latest discoveries in the Upper-Sco \citep{laf08,laf11} and Sco-Cen \citep{jan12} regions, suggests that binary formation extends all the way to planetary masses for wide separations, or at least implies that mass alone is not a clear-cut observational diagnostic for distinguishing between formation mechanisms.

The results of these studies also show that the frequency of wide low-mass companions is higher in star-forming regions than in young moving groups or the field \citep[see, e.g.,][]{laf08,cha10,nie10,jan11}, implying that binary ionization mechanisms in dense environments could be relevant on star-forming timescales $\lesssim$10\,Myr, as predicted by, e.g., \cite{ver09}.
Some of these wide companions, such as 1RXS~J1609b \citep{laf08},  HIP~78530b \citep{laf11}, or ROXS\,42Bb \citep{cur14}, have mass ratios with respect to their parent stars of only $\sim$1\% and due also to their extreme separations (up to $\sim$700\,AU) are hard to explain in any formation paradigm suggested so far, whether based on core accretion \citep[e.g.,][]{mor12} or disk instabilities \citep[e.g.,][]{raf11}. Hence, these objects represent extreme outcomes of their underlying formation mechanism(s), regardless of which it is. 

The bulk of the proposed formation mechanisms can be categorized as \emph{fragmentation}, \emph{fission}, or \emph{capture} scenarios, each with their own predictions of the resulting binary parameters (multiplicity frequency, separation, mass ratio) that must be compared to observations. While it has been found that stellar densities are not high enough to explain the observed overall binary fraction with capture alone \citep{toh02}, the currently two most popular formation channels are direct collapse of a cloud core into multiple components and the fragmentation of a massive disk around a forming stellar object. In these scenarios binaries form with separations on the order of the size of the collapsing core ($\sim$0.01--0.1\,pc) or the size of a disk ($\lesssim$ a few hundred AU), respectively, and the individual objects may continue to accrete from a common envelope or their individual circumstellar material to reach their final mass ratios. 
Since orbital evolution due to stellar encounters in clustered star-forming regions and interactions with disks can be substantial, otherwise possibly identifying signatures of the formation channel will likely be blurred. Accordingly, observations of dynamically young clusters featuring rather primordial binary distributions are essential.

We performed a deep-imaging survey of 72 stars in the Taurus-Auriga molecular cloud, using the ALTAIR/NIRI adaptive optics system on the Gemini North telescope \citep{hod03}. The Taurus-Auriga region is the nearest ($\sim$140\,pc) star-forming region in the Northern sky. It spans a region of approximately 25\,pc diameter, with a depth of about $20$\,pc \citep{ken94,tor12}. The list of currently known likely (albeit not necessarily confirmed) members counts more than 350 mainly low-mass stars. Taurus is characterized by a paucity of high mass ($M \geq 5\,M_{\odot}$) young stars \citep{ken08}. Its proximity and low stellar density make it a very rewarding target for multiplicity studies, with respect to more distant, massive and dense star-forming environments. This explains the large number of studies performed in the last few decades, aiming at defining the stellar population and multiplicity features of this region using a number of different techniques.

Previous studies of Taurus \citep[see, e.g.,][and references therein]{kra11} have found a large number of multiple systems in the region, with a much higher frequency of wide binary companions than in denser clusters and the field \citep[][]{kra07,kra09a}. On the other end, the frequency of binaries in the extremely young proto-stellar stage appears to be even higher \citep{che13}, possibly suggesting destruction of proto-stellar binaries through dynamical interactions.

The new survey presented here has been specifically designed to span a wide range of primary masses and spectral types while achieving high completeness down to the sub-stellar mass regime at separations $>50$\,AU. In fact, our sensitivity to companions reaches well into the planetary mass regime at separations $>100$\,AU. We targeted Taurus in the context of our other recent multiplicity surveys with similar completeness, such as those in Chamaeleon\,I by \citet{laf08}, Sco-Cen \citep{jan13} and Upper-Sco \citep{laf14}.

\section{Sample \& Observations}
\subsection{Sample Selection \& Biases}
The target sample was originally constructed by picking 10--15 targets in each of five equal logarithmic mass bins over the 0.15--3.0\,M$_\odot$ range, originating from a compilation of 275 stars in Taurus with known spectral types and magnitudes $R \leq 13$\,mag \citep{lei93,ghe93,sim95,koh98,har05,luh06,luh06a,gui06}.
A total of 72 stars were observed between fall 2011 and 2012 with a few follow-up observations in 2013 to confirm common-proper motion of the newly-found companion candidates (Sect.~\ref{sec:epoch} for details). A significant fraction of the targeted stars have been part of previous binary surveys. Since one of our goals is an unbiased view on the multiplicity properties of Taurus stars, we do not exclude them from the sample.

The magnitude cut at R$<$13 mag results from the fact that we use adaptive optics to observe the sample. At a distance of $\sim$140\,pc and an age of 2\,Myr, this corresponds to a lower mass cut at 0.17\,M$_\odot$, assuming negligible extinction. As described in Sect.~\ref{sec:membership}, in addition to a young population, we evoke a subpopulation with an estimated age of $\sim$20\,Myr. At this age, the magnitude limit translates to masses $\gtrsim$0.75\,M$_\odot$, making this sub-sample on average more massive than the younger sample. The results in this paper are independent from this skew since singles and binaries will be affected in the same way. However, the deduced statistics as a function of mass will be less precise for the lowest-mass objects because fewer targets enter the lowest-mass bins. At the very faintest end, applying a magnitude limit may introduce a bias for the inclusion of binaries. The inferred multiplicity frequency might accordingly be biased towards higher fractions in the lowest mass bin under consideration. This is further discussed in the respective section (\ref{sec:BF_vs_mass}).

\subsection{Membership\label{sec:membership}}
A detailed analysis of the membership of our target stars in the Taurus cloud is presented in Appendix~\ref{sec:A1}, while a summary of it is given here. To be able to draw conclusions that apply to the Taurus population as a whole and to its very young stars in particular, we thoroughly consider evidence for both membership and youth, for all targets of this study. This results in a necessary split of the sample into \emph{young} Taurus members with ages $\sim$2\,Myr and an \emph{extended sample} that appears to be co-moving with Taurus, i.e., formed within the same association of clouds, but that lacks conclusive evidence for extreme youth. Membership in the \emph{young} category is indicated by the presence of accretion or strong infrared excess originating from a disk in addition to strong Lithium absorption, which indicates an age $\ll$10\,Myr \citep[e.g.,][]{her08}. Stars in the \emph{extended} category are likely part of a population with an age $\sim$20\,Myr, as indicated by their Lithium absorption which appears comparable to or stronger than that of the weakest Li signature found in members of Lower Centaurus Crux/Upper Centaurus Lupus ($\sim$16--17\,Myr) for a given effective temperature \citep{son12}. We take this and other evidence (for the full analysis see Appendix~\ref{sec:A1}) as indication that a burst of star formation about $\sim$20\,Myr ago created this group of stars which was followed by another star formation burst 1--2\,Myrs ago.

We find 36 stars to be classified as \emph{young members}, while another 29 belong to the \emph{extended sample}. A total of eight stars in the original sample lack evidence for youth and/or membership in the Taurus association. We present their measurements for completeness reasons but do not include them in the analysis. The resulting sub-samples together with basic stellar parameters and the applied membership criteria are listed in Tab.~\ref{tab1}. In the course of the paper, the entity of all Taurus members of this study, i.e., the \emph{extended sample} together with the \emph{young members} sample, is referred to as the \emph{combined} sample.

\subsection{Observations and Data Reduction}
 All observations were obtained at the Gemini North telescope with the NIRI instrument \citep{hod03} in $K_s$ band (1.95--2.30 $\mu$m). The ALTAIR adaptive optics (AO) system \citep{her00} was used to obtain diffraction-limited images on NIRI's 1024$\times$1024 Aladdin InSb detector array. The f/32 camera provides a sampling of $\sim$21.9\,mas/pixel and a FoV of $\sim$22\arcsec$\times$22\arcsec. 

The observing strategy was similar to the one described by \citet{laf08}. For sky subtraction and bad pixel correction, each target was observed at five dither positions. At each dither position, one non-saturated short exposure (divided into many coadds) was acquired in high-readnoise mode, followed immediately by a longer (typically 10\,s/dither) saturated exposure in low-readnoise mode. Using the low-readnoise mode for the latter exposure allows us to reach the sky background limit; otherwise, the contrast would be limited by the high readnoise. This strategy results in a high observing efficiency, and a large dynamic range providing sensitivity at both small and large angular separations. 
With typical Strehl Ratios between 15\% and 25\%, the observations were sensitive to companions as faint as K\,$\sim$\,18\,mag (5$\sigma$) at large separations $\gtrsim$2\arcsec\ from the central star.

\begin{figure*}
\includegraphics[angle=0,scale=0.25]{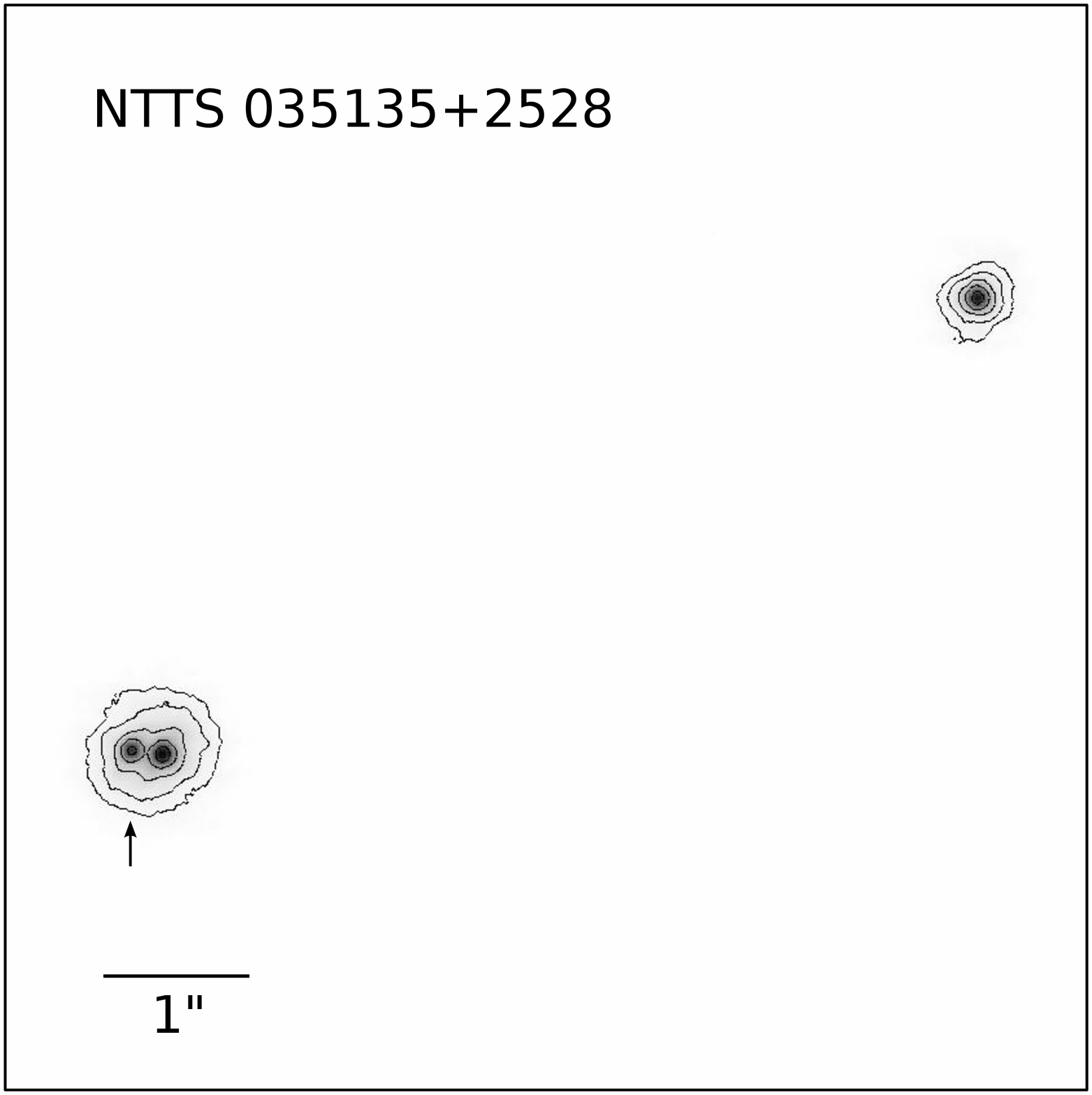}
\hfill\includegraphics[angle=0,scale=0.25]{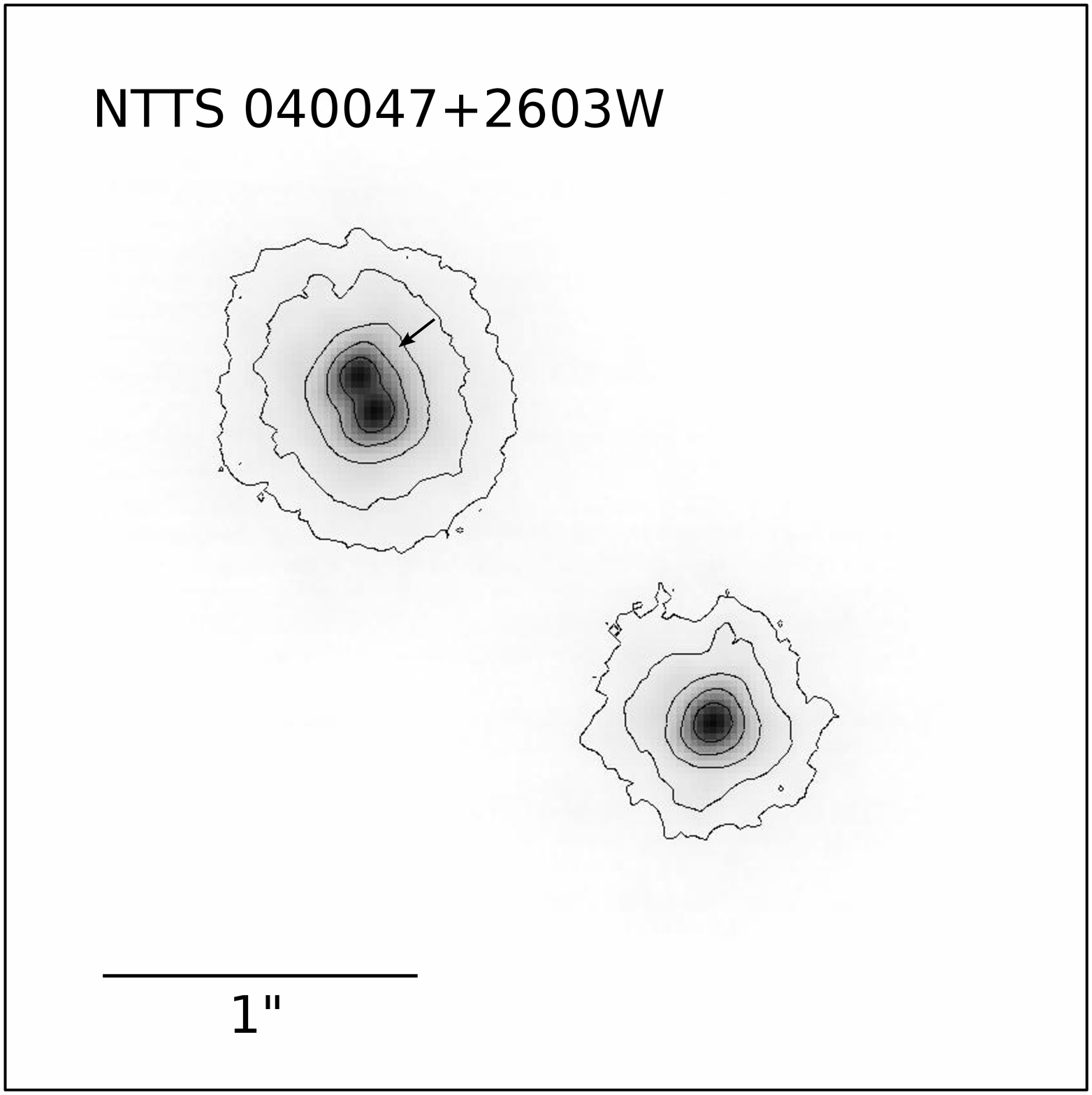}
\hfill\includegraphics[angle=0,scale=0.25]{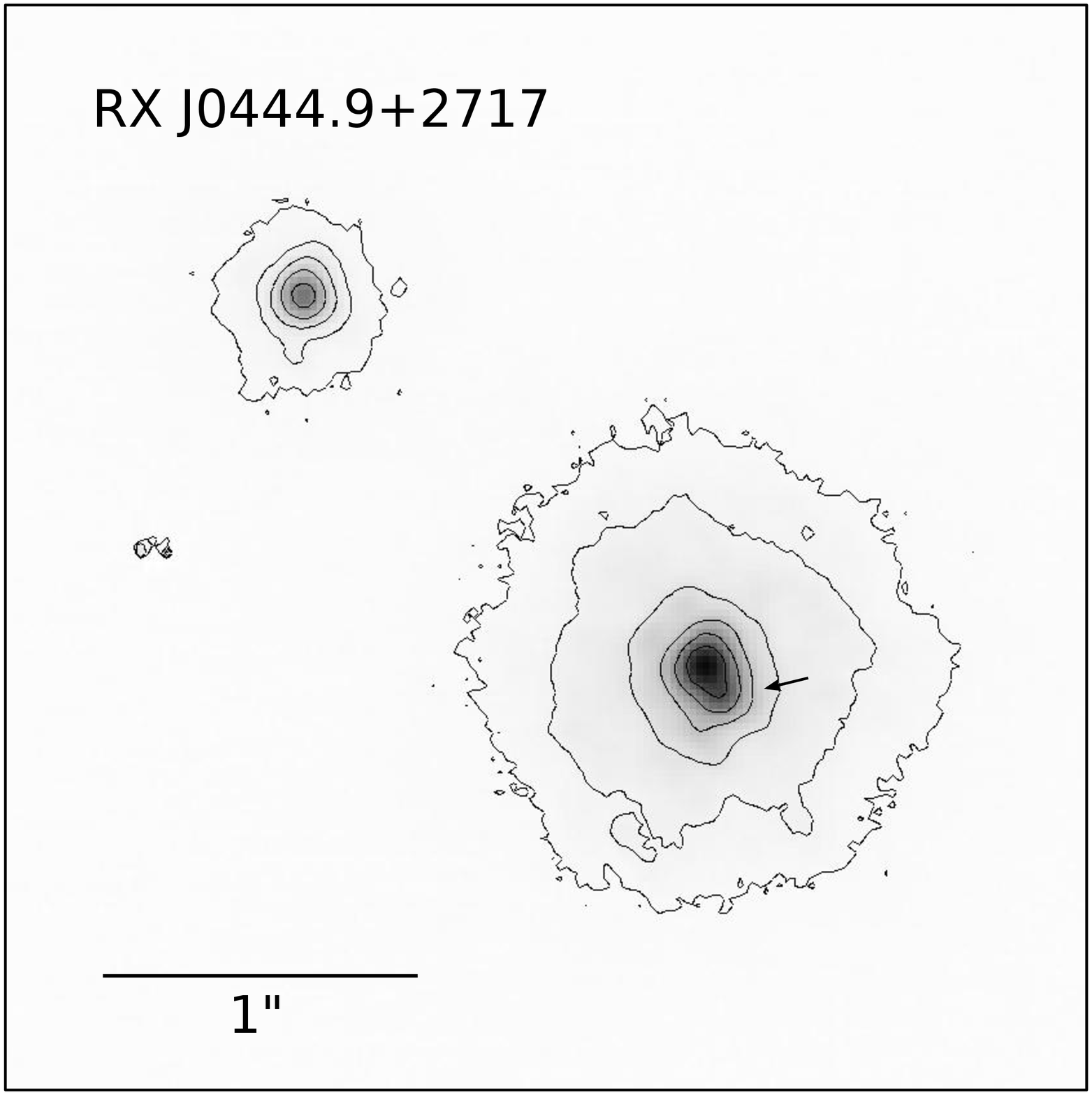}
\caption{Unsaturated Gemini North/NIRI $K$-band images of three close multiple systems, {two of them (NTTS\,035135+2528, RX\,J0444.9+2717) previously unknown,} and their wide tertiary companions. The contours show levels of constant logarithmic intensity. North is up and east is to the left. Note that NTTS\,040047+2603W has been classified as likely a non-member of the Taurus association.\label{fig:new}}
\end{figure*}

We followed standard procedures for data reduction. A sky frame was constructed by taking the median of the dithered images (masking the regions dominated by the target's signal). The individual images were then sky subtracted and divided by a normalized flat field, and the bad pixels were replaced by a median over their good neighbors. For all images, field distortion was corrected as described in \citet{laf14}. Their analysis shows residual systematic uncertainties of 15\,mas, 25\,mas, and 50\,mas at radii of 4\arcsec, 8\arcsec, and 12\arcsec~from the center, respectively, {and a systematic uncertainty of the position angle of 0.15$^\circ$}. These uncertainties do not affect our relative astrometric measurements with NIRI, but may result in systematic separation offsets for wide companions when compared to observations with other instruments (Sect.~\ref{sec:epoch}). Fig.~\ref{fig:new} shows fully reduced, unsaturated images of three discovered close-in bona-fide companions. 

\section{Analysis}
\subsection{Photometry \& Astrometry\label{sec:photometry}}
We identified a total of 74 companion candidates through visual inspection of the individual shallow and deep exposures {of the targeted 64 Taurus members}. Our attempts to remove the static speckle pattern by subtracting a PSF from a different exposure taken close in time did not reveal any additional companion candidates.
The companions' positions and magnitudes relative to the target stars were determined with simultaneous PSF photometry using the \emph{IRAF} task {\tt daophot}. The required reference PSFs were constructed from the target stars themselves, within a radius of 150 pixels. If one or more of the companions appeared within this radius, they were removed by replacing the contaminated section with a copy of the equivalent region on the other side of the reference star's PSF, rotated by 180$^\circ$. We confirmed that this procedure introduces negligible photometric and astrometric uncertainties by comparing the resulting photometry of an uncontaminated PSF with and without substitution.

Whenever possible, the individual shallow images were used for photometry/astrometry. If the companion was too faint to be reliably detected in the shallow images, a combination of the deep and shallow exposures, which restores the natural PSF shape in the central saturated core of the central star in the deep exposures, was used. This combination replaces the innermost 8--10\,pix with the unsaturated PSF core from a shallow exposure of the same star, scaled to the same exposure time. The introduced uncertainties are $\sim$0.05 pixels for the measured separations and 0.1\,mag for the relative photometry. 

We determined astrometric and photometric uncertainties from the statistical fluctuations between the individual dither images, the residual flux after PSF subtraction (typically $<$5\%), and the deep+shallow combination noise. The typical full width at half maximum (FWHM) of the measured PSFs was $\sim$0\farcs08 assuming a pixel scale of 0.0219 arcsec/pixel of NIRI. The resulting photometry is listed in Table~\ref{tab2}.

The photometry and astrometry of the components of RX~J0444.9+2717 required a special routine considering that the central object is a close binary with significantly different component magnitudes and the wide companion is not bright enough to reliably serve as a PSF reference (see Fig.~\ref{fig:new}). We used aperture photometry of the isolated wide and the combined close components, respectively. The resolved component photometry was inferred from the peak flux at the position of the individual components (after correction for the contribution of the other PSF wings), using a power law fit to the correlation between peak and total flux. The relative positions were measured from the centroids of the individual peaks. The location of the faint peak in the wing of the central object is not expected to be strongly biased, since two individual peaks can be identified at a separation larger than the FWHM of the PSF.

\subsection{Detection Limits \& Completeness\label{sec:detection_limits}}
The 5$\sigma$ sensitivity limits for the detection of companions to all targeted stars {are tabulated in Appendix~\ref{sec:appC} and are} shown in Fig.~\ref{fig:sensitivity}. They are derived from our companion-free exposures (see Sect.~\ref{sec:photometry}). To obtain the curves, the noise in 5$\times$5 pixel regions is measured around every pixel in the image, multiplied by a factor of five, and averaged in annuli of width $w=1\times$FWHM around the central star. Since all companions are detected in all five dither images of an observation, the displayed effective sensitivity is a factor of $1/\sqrt{5}$ ($\approx$0.87\,mag) better than that of the individual exposures.
\begin{figure*}
\epsscale{0.8}
\plotone{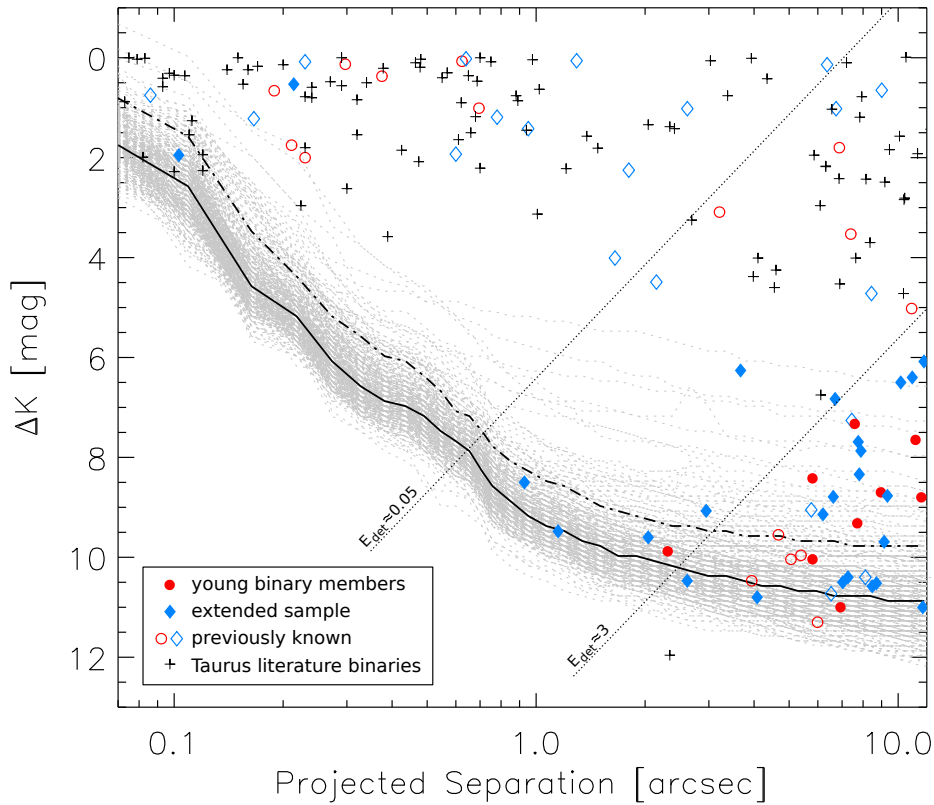}
\caption{Sensitivity of our observations to the detection of companions with a significance of 5$\sigma$. The two bold black curves mark the range of companions that would be detected around 50\% and 90\% of all stars, individual curves are shown in gray. Filled symbols show our new discoveries in the young (circles) and extended (diamonds) samples, open symbols show previously reported companion candidates in the two samples, respectively. Plus signs show the position of other known binary companions found in the literature (for references, see text). {The dotted curves show the approximate locus of targets with $E_\mathrm{det}\approx0.05$ and $E_\mathrm{det}\approx3$, equivalent to a 95\% chance of being bound to their host star or a 95\% chance of being an unrelated background star, respectively (see Sect.~\ref{sec:background}).} 
\label{fig:sensitivity}}
\end{figure*}
The separation and brightness distribution of the discovered binary companions relative to our sensitivity limits (Fig.~\ref{fig:sensitivity}) suggests that observations are better than 90\% complete for binary companions at separations $>$1\farcs0 and brighter than $\Delta K\approx8$\,mag.
For reference, Fig.~\ref{fig:sensitivity} shows all literature binary companions detected in the Taurus region within the same separation limits 0\farcs05--12\arcsec\ \citep{har12,dah11,kra11,duc10,luh10,reb10,tod10,kra09b,kra09a,kra07,ito08,cor06,duc04,cut03,whi01,woi01,kor00,koh98,lei97,lei93,ghe97,ghe93,ghe91,sim96,thi95,har94,mon91}.

\subsection{Masses \& Mass ratios}
Since none of the presently available pre-main sequence evolution models cover the large range of spectral types of the binary components in Taurus, masses are estimated from a combination of model isochrones ($T_\mathrm{eff}<2700$\,K: \emph{DUSTY}, \citealt{cha00}; $2700\,\mathrm{K} \le T_\mathrm{eff} < 3400$\,K: \emph{BCAH98}, \citealt{bar98}; T$_\mathrm{eff}\ge3400$\,K: \citealt{sie00}). Ages assumed for the individual subsamples were 2\,Myr for the confirmed \emph{young members} sample and 20\,Myr for the \emph{extended sample}. 
Since we cannot exclude the possibility that a few targets of the \emph{extended sample} are actually very young $\ll$10\,Myr (see Appendix~\ref{sec:A1}), part of the inferred masses may be overestimated, because the dependence of spectral type depends only weakly on effective temperature which itself is higher at young ages for a star of the same mass. 

Primary masses were derived directly from their effective temperatures, which were inferred from spectral types according to the transformations in \citet[SpT later than M0]{luh03} and \citet[earlier SpT]{sch82}. If no individual component spectral types were known, system spectral types were assigned to the primary assuming that the most massive component dominates the system luminosity. 
Following the scheme in \citet{laf08}, secondary masses were derived from the \emph{model} magnitude of the respective primary according to its assumed spectral type and the magnitude difference between the primary and the secondary, if known. Under the assumption that extinctions are similar for both components of a binary, this is expected to return an extinction-independent model magnitude that can be converted to mass, again with the same isochrones. It has been shown that the inferred mass ratios $q=M_\mathrm{B}/M_\mathrm{A}$ do not depend strongly on the used pre-main sequence models \citep{cor13}, suggesting that the derived values are reliable. Calculated primary and secondary masses are listed in Tab.~\ref{tab2}.

\section{Results}

\subsection{Are the companion candidates physically bound?\label{sec:background}}
In a sample of 64 observed Taurus members, we found a total of 74 companion candidates (28 in the young and 46 in the extended sample) to 40 stars in the separation range of $\sim$0\farcs06 to 12\arcsec\ and within the sensitivity limits shown in Figure~\ref{fig:sensitivity}. Of the detected companion candidates, 40 were previously known and 34 are new discoveries. Since we have no photometry in other filters and second epoch observations for only a fraction of the new companion candidates (see Sect.~\ref{sec:epoch}), we evaluate the probability that a particular companion candidate is physically bound from the density of background stars in the Taurus region.
{The procedure closely follows the analysis presented in \citet{laf14}, who use the number of expected chance alignments with unrelated background stars $E_\mathrm{det}$ to quantify whether a source is likely bound or not. $E_\mathrm{det}$ is calculated as the sum over all target stars of the number of expected background sources that are as bright or brighter and within the separation of the detected companion. The number of background stars is estimated from the density of objects in the 2MASS point source catalog \citep{cut03} within 15\arcmin\ from each star using Poisson statistics. Since many of our detections are fainter than the limiting magnitude of the 2MASS catalog of K$\approx$15\,mag, we extrapolate the almost linear increase of sources with magnitudes between K=7 and K=14 to fainter magnitudes. This approach is validated by the fact that the number $N_\mathrm{det}$ of actual detections of similar sources in all our images is always of the same order as $E_\mathrm{det}$. Both values are presented in Table~\ref{tab2}.

We consider all sources with $E_\mathrm{det}\le0.05$ as bound, which equals to a probability of 95\% that they are indeed bound, according to Poisson statistics. Similarly, we classify all sources with $E_\mathrm{det}\ge3$ as background with 95\% probability that this is true. The status of sources with $0.03<E_\mathrm{det}<3$ cannot be decided statistically and follow-up observations are required (see next Sect.~\ref{sec:epoch}). The statistically classified number of sources in each category bound:unbound:inconclusive is 22:36:16.}

\subsection{Common Proper Motion and Colors\label{sec:epoch}}
\begin{figure*}
\includegraphics[angle=0,scale=.32]{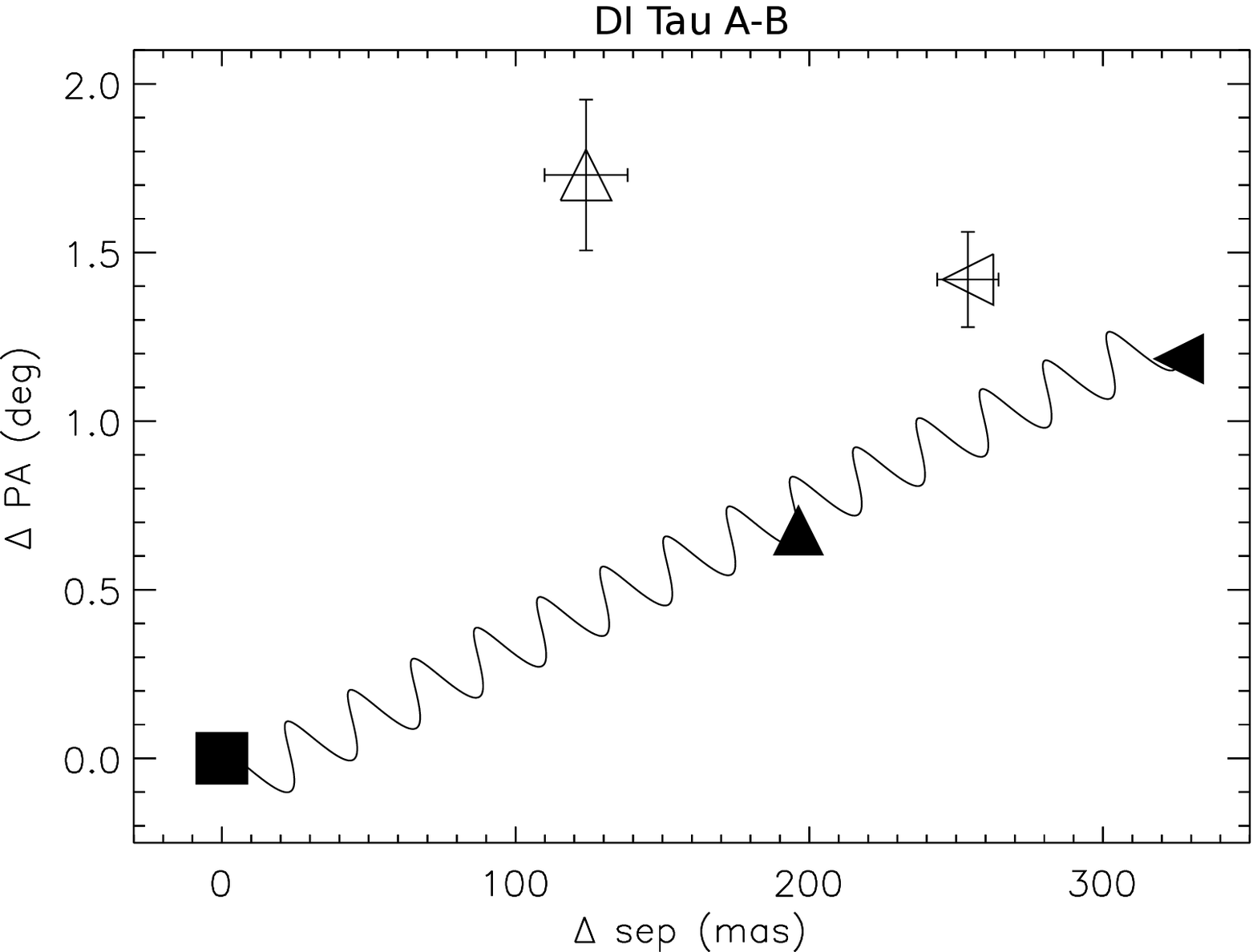} 
\hfill\includegraphics[angle=0,scale=.32]{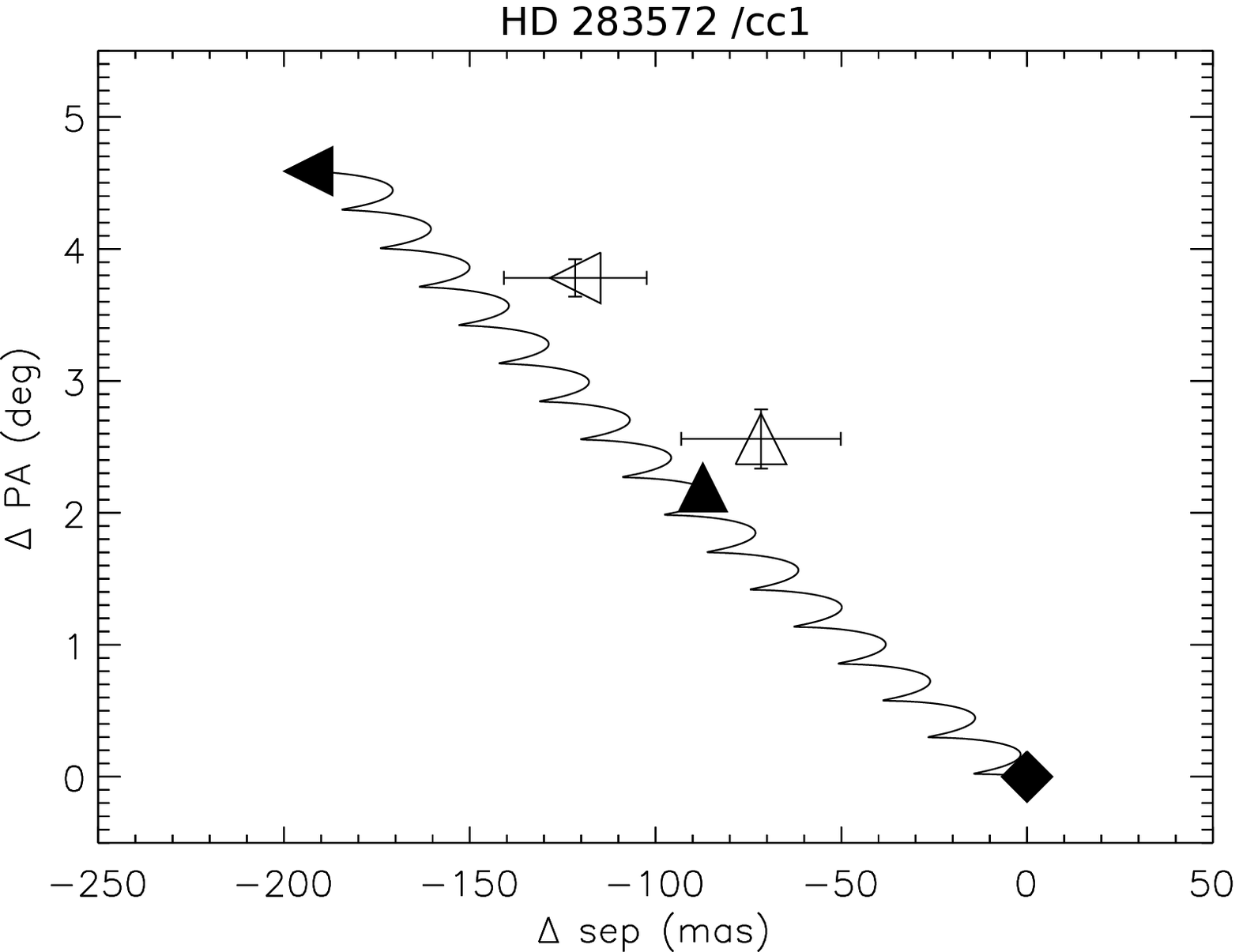} 
\hfill\includegraphics[angle=0,scale=.32]{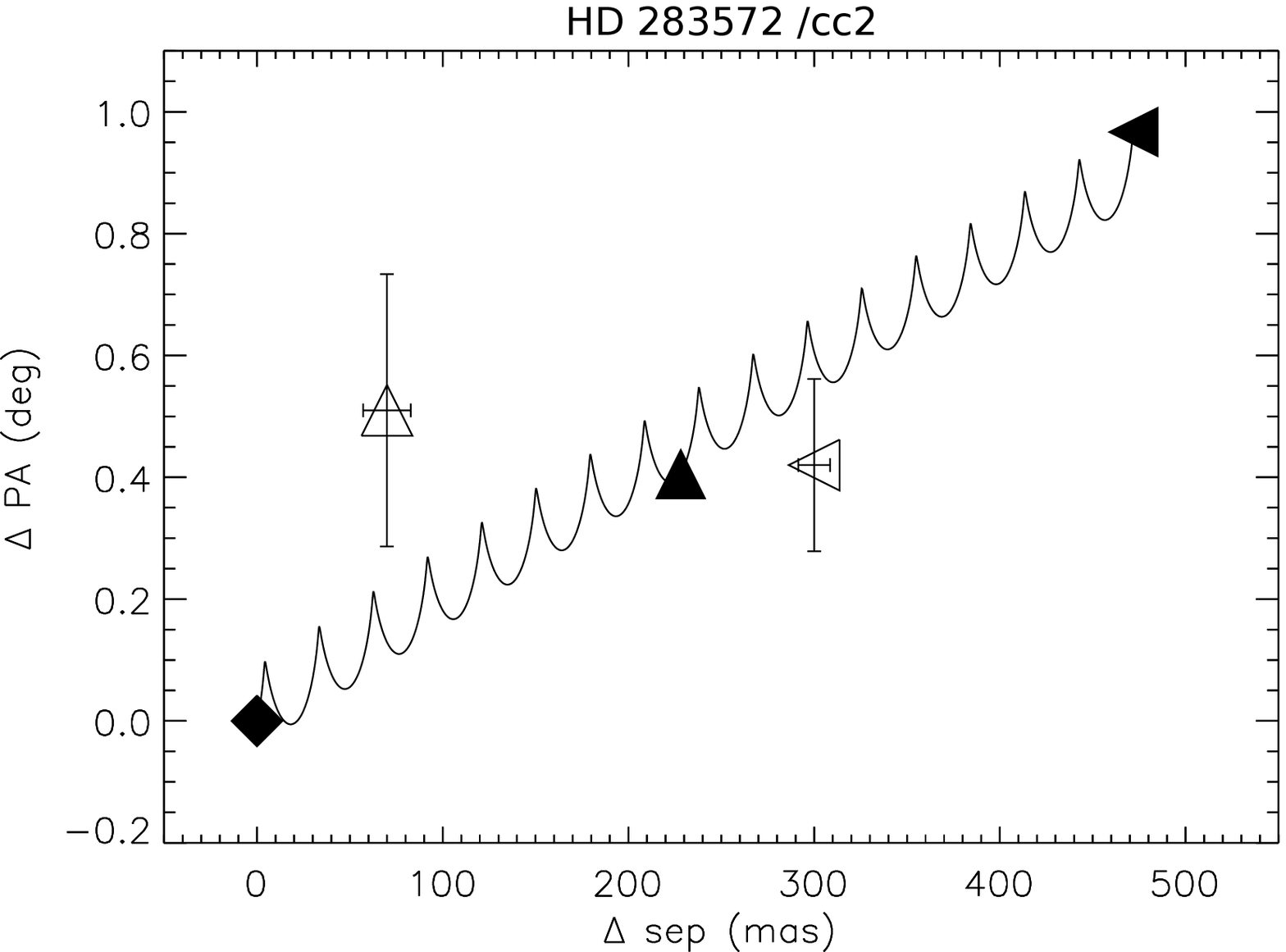}\\[0.4cm]
\includegraphics[angle=0,scale=.32]{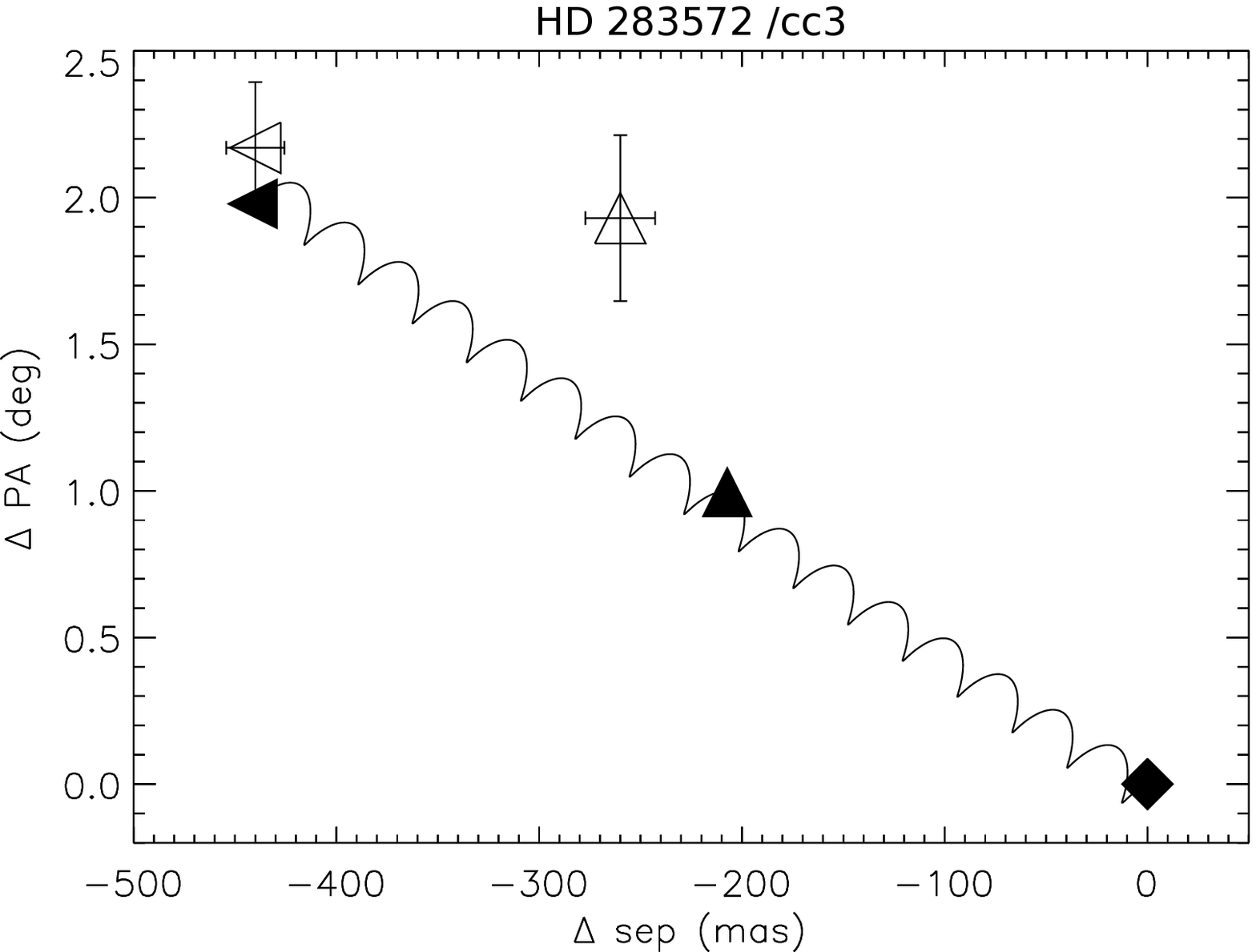} 
\hfill\includegraphics[angle=0,scale=.32]{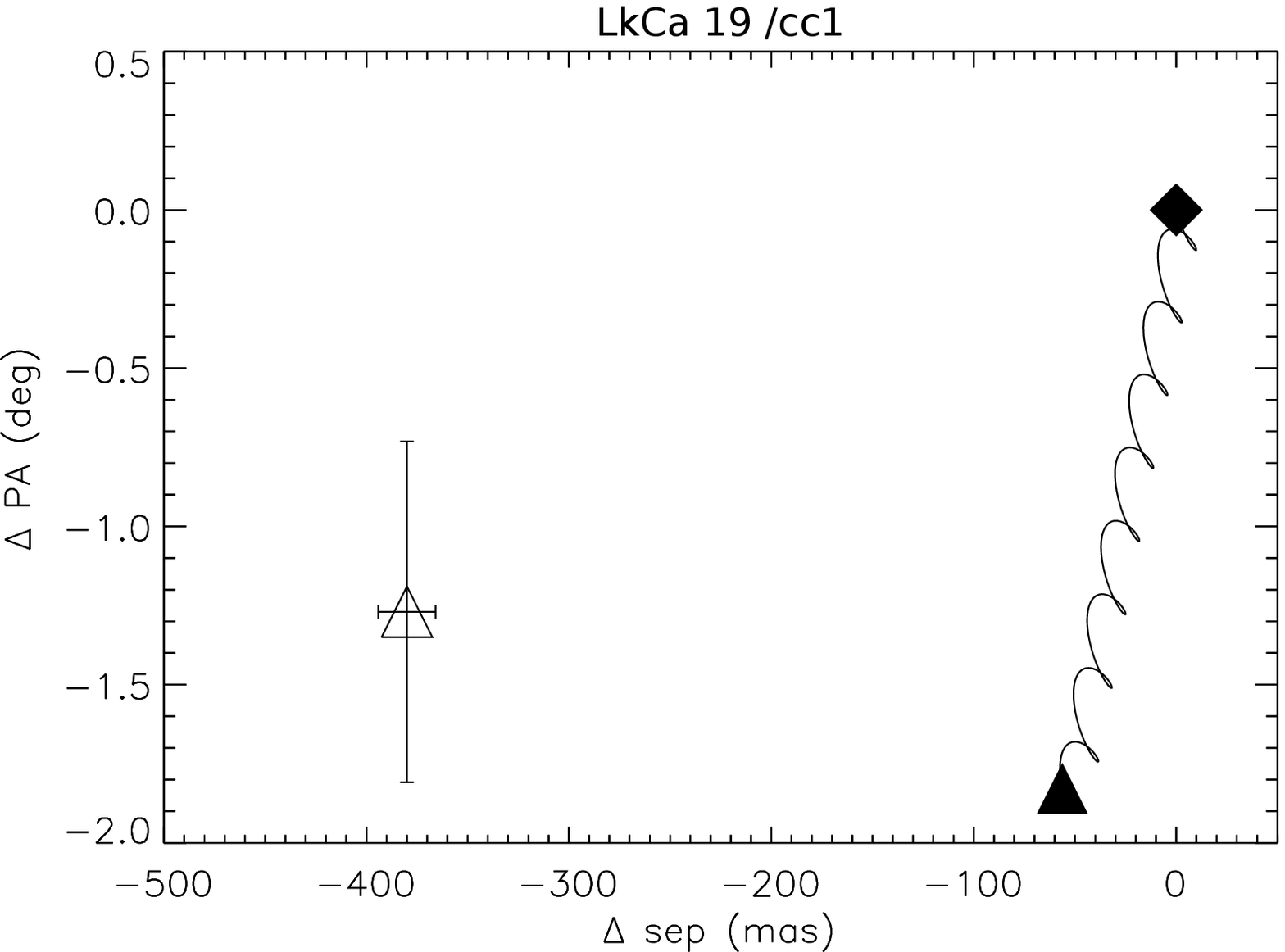} 
\hfill\includegraphics[angle=0,scale=.32]{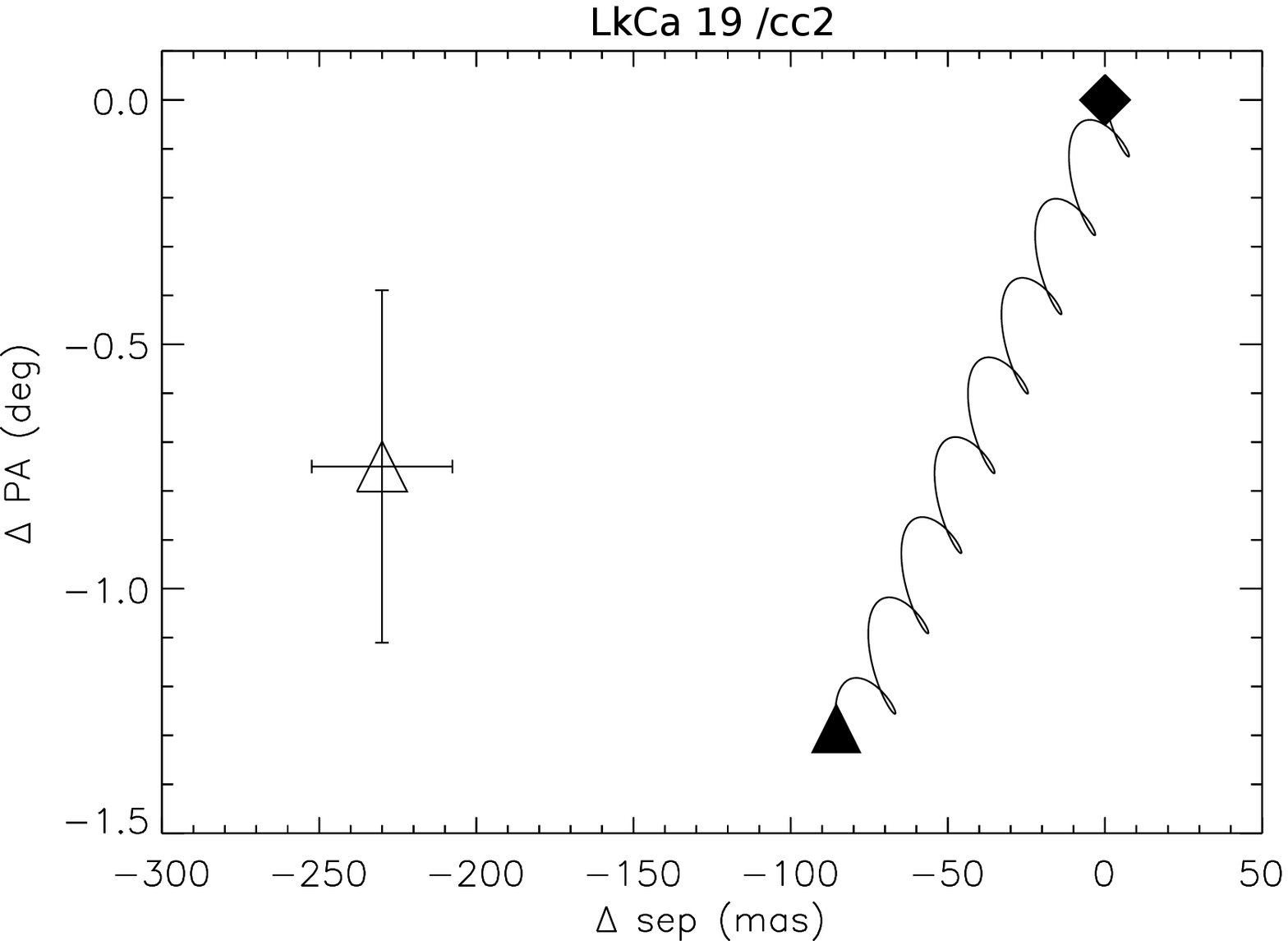}\\[0.4cm]
\includegraphics[angle=0,scale=.32]{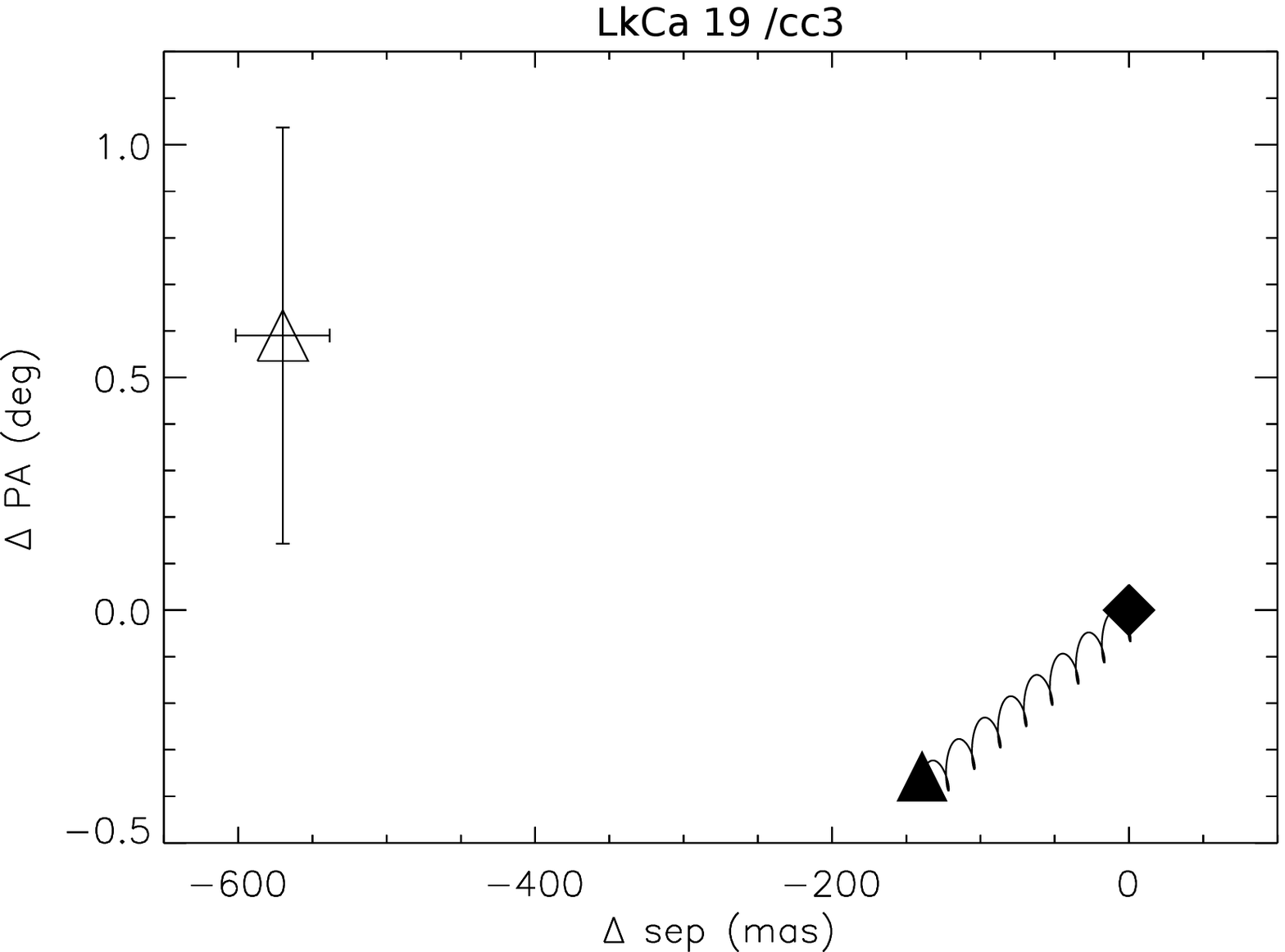} 
\hfill\includegraphics[angle=0,scale=.32]{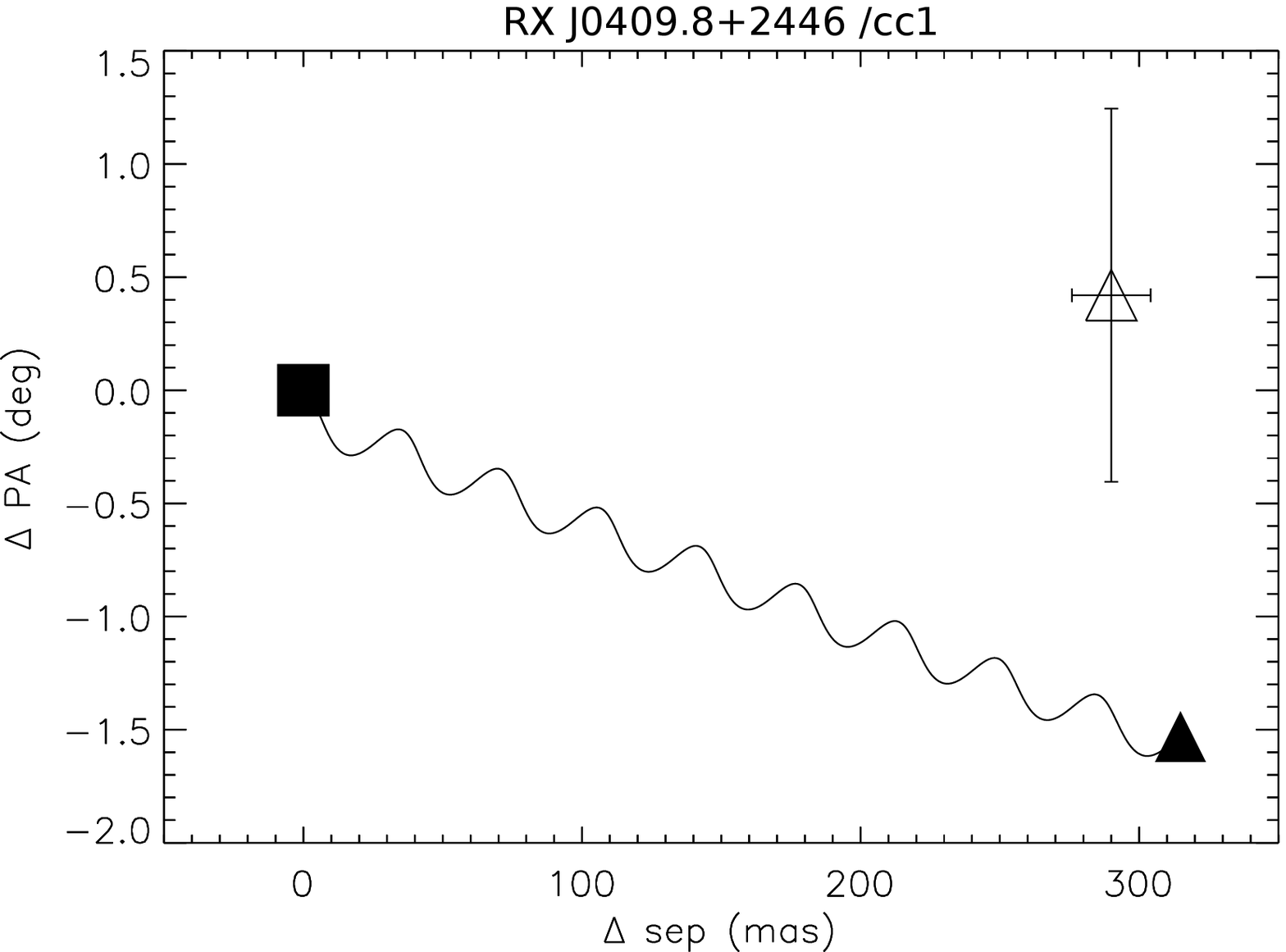} 
\hfill\includegraphics[angle=0,scale=.32]{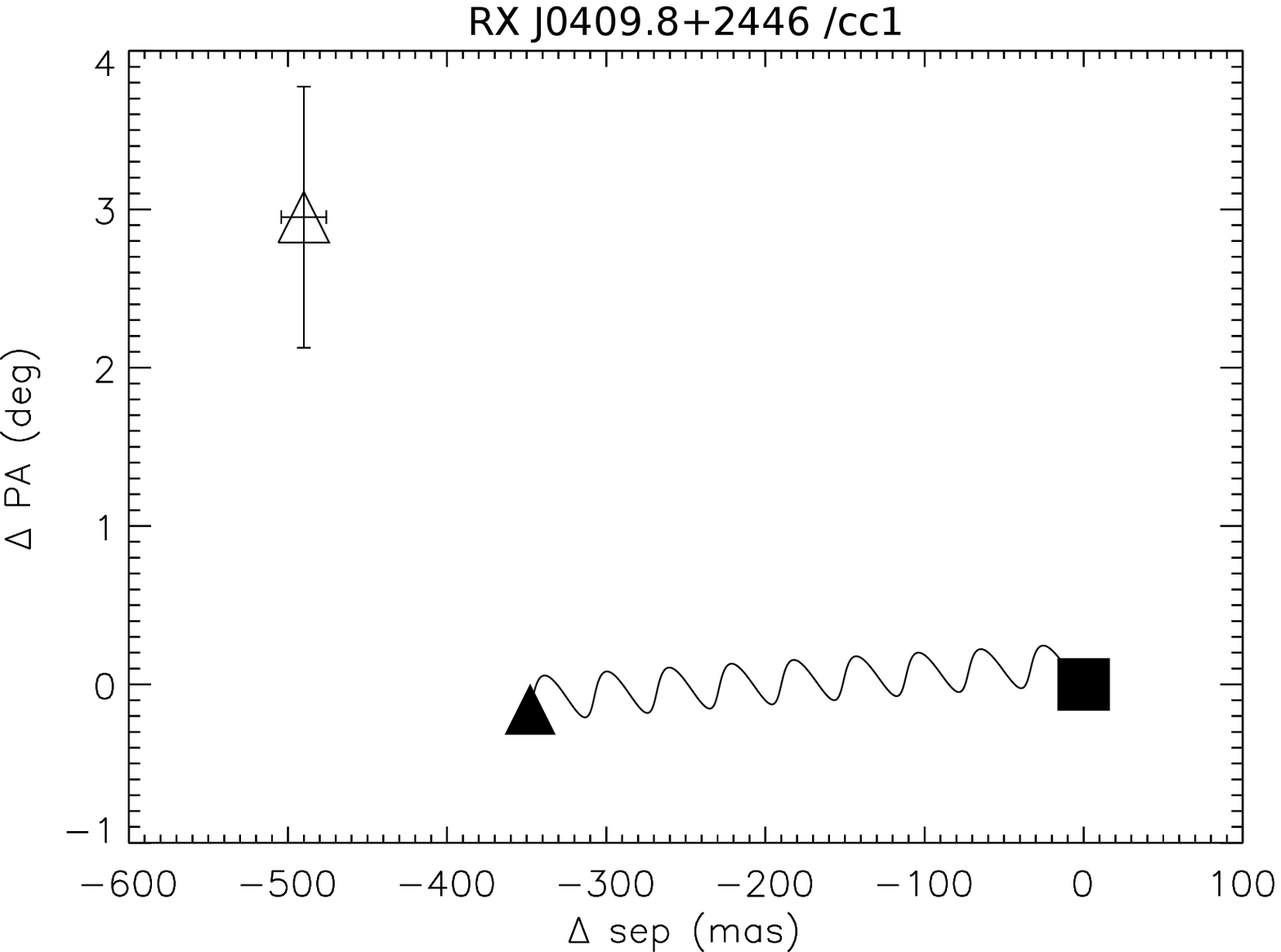}
\caption{Evaluation of common proper motion for targets in our sample that were also observed by \citet{ito08}. Each panel shows the relative change of the separation and position angle of a companion between a previous observations ($\triangle$: \citealt{ito08} $H$-band, $\triangleleft$: HST) with respect to our observations ($\diamond$: 2011B, $\blacksquare$: 2012B). The continuous line shows the motion of a background star and the filled symbols the position of the companion if it was a background star. 
If the star and companion are co-moving, the open symbol with error bars should be close to the filled symbol in the origin. If the companion candidate does not move with respect to the background, then the open symbol is consistent with the location of the identical filled symbol.
\label{fig:CPM}}
\end{figure*}

Fig.~\ref{fig:sensitivity} illustrates that {most} detections with $\Delta K\gtrsim5$\,mag have low probabilities of being bound. A few of these (a total of 10 companions to DI Tau, HD~283572, LkCa~19, RX~J0409.8+2446, RX~J0420.8+3009) were previously observed with deep $H$-band AO imaging by \citet{ito08} and both relative proper motion and color information can be used to assess their nature. Furthermore, ten target stars were re-observed with a time lapse of $\sim$1 or 2 years as part of our survey. If the companions are found to be co-moving with their parent stars, physical association can be assumed. Proper motion checks for all companion candidates with multiple-epoch observations are shown in Appendix~\ref{sec:A2}, and the comparison with data from \citeauthor{ito08} in Fig.~\ref{fig:CPM}.

We find that almost half of the companion candidates with second epoch observations are likely real companions. These are mostly bright close-in companions that also exhibit a high probability of being bound {according to their $E_\mathrm{det}$ value}. Some of these show evidence for orbital motion. We detect four companion candidates to be consistent with being background objects. The rest are inconclusive, either because the uncertainties of the measurements are larger than the proper motion (e.g., DQ~Tau), or because they are neither consistent with being background nor co-moving. This suggests significant motion relative to both the background and the respective target star, which can be the case if the faint source is a foreground star or a high-relative-motion member of the Taurus cloud. 

The mostly inconclusive results in Fig.~\ref{fig:CPM}, when combining the observations from this survey with $H$-band adaptive optics and HST observations from \citet{ito08} lead us to suspect that systematic offsets of relative position angles and/or separations wash out possible signatures. These may be partly caused by the insufficiently constrained distortion correction of NIRI. An attempt to support the proper motion measurements with $H$$-$$K$ colors failed due to large systematic uncertainties originating from possible extinction through an edge-on disk, foreground nebulosity, or photometric variability.
Additional attempts to use the near-infrared colors of widely-separated companions that have been cataloged by the UKIDSS \citep{law07} and SDSS\footnote{\url{http://www.sdss3.org/}} surveys did not provide further indications for or against physical association of the companion candidates.
 Since both the currently available astrometric and photometric measurements suffer from significant systematic uncertainties, new simultaneous observations in different NIR bands will be required to confirm physical association of the faint companion candidates. 
 
Taking into account all of the available evidence, we find that 25 of the 74 companion candidates are likely physically bound to their primaries -- confirmed either by proper motion or {$E_\mathrm{det}\le0.05$} -- while 45 other candidates still need confirmation through future measurements. The remaining four could be ruled out as likely background objects (Appendix~\ref{sec:A2}). Two of the confirmed targets are likely of sub-stellar nature. The interesting faint, co-moving sub-stellar companion at a separation of $\sim$3\farcs7 from HD~284149 is described in detail in a separate publication \citep{bon14}. Companion candidate RX\,J0444.9+2717/cc4 will be part of further investigations.

\subsection{Multiplicity Fractions}
We find 25 companions with a high chance of being bound around 21 of our 64 observed Taurus members. This is equivalent to a raw multiplicity fraction of $M\!F_\mathrm{raw}  = 32.8^{+6.3}_{-5.2}$\%. The sensitivity for the detection of faint companions, however, is variable between the stars due to different target brightnesses and variable quality of the AO correction. More meaningful quantities can be derived using only those companions between 10 and 1500\,AU that could have been detected around 90\% of the stars and taking into account only those stars whose sensitivity curve is equal to or better than that of 90\% of the targets. We further base our statistics only on those companions with either statistical or other evidence that they are bound. The resulting multiplicity fractions are lower limits, as it is likely that some of the inconclusive companion candidates may eventually turn out to be real companions.

Among 57 stars meeting the above criteria, we find 14 binaries and 1 triple system with a high probability of being physically bound. This is equivalent to a multiplicity fraction of $M\!F=\frac{\mathrm{B}+\mathrm{T}+\dots}{\mathrm{S}+\mathrm{B}+\mathrm{T}+\dots}=26.3^{+6.6}_{-4.9}$\% assuming Binomial uncertainties and stating 68\% confidence intervals. The companion star fraction is $CSF=\frac{\mathrm{B}+2\mathrm{T}+3\mathrm{Q}+\dots}{\mathrm{S}+\mathrm{B}+\mathrm{T}+\dots}=28.1^{+6.0}_{-5.3}$\% assuming Poisson statistics. When separating into the young and extended subsamples, multiplicity fractions are $M\!F(\mathrm{young})=16.1^{+8.7}_{-4.5}$\%, $M\!F(\mathrm{extended})=38.5^{+10.0}_{-8.4}$\%.
The difference between the two subsamples is consistent with the fact that the extended sample has a significantly larger fraction of stars with masses $>$1\,M$_\odot$ ($(N_{\!M>1M_\odot}$/$N_{\!M<1M_\odot})_{\mathrm{extended}}=3$) than the sample of young stars ($(N_{\!M>1M_\odot}$/$N_{\!M<1M_\odot})_{\mathrm{young}}=0.6$) and that we detect a positive correlation between multiplicity fraction and mass (Sect.~\ref{sec:BF_vs_mass}).

\subsubsection{The Frequency of Triples and Higher Order Multiples}
Based on our statistics and proper motion measurements, we find only one bona fide triple star, RX\,J0420.8+3009, in our sample. As part of a hierarchical system with a projected separation ratio $\mathrm{sep}_2/\mathrm{sep}_1\sim$30, dynamical stability is implied. No higher-order systems with four or more components with a high probability of being bound were detected within the 90\% sensitivity limits of the survey. Thus we derive a triple star frequency $T\!F = \frac{\mathrm{T}+\mathrm{Q}+\dots}{\mathrm{S}+\mathrm{B}+\mathrm{T}+\dots} = 1.8^{+4.2}_{-1.5}\%$.
This estimate is a lower limit to the true higher-order multiple star frequency of Taurus members, because a) this number does not take into account the finite probability that any additional candidates {that we detected with $E_\mathrm{det}>3$} are in fact real, b) any of the observed components can turn out to be close multiples, and c) a binary detected here may be accompanied by a wide tertiary component outside the survey limits.

A correction for the close and wide tertiary companion biases cannot be accomplished with high confidence, since not all stars in our survey have been observed for the existence of spectroscopic or very distant companions. While consultation of the spectroscopic survey of \citet{ngu12} reveals that at least five of the stars in the survey can be associated with spectroscopic binarity, the overlap between their survey and the one presented here is not complete.

\subsection{Mass ratio and separation distributions\label{sec:dists}}
\begin{figure*}
\includegraphics[angle=0,scale=.4]{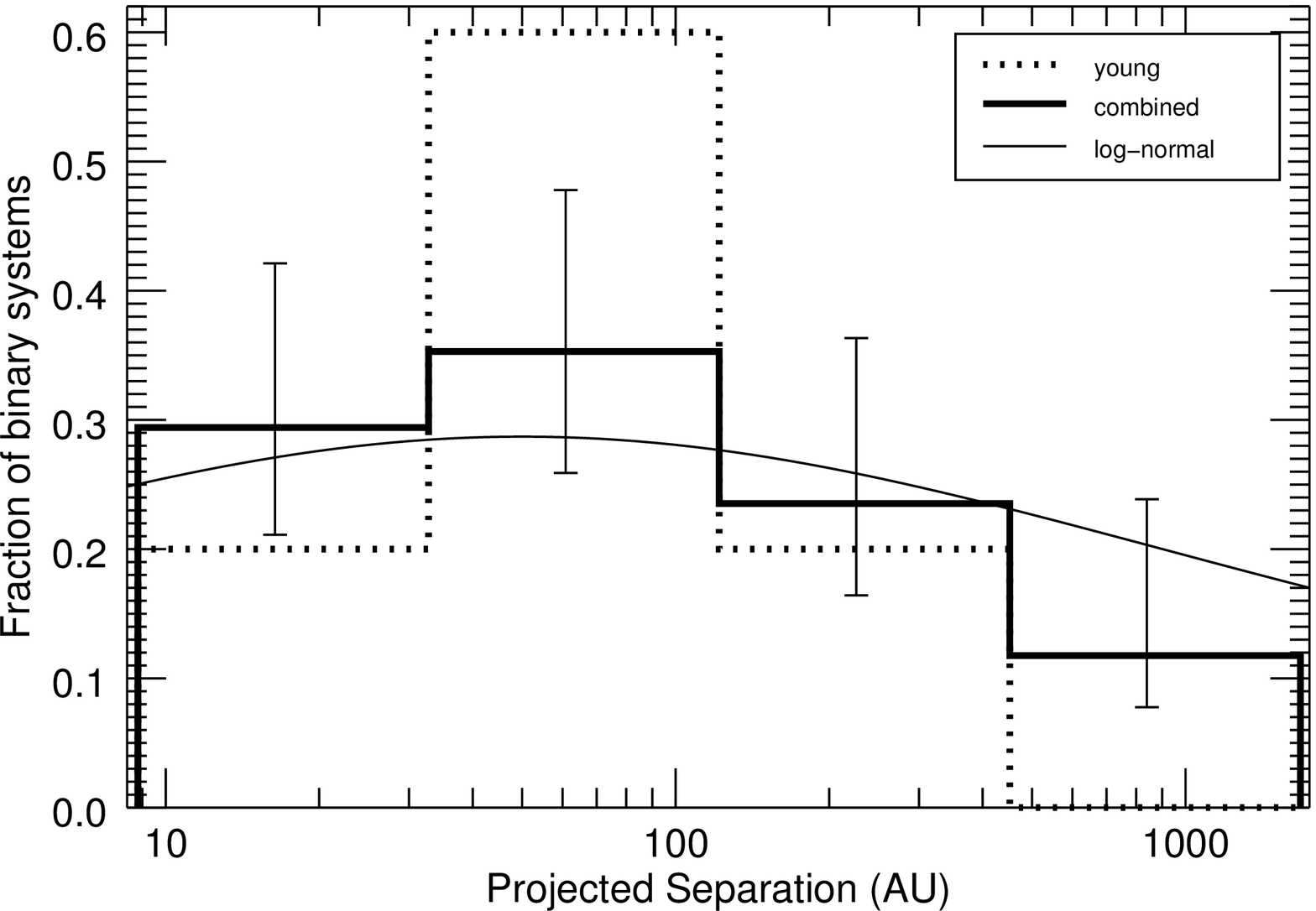}
\hfill\includegraphics[angle=0,scale=.4]{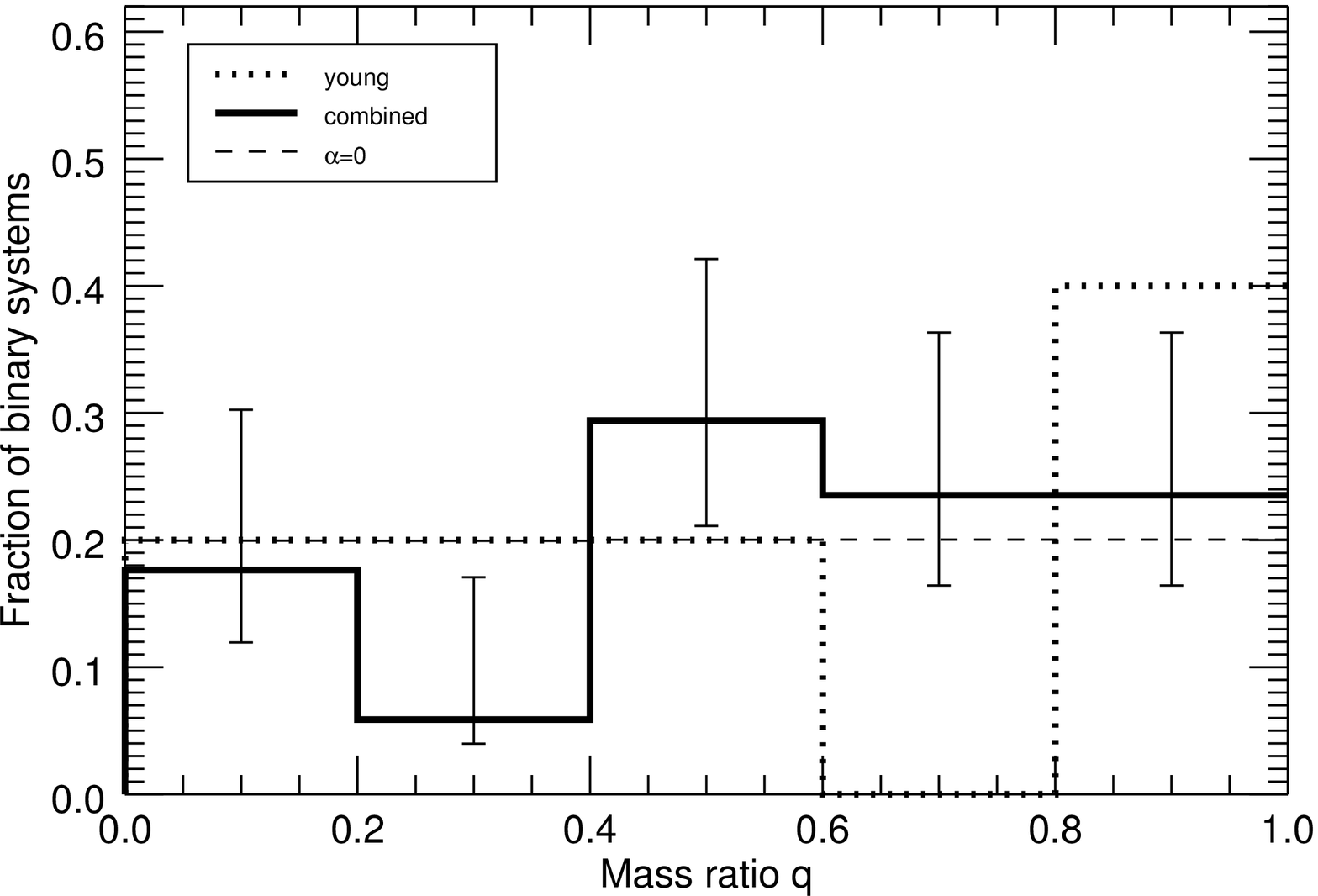}
\caption{Measured separation distribution (left) and mass ratio distribution (right) for the young and combined subsamples. Overplot are the theoretically assumed power-law function ($\alpha$=0 for linear-flat) and the log-normal separation distribution of solar-type field stars by \citet{rag10}.\label{fig:dists}}
\end{figure*}

Fig.~\ref{fig:dists} demonstrates that our data are compatible with the same binary parameter distributions as field stars, i.e., a flat mass ratio distribution and a log-normal separation distribution \citep{rag10}.
While Kolmogorov-Smirnov \mbox{(K-S)} tests suggest that the assumed and measured distributions cannot be distinguished ($P_\rho$=98.1\%, $P_\mathrm{q}$$>$99.99\%), we caution that a log-normal separation distribution as seen in the field might not accurately represent young stars because dynamical evolution can change separations on timescales of up to 100\,Myr \citep{rei12}. 
An alternatively suggested log-flat separation distribution of companions in Taurus \citep{kra11,kra12a} cannot be ruled out by our data. We find, however, a comparably low \mbox{K-S} probability of $P_\rho=0.46$ for this scenario. Since the \citeauthor{kra11} results were derived from shallower photometric data and with a different binary separation coverage, a direct comparison is not straight-forward and reanalysis of their data is beyond the scope of this study. Accordingly we will assume in the following a log-normal separation distribution like the one observed for the field, as it appears statistically more likely and in general more physical than a log-flat distribution.

\section{Discussion}
\subsection{Multiplicity in Taurus\label{sec:mult}}
The young age and low stellar density of the Taurus region implies that it is in a dynamically young state and its binary population is closer to being primordial than denser regions like the Orion Nebula Cluster (ONC) or the dynamically strongly processed population of field stars. Thus, measuring binary parameters in Taurus can shed light on the binary population as it is generated during the embedded star-formation phases before external influences become significant or even wash out the primordial distribution altogether.

The derived multiplicity frequency of $M\!F=26.3^{+6.6}_{-4.9}\%$ is in line with previous multiplicity studies which consistently find that Taurus has a larger binary population than the field and other star-forming regions \citep{lei93,pet98,duc99,kra11}. {This can be seen when comparing with the large survey of Solar-type field stars by \citet{rag10}. When limiting their catalog to companions between 10 and 1500\,AU, we find a multiplicity fraction of 27$\pm$2\% for field stars with masses between 0.7 and 1.4\,$M_\odot$. This is a factor of 1.63$\pm$0.10 smaller than the overall field-star MF 44$\pm$2\% in the full separation range between $\sim$0.01\,AU and $\sim$10$^5$\,AU covered by the survey. We use this factor as a completeness correction for our Taurus sample, under the assumption that field-like mass ratio and separation distributions apply (Sect.~\ref{sec:dists}). In the identical mass and separation range, we find that 38$^{+9}_{-8}$\% of all stars in our Taurus sample are multiple, $\sim$1$\sigma$ larger than in the field. This corrects to a completeness-corrected multiplicity fraction of $\sim$62$\pm$14\% for Taurus stars between 0.7 and 1.4\,$M_\odot$.%

This indicates that the total multiplicity fraction of Taurus is larger than that of field stars in the same mass and separation range. This is likely due to dynamical processing in dense clusters and agrees well with the proposal that most field stars were born in dense regions like the ONC \citep{lad03}. Our findings are in agreement with previous surveys by, e.g., \citet{kra11}, who find a multiplicity frequency of $2/3$--$3/4$. The fact that our deeper observations return a similar multiplicity frequency is due to two factors. On one hand, we could not yet confirm the physical nature of a large number of our candidate sub-stellar companions. On the other, the abundance of low-mass companions is very small, as further discussed in Sect.~\ref{sec:comp}.}

In the process of star formation not only binaries form frequently, but a significant fraction of the emergent systems consists of three or more components. While there exists no good theory of how many triples, quadruples, etc., must form as a result of core fragmentation, observations of field binaries show that multiples with $n$ (=2,3,4,5,...) components are on average four to five times as abundant as multiples of the next higher level $n+1$ \citep{egg08}. With respect to this finding, the present observations of Taurus show a relatively low abundance of triple stars of $f_3$=T/B=1/14=$0.07_{-0.02}^{+0.13}$. Rather than pointing to a physical property of Taurus' multiples, this value is likely a consequence of the hierarchical nature of stable multiple systems: since all detected stellar companions must have separations within the survey's sensitivity limits, and most stable configurations require a ratio of semi-major axes of at least 3--4 \citep{har72}, there is a significant chance that either the close pair or wide tertiary remain undetected. This conclusion is supported by the fact that the frequency of triple and higher-order multiples among field stars of $\sim$11\% \citep{rag10} reduces to $2.6\pm0.8$\% when limiting separations to 10--1500\,AU, equivalent to a reduction of $f_3$ from $0.27^{+3.8}_{-3.2}$ to $0.10^{+3.6}_{-2.1}$. We accordingly find no significant difference between the abundance of triple stars in Taurus and the field.

\subsection{Multiplicity as a Function of Mass\label{sec:BF_vs_mass}}
Previous observations of the multiplicity fraction of stars on, e.g., the Main Sequence, often revealed a small number of binaries with low-mass primaries compared to higher-mass systems \citep{rag10,jan12}. 
If mostly due to dynamical processing, this signature should be less pronounced in Taurus with its short lifetime compared to dynamical timescales and low stellar density.

\begin{figure}
\epsscale{1.15}
\plotone{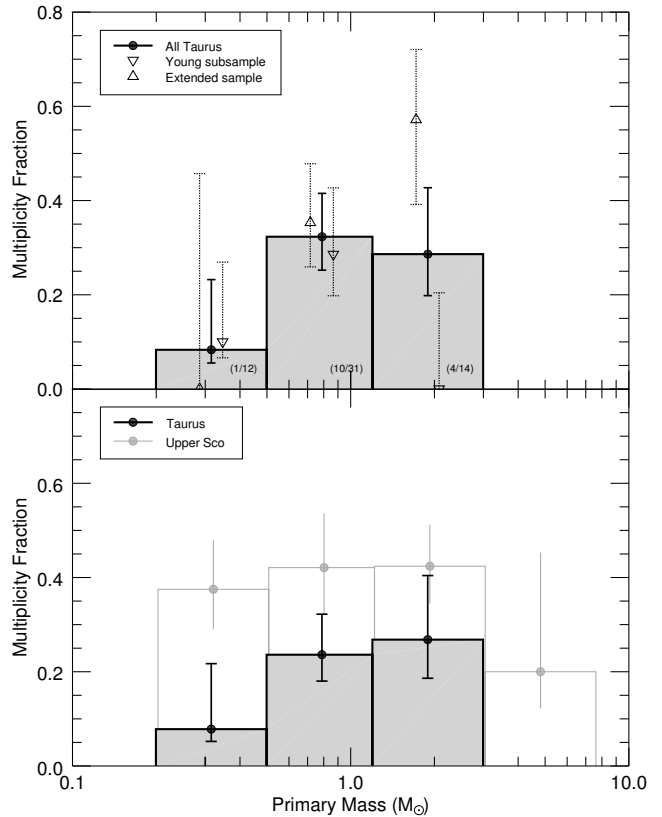}
\caption{{Top: Multiplicity fraction as a function of primary mass in three logarithmic mass bins for the combined sample of Taurus members (shaded histogram) and the young and extended subsample (Triangles). Numbers are derived within the 90\% completeness limits for separations $<$1500\,AU. The numbers in parentheses show the number of binaries and total number of stars in each bin. Bottom: Multiplicity frequency in Taurus compared to a similar study in the Upper Scorpius region \citep{laf14}. A uniform sensitivity was assumed which is equal to the 90\%-complete curve in \citet{laf14}, including only companions within $<$5\arcsec. To minimize uncertainties introduced by stellar evolution models, Upper Scorpius data were reanalyzed with masses derived after the current scheme using 5\,Myr isochrones.}
\label{fig:BF_vs_mass}}
\end{figure}

The top panel in Fig.~\ref{fig:BF_vs_mass} shows the multiplicity fraction divided into three equal logarithmic primary mass bins between 0.2 and 3\,$M_\odot$. 
Fitting a linear correlation to the data points we find a slope of $\alpha$=0.26$\pm$0.19, i.e., $\sim$1$\sigma$ evidence for a positive correlation of multiplicity frequency with primary mass. A positive slope is particularly evident ($\sim$3$\sigma$) for the extended subsample with $\alpha$=0.73$\pm$0.24 while young Taurus members show no such correlation ($\alpha$=$-0.12$$\pm$0.12). The positive slope for the slightly older systems in Taurus agrees with the correlation of multiplicity frequency with mass seen in previous observations of the field \citep{duc13} and cluster formation simulations \citep{bat12,kru12}. For example field stars show an increase of multiplicity from about 20--30\% close to the Brown Dwarf boundary to $>$50--60\% at 2--10\,M$_\odot$ \citep{duc13} which is consistent with a slope $\alpha_\mathrm{field}\approx0.37$ between 0.1 and 2.7\,M$_\odot$. It is accordingly possible that we see an indication for evolution from a flatter distribution at young ages $\ll$10\,Myr to a steeper curve at older ages which would be in agreement with dynamical evolution destroying preferably low-mass binaries due to their lower binding energy. 
A comparison between the current study and previous results in the field must, however, be regarded with caution as it is possible that binaries with separations outside our sensitivity range follow a different multiplicity--mass relation.

Nevertheless, no strong biases should exist when comparing our young and extended samples. The possibility that our lowest mass-bin may be biased toward a higher multiplicity frequency in this magnitude-limited survey does not significantly change the slope difference even if all tentative companion candidates turn out to be physically bound. 
It thus appears that we see significant evolution of binary characteristics during the first 20\,Myr of stellar evolution. This can be caused by the evolution of either the multiplicity frequency through, e.g., ejection of stellar companions, or the separation distribution. Evidence for changes of the latter can neither be detected nor excluded with confidence in our sample: Gaussian fits to the log-separation histograms of young vs.\ extended binaries find a shift of the peak from $-0.34\pm1.93$ to $-0.04\pm1.93$ between the samples and a KS-test returns an insignificant probability of $P$=0.34 that the two distributions are different. The third possibility, significant evolution of mass ratios, is unlikely at this late evolutionary stage where the mass accretion rates from a circumstellar disk are low, typically $\sim$10$^{-10}$--10$^{-7}$\,M$_\odot$\,yr$^{-1}$, and disk lifetimes are limited to a few more Myr \citep{nat06,jay06}.

We perform a quantitative comparison of the multiplicity frequency as a function of mass with other surveys of young star-forming regions by applying uniform sensitivity limits. 
In the bottom panel of Fig.~\ref{fig:BF_vs_mass} we compare our results to observations in the 5\,Myr-old Upper Scorpius star-forming region. The $M\!F$ vs.\ primary mass distribution in both regions is qualitatively similar, but with a slightly smaller ($\lesssim$1\,$\sigma$ significance) multiplicity fraction of Taurus stars in each mass bin. Log-linear slopes in the three populated mass bins are not significantly different ($\alpha_\mathrm{USco}$=0.06$\pm$0.17, $\alpha_\mathrm{Tau}$=0.25$\pm$0.17). \citet{laf14} further find that their Upper Scorpius data are consistent with a previous study of stars in the 2--3\,Myr-old Chamaeleon\,I star-forming region \citep{laf08}. The fact that these distributions of young stars appear relatively flat -- within the current sensitivity limits -- might thus be a universal feature of star formation.

\subsection{Sub-stellar companions in Taurus\label{sec:comp}}
Interestingly, only very few companion candidates were detected with magnitude differences with respect to the primary of 5\,mag\,$<$\,$K$\,$<$\,8\,mag at separations $<$6\arcsec. This magnitude range deserves particular attention, because it is roughly consistent with the sub-stellar mass range between the deuterium and hydrogen burning limit at young ages. Since the magnitude limits of our survey are for the most part significantly fainter than the faintest observed companions at separations $>$0\farcs2, this void is apparently a real feature of the binary population\footnote{To double check our sensitivity to companions in this region of the diagram, we assumed a companion of the brightness of the companion to HD~284149 ($\Delta K=6.3$\,mag) at various separations in the questioned region and found that it would be easily detectable.}. 

{With two candidate brown dwarfs discovered with a high chance of being bound, we calculate a frequency of brown dwarf companions of $M\!F(\mathrm{BD})$=3.5$^{+4.3}_{-1.1}$\% within our 90\% sensitivity limits between effectively $\sim$20--70 and 1500\,AU. As there are a three additional companion candidates with currently unclear status ($0.05\le E_\mathrm{det}\le3$), an upper limit of 8.8$^{+5.2}_{-2.5}$\% can be estimated when assuming all of these turned out to be true companions.
This result is consistent with previous estimates in Taurus \citep[$>$3.9$^{+2.6}_{-1.2}$\% between 5 and 5000\,AU;][]{kra11} as well as young stars in Upper Scorpius with frequencies of 1.8\% and $\lesssim$4.0$^{+3.0}_{-1.2}$\% for sub-stellar companions at separations of 50--250\,AU and 250--1000\,AU respectively \citep{laf14}. Furthermore, \citet{met09} find a brown dwarf companion frequency of $\sim$3.6\% to solar-type field stars between 28 and 1590\,AU. Thus, brown dwarf companions with separations of a up to several hundred AU appear with frequencies of a few percent at all evolutionary stages.}

We want to know whether the low density of brown dwarf and planetary companions is consistent with being drawn from common separation and mass ratio distributions between the stellar and sub-stellar mass regime. We test this by determining whether the observed number of Brown Dwarf companions can be reproduced with the observed separation and mass ratio distributions (Fig.~\ref{fig:dists}), which are based on an almost entirely stellar (except one Brown Dwarf) companion population.
We simulated and analyzed 10\,000 sets of synthetic observations. These were used to determine the frequency of cases where the supposedly under-dense region harbors at most one companion. The individual data sets are generated according to the following algorithm: 1) Draw a random mass from our observed sample (Table~\ref{tab2}). 2) Create a companion with a separation to the host star randomly drawn from the log-normal separation distribution in Fig.~\ref{fig:dists}. 3) Assign a companion mass assuming a linear-flat mass-ratio distribution and the primary star's mass. 4) Repeat 57 times (= the original binary sample size). We find that in 68\% of the cases, the examined region has at most as many stars as were observed here. Accordingly, the observed density is highly compatible with the assumed distributions.

The implications are two-fold. On one hand, we do not find evidence for an extension of the so-called {\em brown-dwarf desert} at wide separations, i.e., a dearth brown dwarf companions below what is expected from extrapolation of the distribution functions of higher-mass companions. The term brown dwarf desert was originally coined to describe the significant lack of close-in ($\lesssim10$\,AU) brown dwarf companions around solar-type stars in radial velocity surveys \citep{mar00} and was suspected to possibly also hold for wider companions. On the other hand, the observed low frequency of brown dwarf companions is consistent with being the low-mass tail of the stellar companion mass distribution. Accordingly their properties require neither a discontinuity of the mass distribution (e.g., a ``break mass'') nor of the underlying dominant formation mechanism of stellar and brown dwarf companions. These results for a young star-forming region such as Taurus appear to be in line with results from the study of much older stars by \citet{met09} who reach similar conclusions from their analysis of solar-type stars in the field.

\section{Conclusions}
We present results from a large, near-infrared high-spatial resolution search for stellar and sub-stellar companions in the Taurus-Auriga star-forming region with the Gemini/NIRI instrument. The sample covers 64 stars, equally distributed with masses between $\sim$0.2 and $\sim$3\,M$_\odot$, to obtain an unbiased view of the multiplicity properties of this young, low-stellar density, and thus dynamically nearly pristine environment. With these data, we were able to measure the frequency of binaries and their orbital separation distribution, and inferred mass ratios shortly after their formation. 

We split our targets into two similar-sized sub-samples, a \emph{young} sample of confirmed members with ages $\sim$2\,Myr and an \emph{extended sample} of stars which are likely members of the Taurus region, but at a possibly older age of $\sim$20\,Myr. Based on statistical arguments and follow-up observations, we find 22 bona fide stellar companions (8 young and 14 extended) to 19 stars and {two likely brown dwarf companions including HD\,284149b \citep{bon14}}. Out of all likely bound new companions, 3 were previously unknown. Another 49 faint companions were detected which, however, have significant probabilities of being background stars. While six of these companions were rendered as likely co-moving with their host stars by second epoch observations, others will require future follow-up observations to confirm or rule out physical association with their host stars.

The inferred properties of multiple stars in Taurus are summarized in the following:

\begin{enumerate}
  \item We find a multiplicity fraction of $M\!F=26.3^{+6.6}_{-4.9}\%$ in the separation range of ~$\sim$10--1500\,AU within our 90\% completeness limits.
  \item We estimate a total multiplicity frequency of $\sim$62$\pm$14\% for stars with primary masses of $M_*$=0.7--1.4\,$M_\odot$ and assuming that the underlying separation and mass ratio distributions are identical to the field. This is comparable to results from previous multiplicity studies in Taurus.
  \item The fraction of Taurus star systems consisting of three or more members is $T\!F=1.8^{+4.2}_{-1.5}$\%, comparable to the field value of solar-type stars in the same separation range.
  \item The multiplicity fraction as a function of primary mass was shown to be qualitatively consistent with that of other young regions such as Upper Scorpius (5\,Myr). When split into the young and extended populations, it appears to support an evolution with age from a flat distribution at $\sim$2\,Myr to a comparably steep function similar to the field at $\gtrsim$20\,Myr. We interpret this as seeing evolution of the binary population in terms of binary separation or number of bound companions within the first $\sim$20\,Myr of cluster evolution.
  \item Brown dwarf companions were found around 3.5$^{+4.3}_{-1.1}$\% of the stars. 
This fraction appears to be consistent with extending the best-fit stellar companion mass ratio (linear-flat) and separation distributions (log-normal) to sub-stellar mass companions. This is consistent with a common dominant formation channel of stellar and sub-stellar companions, i.e., that brown dwarf and stellar companions form alike.
\end{enumerate}

\acknowledgments
We sincerely thank the unknown referee for the effort invested in the review of the manuscript which ultimately led to a significantly improved paper. 
The authors would like to thank James Owen and Thayne Currie for fruitful discussions. This work was supported by grants from the Natural Sciences and Engineering Research Council of Canada and the University of Toronto McLean Award to R.J. 
Based on observations obtained at the Gemini Observatory, which is operated by the Association of Universities for Research in Astronomy, Inc., under a cooperative agreement with the NSF on behalf of the Gemini partnership: the National Science Foundation (United States), the National Research Council (Canada), CONICYT (Chile), the Australian Research Council (Australia), Minist\'{e}rio da Ci\^{e}ncia, Tecnologia e Inova\c{c}\~{a}o (Brazil) and Ministerio de Ciencia, Tecnolog\'{i}a e Innovaci\'{o}n Productiva (Argentina).
This research has made use of the SIMBAD database and VizieR catalogue access tool, operated at CDS, Strasbourg, France.

\clearpage
\appendix
\section{A) Membership}
\label{sec:A1}
\begin{figure}
\centering
\includegraphics[angle=0,scale=.49]{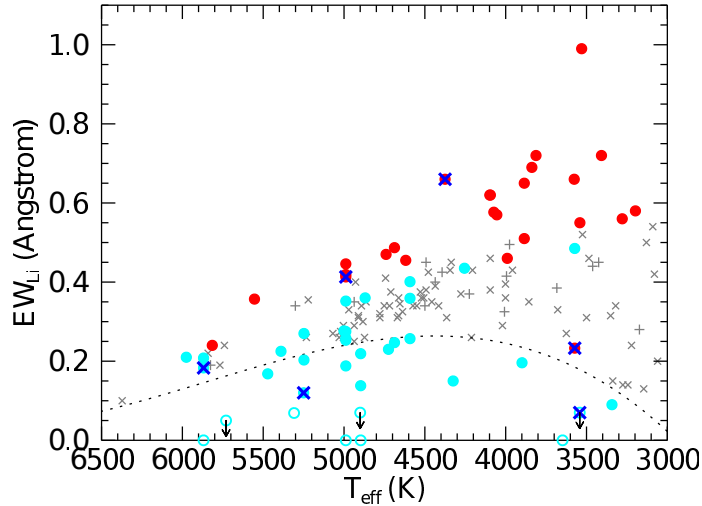}
\caption{Equivalent width of lithium absorption as a function of effective temperature for all targets with available Lithium measurements (circles). Red filled circles are part of the young sample, filled cyan symbols are in the extended sample. Members of the 16--17\,Myr-old Lower Centaurus Crux and Upper Centaurus Lupus region are marked with gray crosses and plus-signs, respectively \citep{son12}. Their lower envelope is sketched with the dotted line. Objects that we classify as significantly depleted (see text) are marked with open symbols. Stars in the proper motion sub-group at pmRA$\approx$20\,mas/yr and pmDEC$\approx$$-$35\,mas/yr in Fig.~\ref{fig:propermotion} are marked with bold blue crosses.\label{fig:lithium}.}
\end{figure}

We use several indicators to assess which of our targets are members of the Taurus region and whether they additionally qualify as being very young. Since primordial circumstellar dust disappears on timescales of $\lesssim$10\,Myr \citep[e.g.,][]{her08}, we use the presence of infrared excess as indication for extreme youth \citep[22 stars;][]{eva09,wah10,reb10}. An age $<$10\,Myr serves as sufficient evidence for membership in the Taurus star-forming region and the thus selected targets are listed in the \emph{young members} sample in Table~\ref{tab1}. This is complemented with another 13 targets that show Lithium absorption stronger than that of the 16--17\,Myr-old Upper Centaurus Lupus and Lower Centaurus Crux regions \citep[][]{son12} at a given effective temperature (Fig.~\ref{fig:lithium}). 

We invoke a second sub-sample, which we call \emph{extended sample}, that contains all stars with evidence for membership in the Taurus association, but no clear evidence for extreme youth. On one hand, this selection comprises targets without sufficient observations to confirm their youth and on the other targets that are likely part of a population which formed in a burst of star formation about $\sim$20\,Myr ago \citep{ses08}. All stars in this category show strong Lithium absorption comparable to, but not stronger than Upper Centaurus Lupus and Lower Centaurus Crux. Since binarity can dilute the lithium signature to 1/2 of its original value (if the relative motion of both binary companions is larger than the line width of Li), we include all stars that have at least half of the minimum Lithium absorption strength found in these clusters (Fig.~\ref{fig:lithium}). This is to prevent biasing our target selection against the inclusion of binary stars. The described criterion leads to the necessary exclusion of eight stars of the original sample that show Lithium absorption incompatible with being young (listed in Tab.~\ref{tab1}). We exclude these from further consideration.

\begin{figure}
\centering
\includegraphics[angle=0,scale=.49]{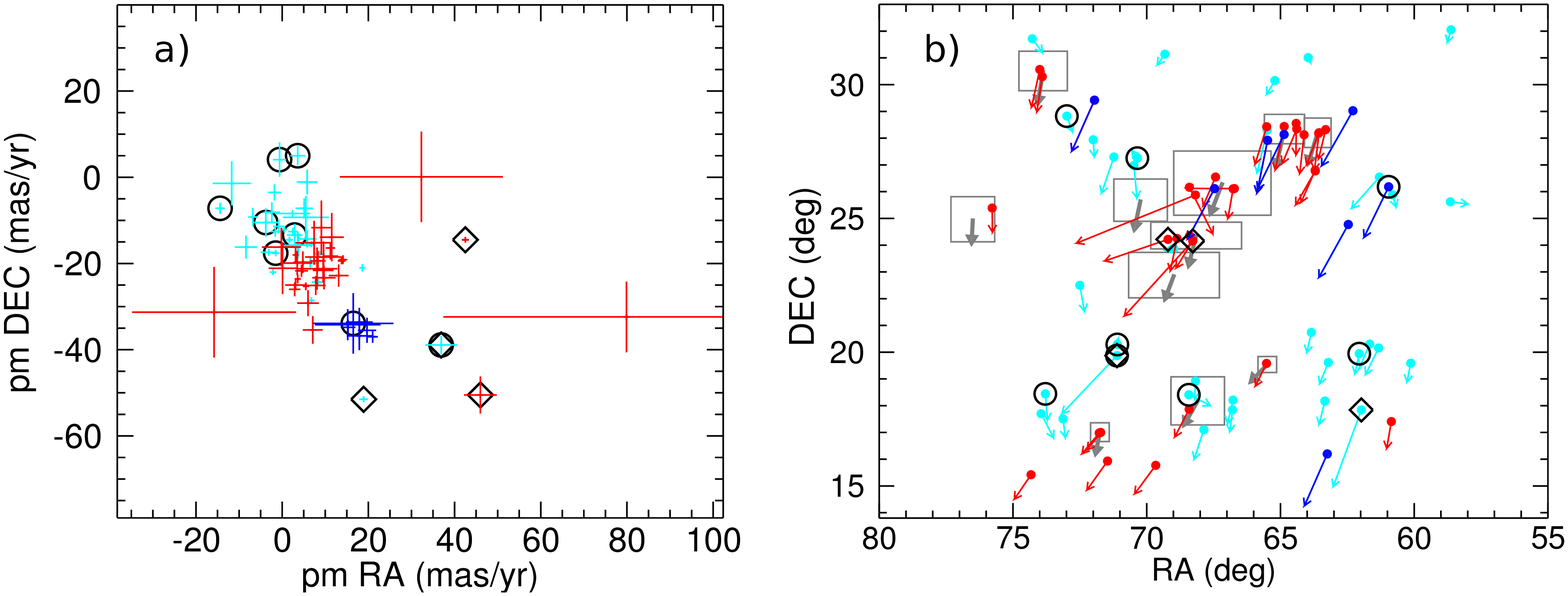}
\caption{{\bf a)} Proper motions and uncertainties in RA and DEC from the UCAC4 catalog \citep{zac13}. Colors as in Fig.~\ref{fig:lithium}, i.e., red and cyan symbols for the young and extended sample, respectively. The proper-motion sub-group of the extended sample at pmRA$\approx$20\,mas/yr and pmDEC$\approx$$-$35\,mas/yr is colored in blue. Lithium outliers from Fig.~\ref{fig:lithium} are circled. The four proper motion outliers are highlighted with a diamond. {\bf b)} Position in the sky and proper motion vector of all targets. Colors and highlights as in a). The gray boxes and bold arrows define the extent and average motion of Taurus subgroups as described in \citet{luh09}. \label{fig:propermotion}}
\end{figure}

As an independent test for membership, we explore proper motions and spatial distribution of the thus selected target stars in Fig.~\ref{fig:propermotion}. We see that the positions of the young sample are strongly clustered around the previously determined average motion of Taurus members of $\mu_\alpha=7.16\pm8.55$\,mas/yr, and $\mu_\delta=-20.91\pm10.31$\,mas/yr \citep{ber06}. The extended sample appears spatially more distributed but mostly compatible with these motions, underlining their likely association with Taurus. We do find four significant proper motion outliers (marked with diamonds) which we, however, do not exclude from the target sample since their peculiar proper motion can be caused by binarity. Interestingly, there appears to be a sub-group of stars at pmRA$\approx$20\,mas/yr and pmDEC$\approx$$-$35\,mas/yr that is clearly separate from the bulk of Taurus members. While this suggests a common origin, these stars do not occupy a distinct location in the Taurus region (Fig.~\ref{fig:propermotion}\,b) and show a large range of lithium equivalent widths (Fig.~\ref{fig:lithium}). As we have no evidence that they do not belong to the Taurus region (their proper motions are compatible with the scatter of proper motions of young Taurus members), we do not exclude them from the \emph{extended} sample and treat them as all other members. The spatial distribution of the target stars shows a strong clustering of the young sample around previously identified regions of recent star formation, which can be associated with the location of dusty filaments in the Taurus-Auriga region \citet{luh09}. The extended population appears more distributed, which is consistent with earlier formation in the Taurus cloud and subsequent dispersal. 
In additional support of membership of the extended sample, some stars of this group have measured Hipparcos parallaxes consistent with the distance of the Taurus cloud (e.g., HD\,284149, 155\,pc; HIP\,20782, 130\,pc or, after reanalysis by \citet{van07}, 108\,pc and 165\,pc, respectively), have no bright companions biasing the photometry, and are inferred to have ages between 15 and $\sim$21 Myr in an HR diagram. We take this as additional evidence for a burst of star formation in Taurus about $\sim$20\,Myr ago that created this group of stars which was followed by another star formation burst 1--2\,Myrs ago to create the young sample.

\section{B) Relative Motion of Candidate Companions}
\begin{figure*}
{\includegraphics[angle=0,scale=.32]{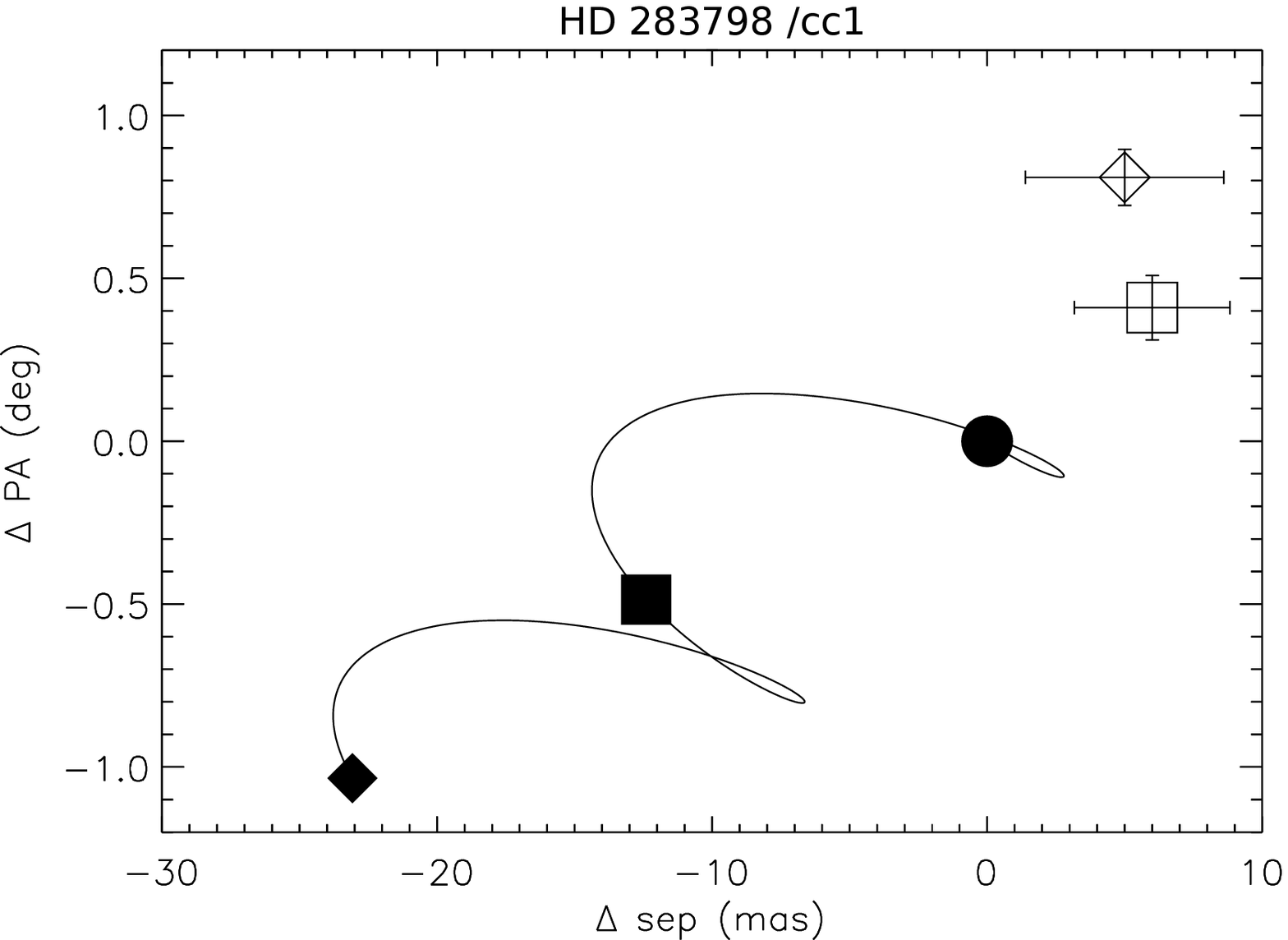}} 
\hfill{\includegraphics[angle=0,scale=.32]{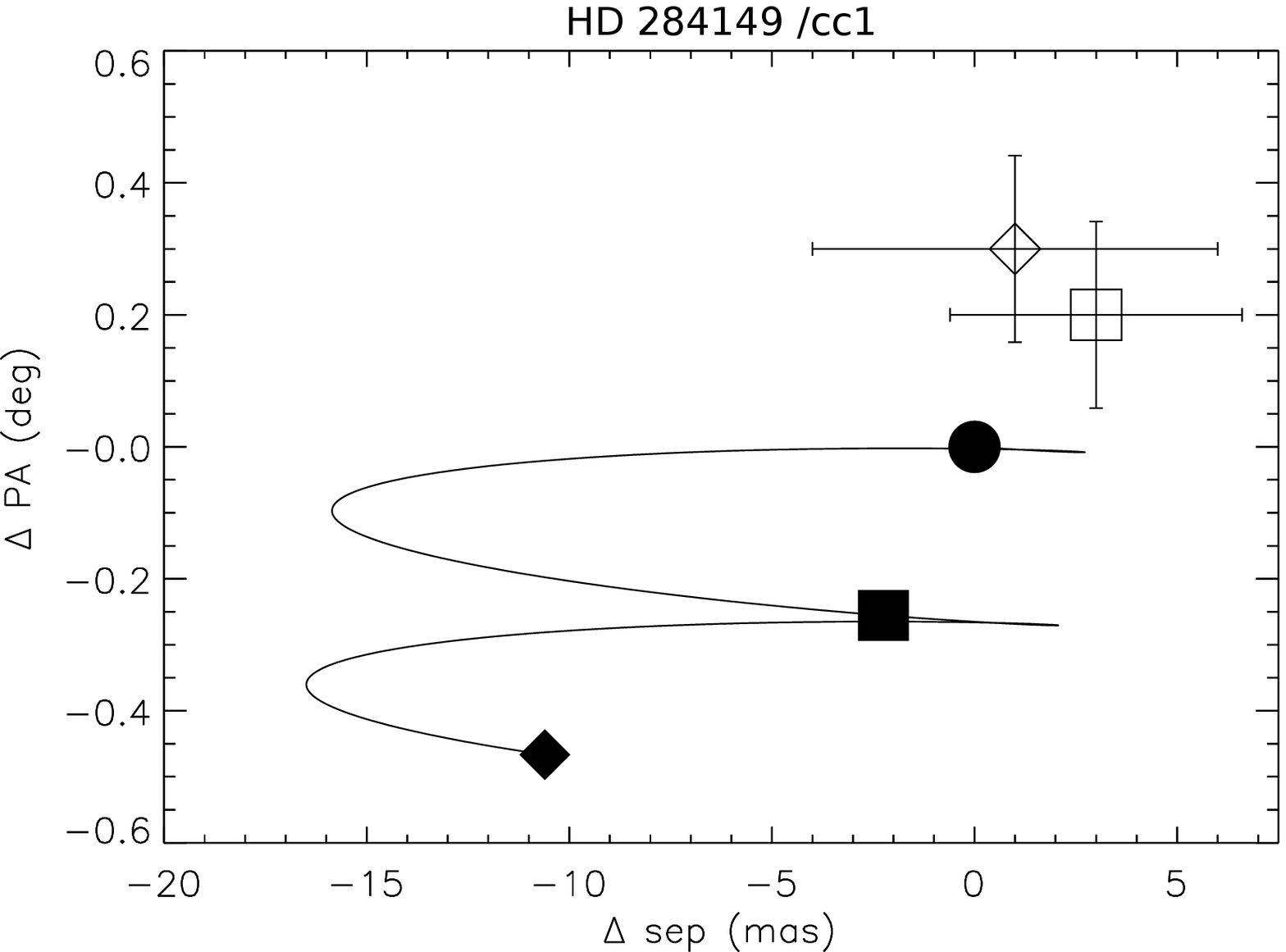}} 
\hfill{\includegraphics[angle=0,scale=.32]{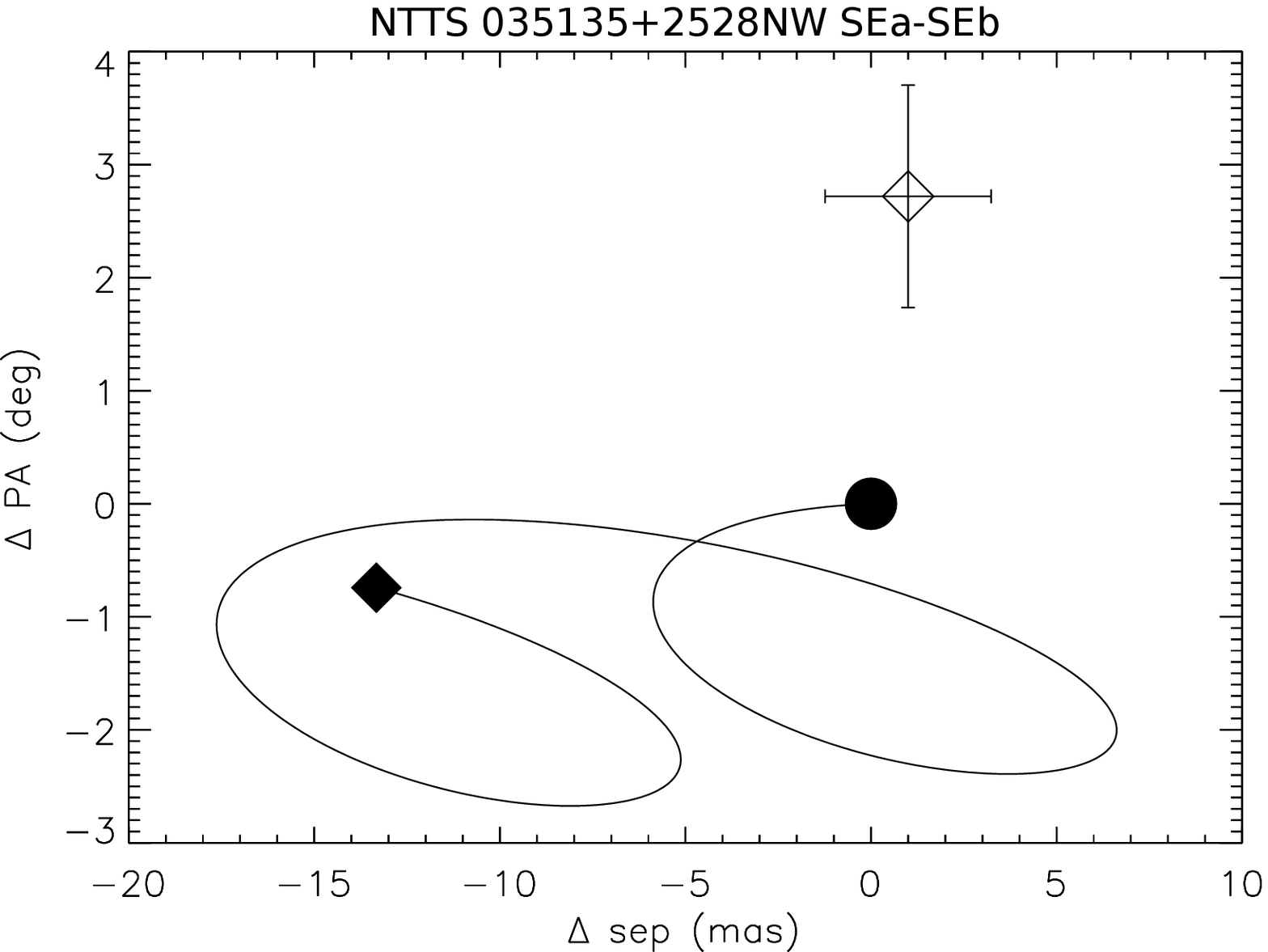}}\\[0.4cm]
\hfill{\includegraphics[angle=0,scale=.32]{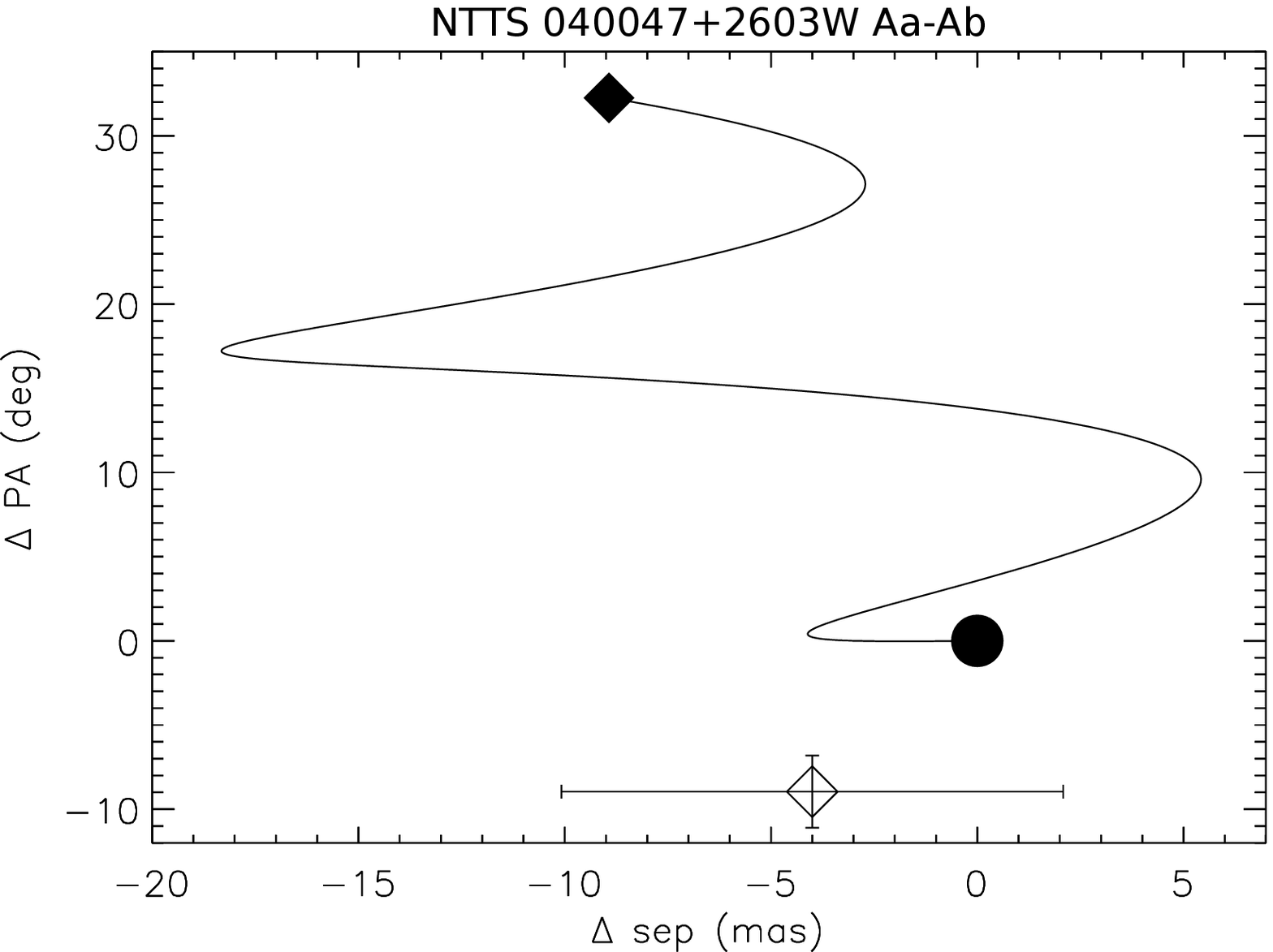}}
\hfill{\includegraphics[angle=0,scale=.32]{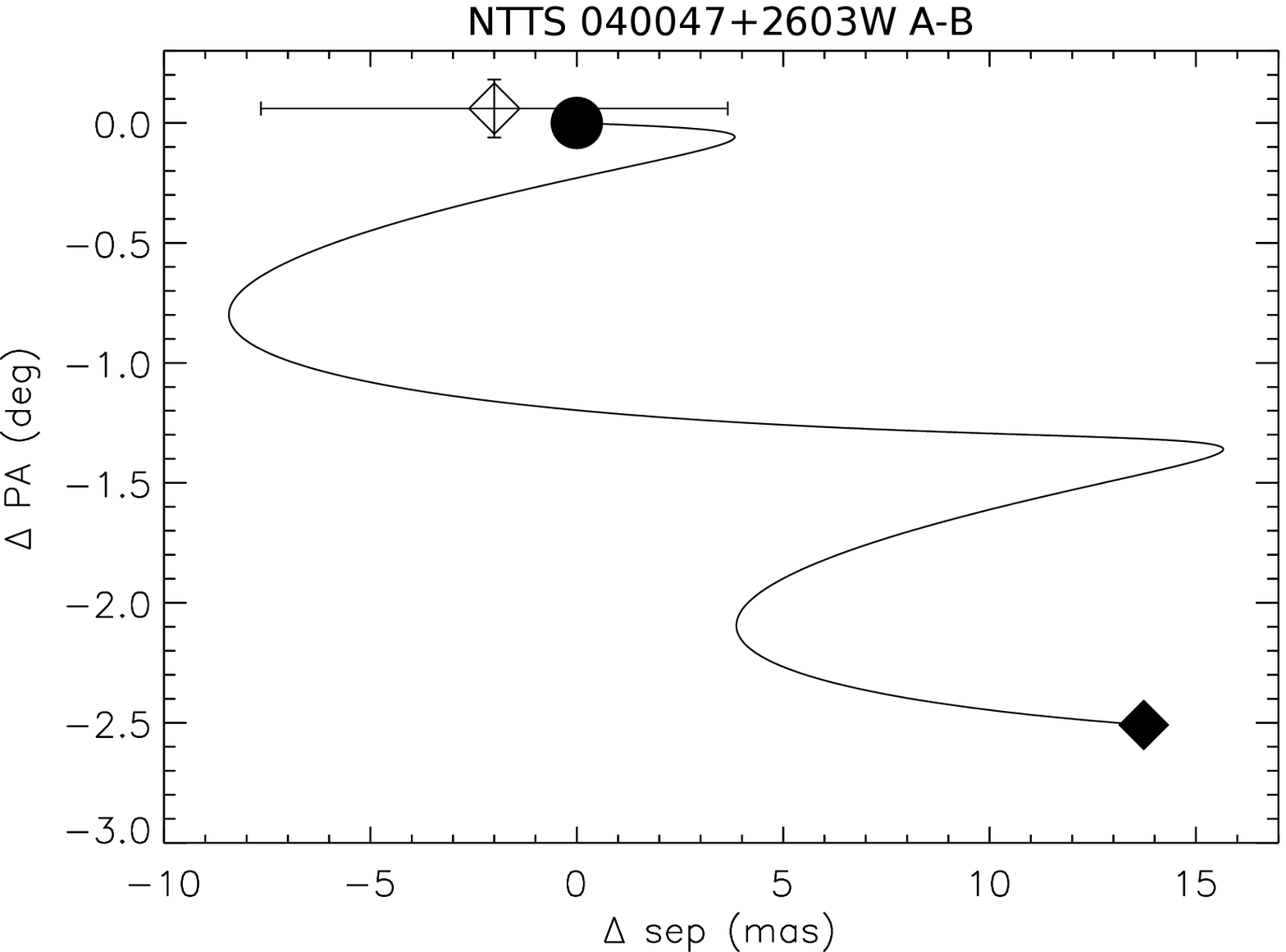}}
\hfill{\includegraphics[angle=0,scale=.32]{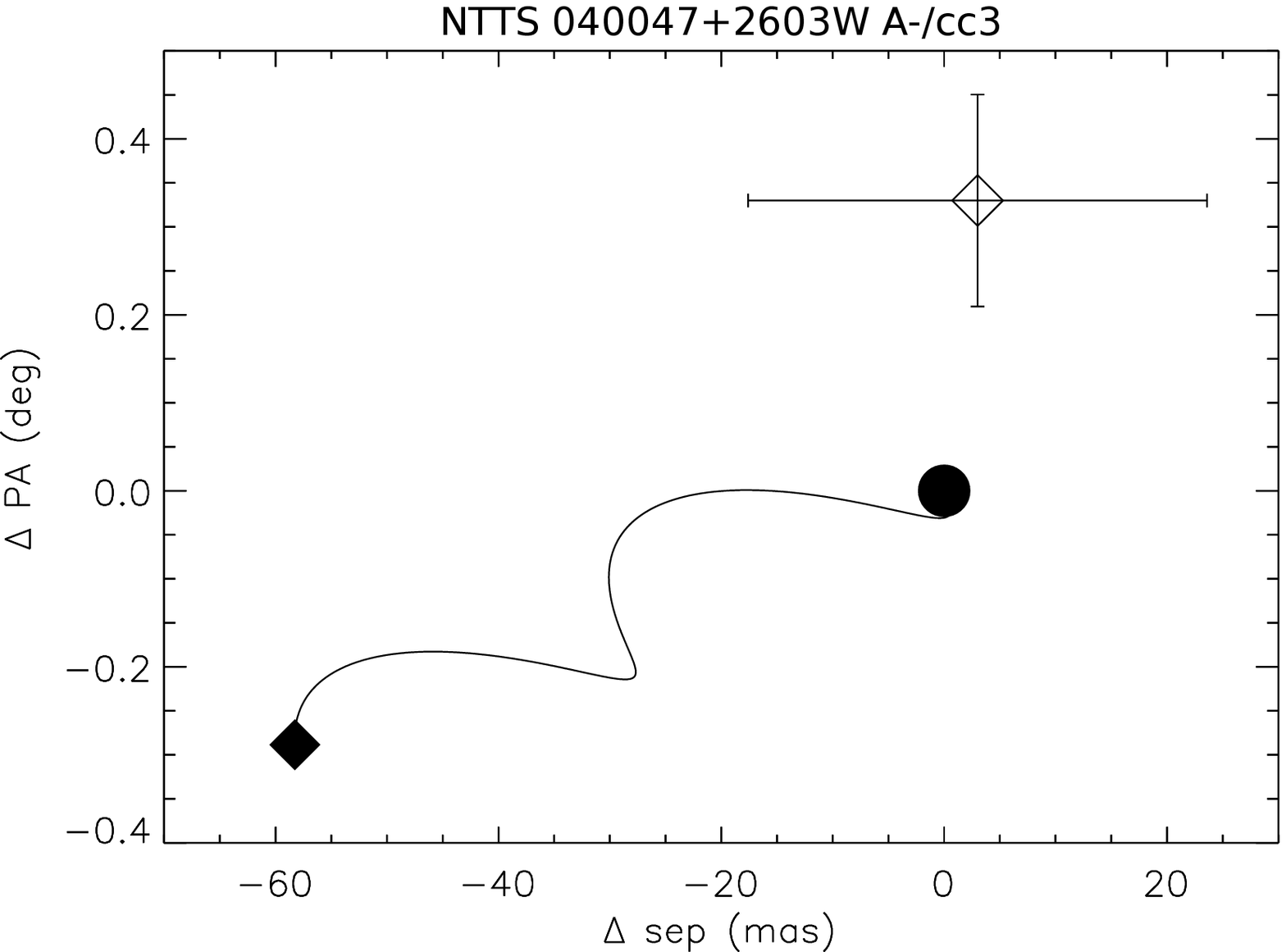}}\\[0.4cm]
\hfill{\includegraphics[angle=0,scale=.32]{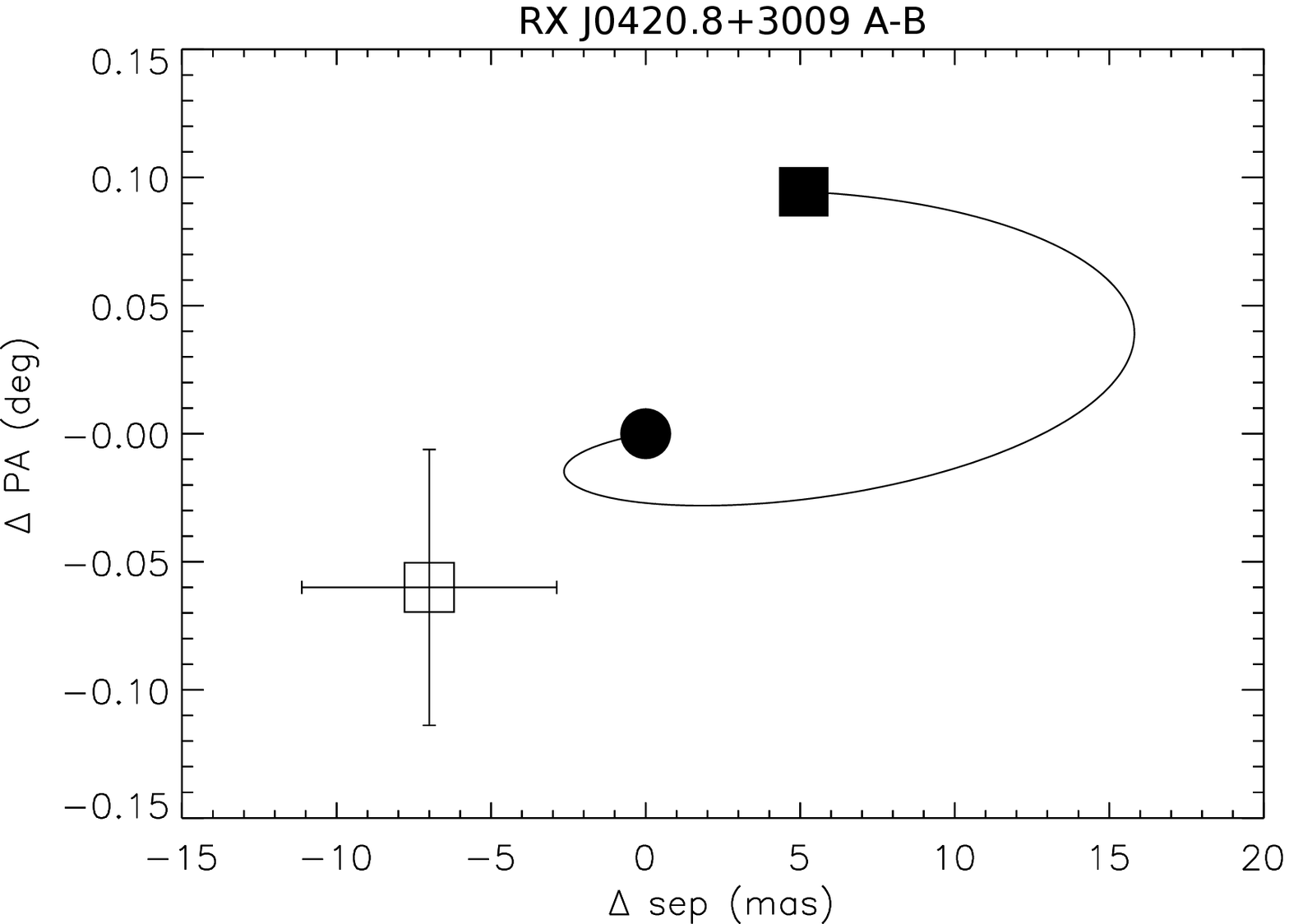}} 
\hfill{\includegraphics[angle=0,scale=.32]{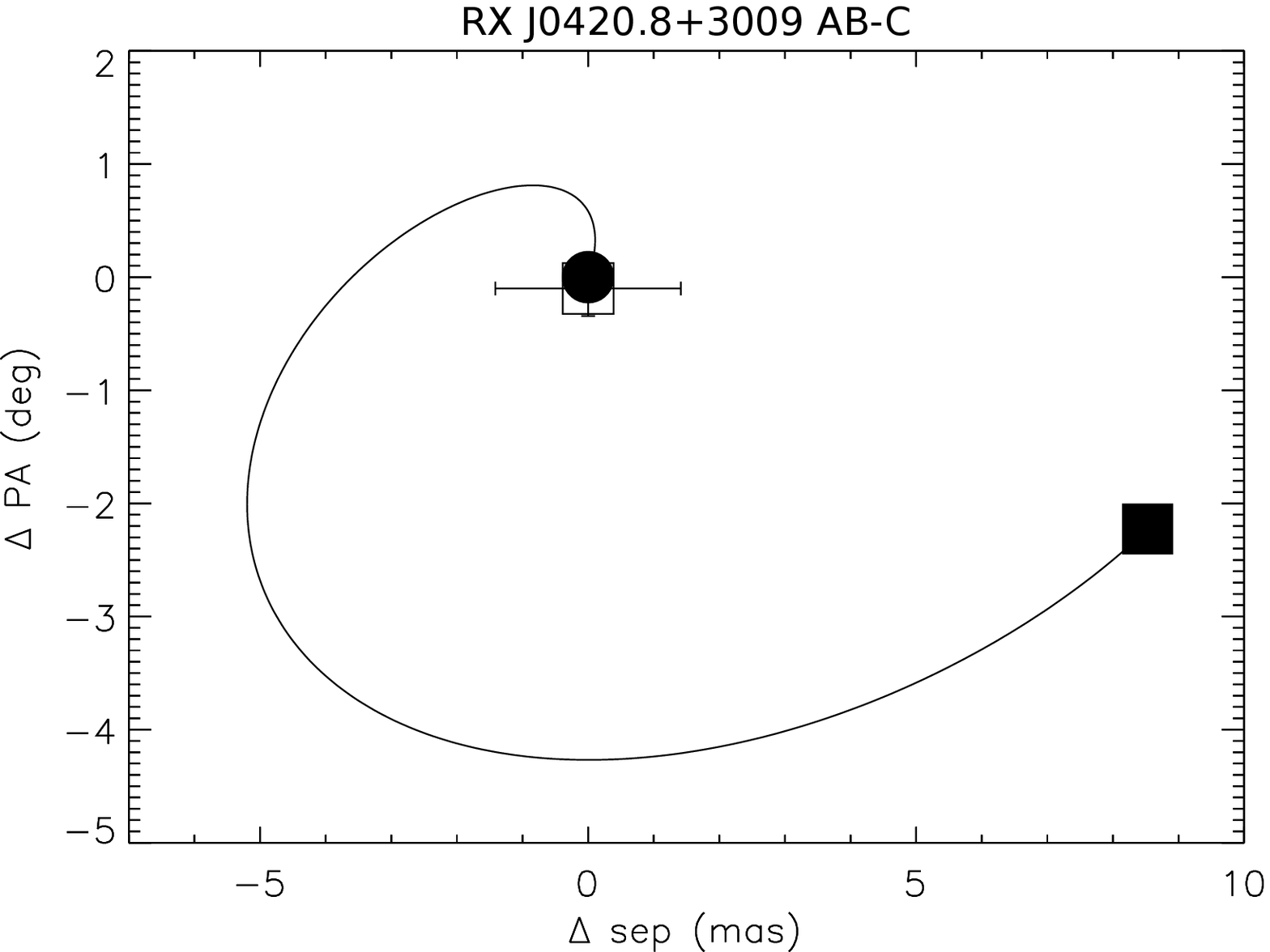}}
\hfill{\includegraphics[angle=0,scale=.32]{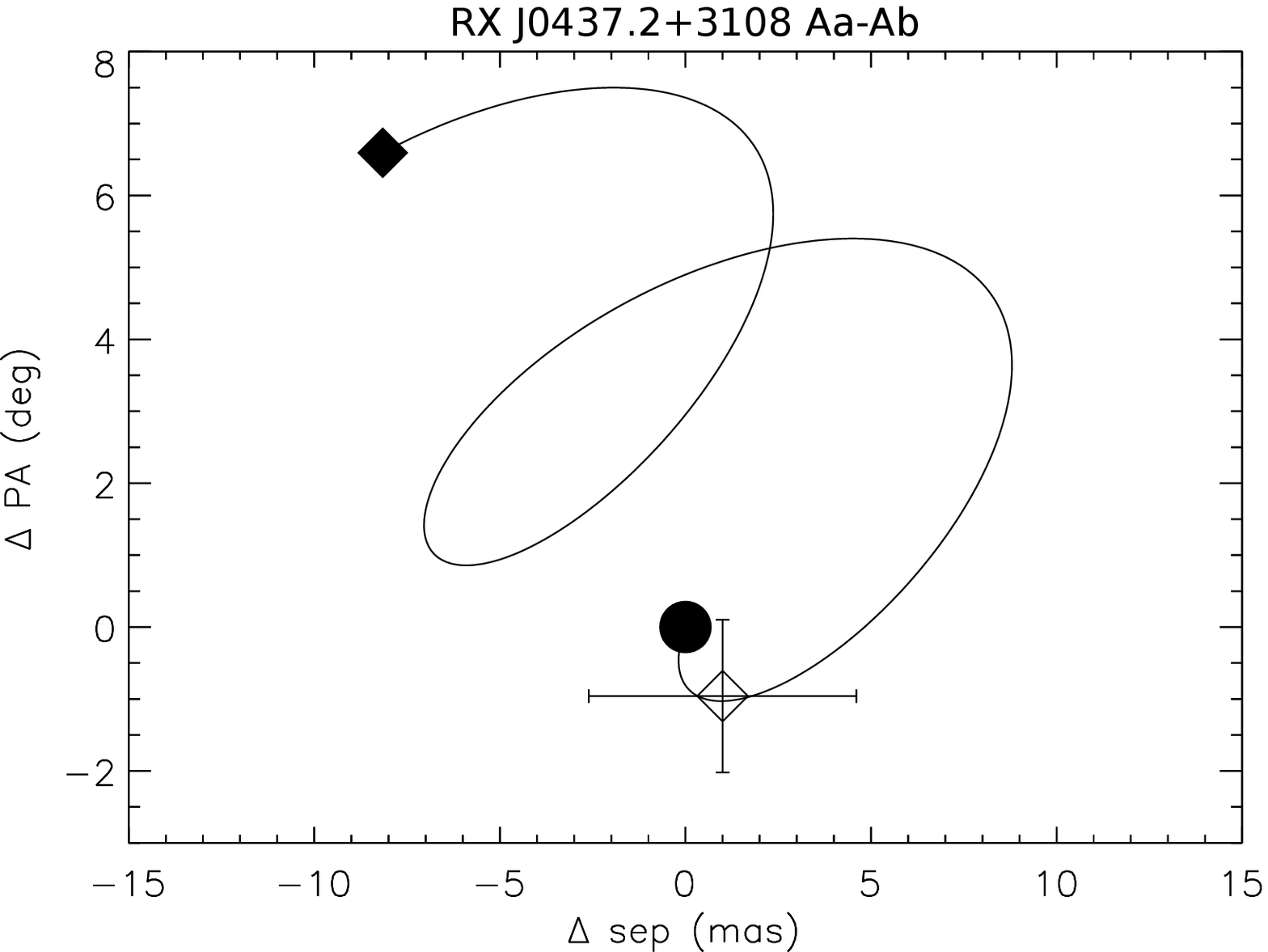}}\\[0.4cm]
\hfill{\includegraphics[angle=0,scale=.32]{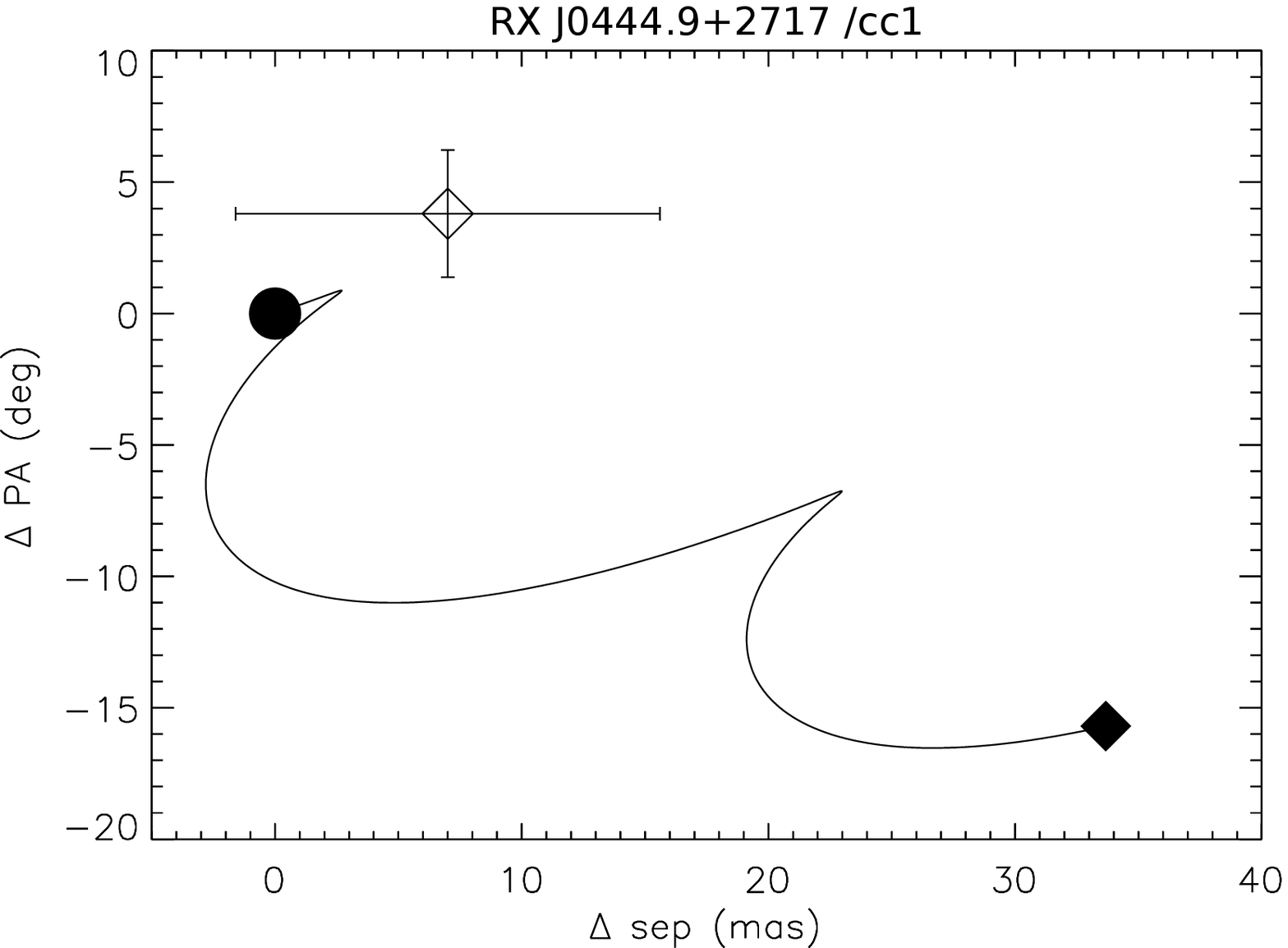}}
\hfill{\includegraphics[angle=0,scale=.32]{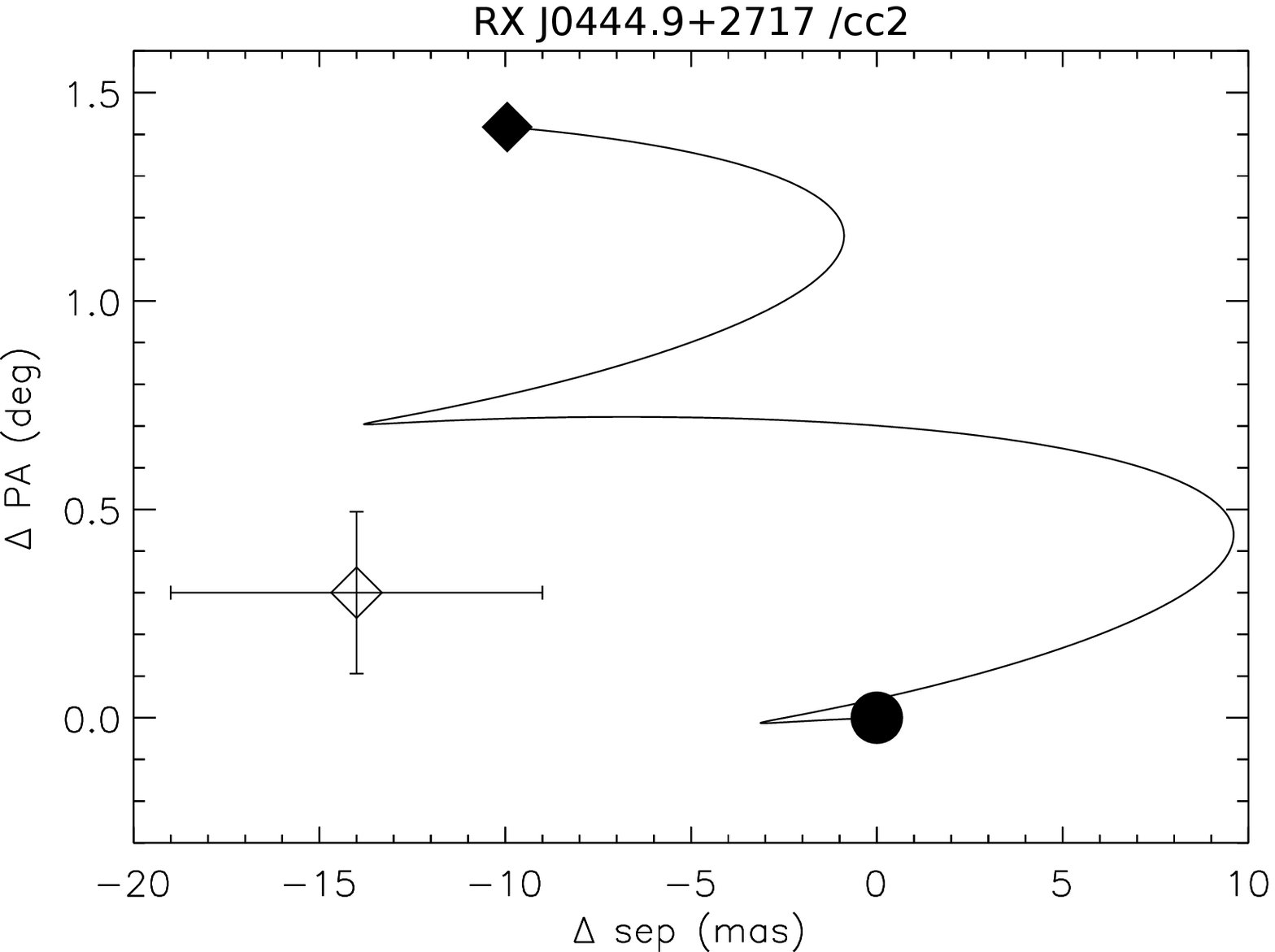}} 
\hfill{\includegraphics[angle=0,scale=.32]{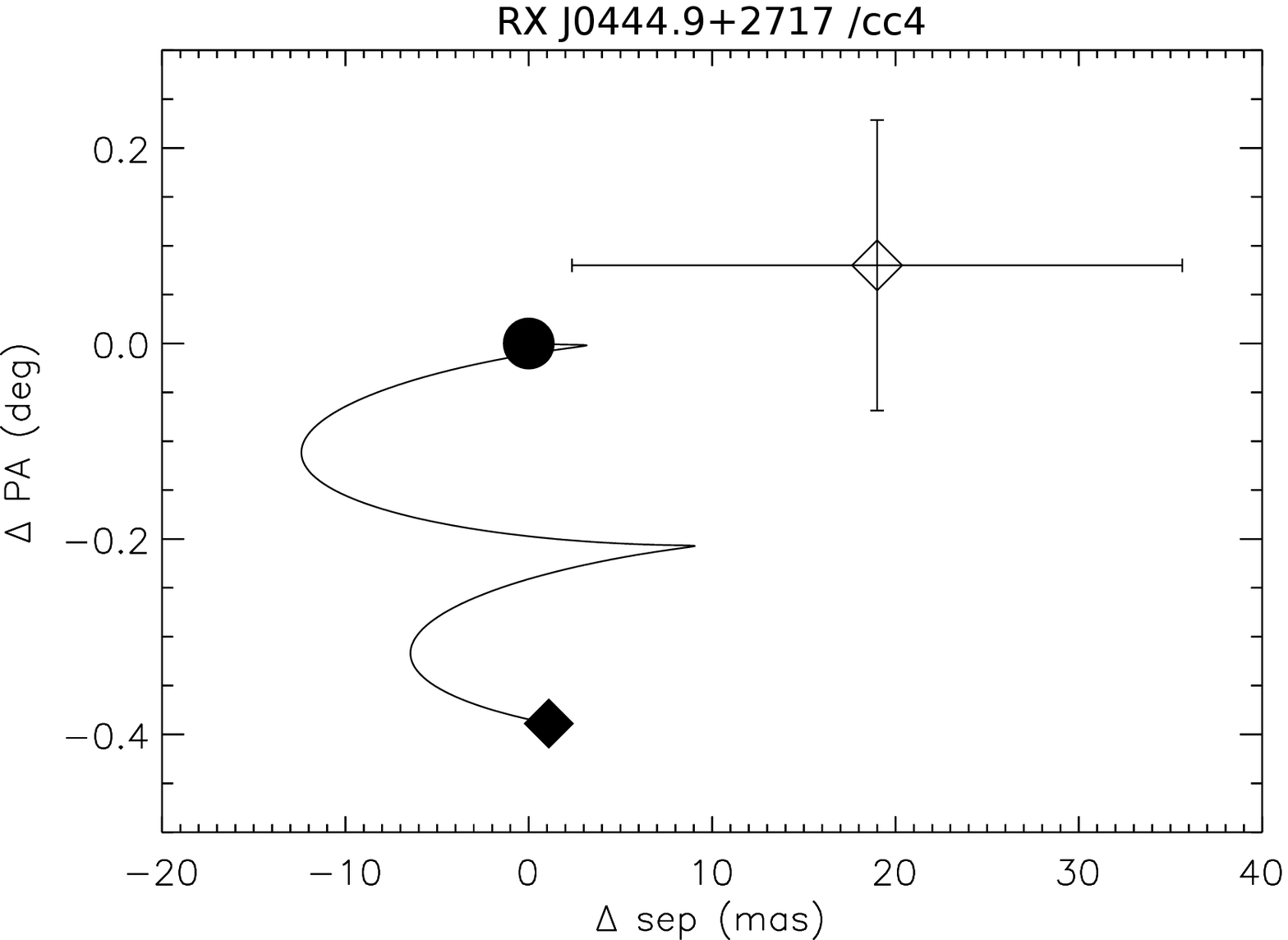}}
\caption{Evaluation of common proper motion as in Fig.~\ref{fig:CPM} for all companions classified as co-moving. Epochs are $\diamond$: 2011B, $\square$: 2012B, \CIRCLE: 2013B (=reference). \label{fig:CPMours1}}
\end{figure*}

\begin{figure*}
{\includegraphics[angle=0,scale=.32]{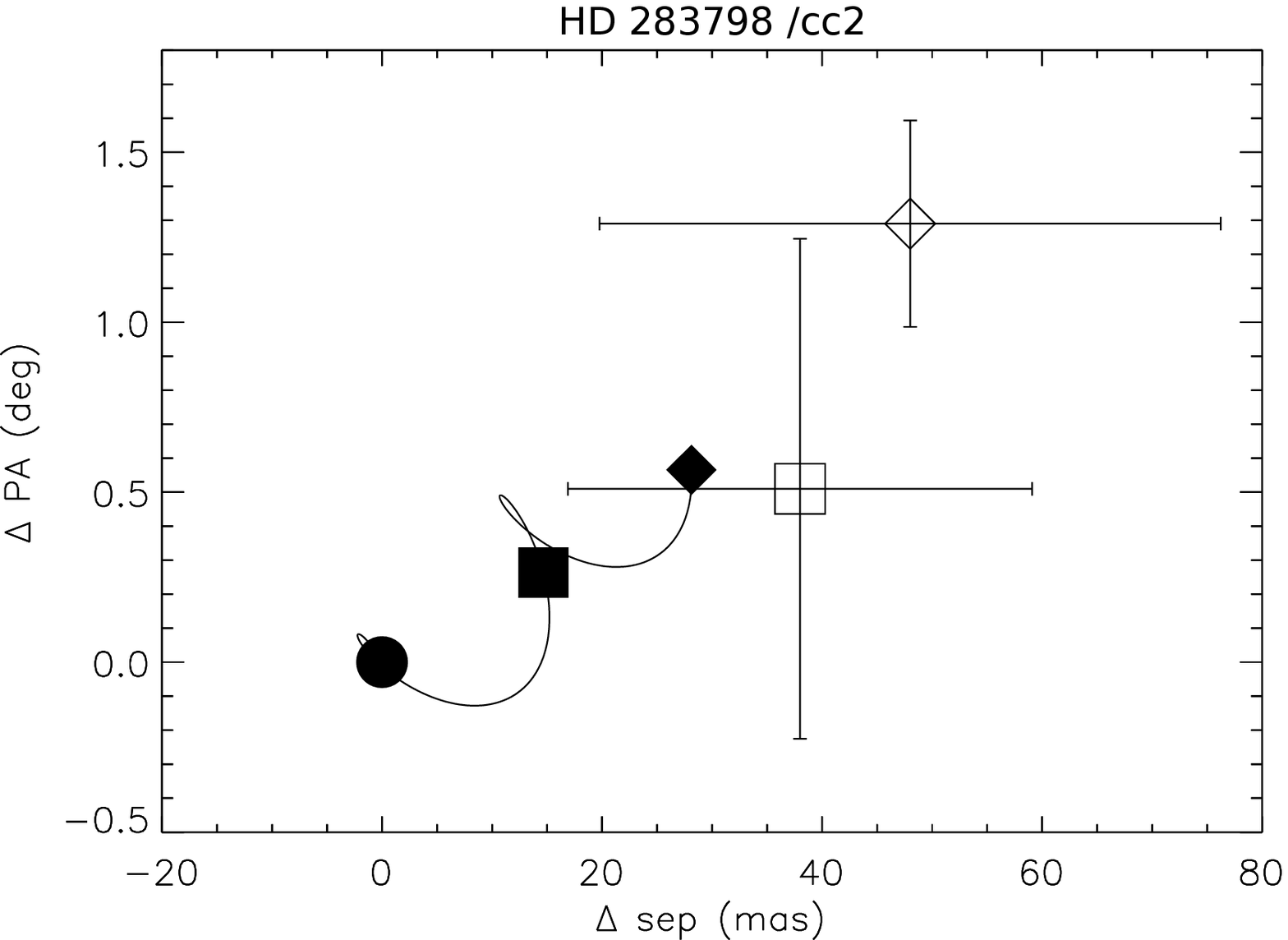}}
\hfill{\includegraphics[angle=0,scale=.32]{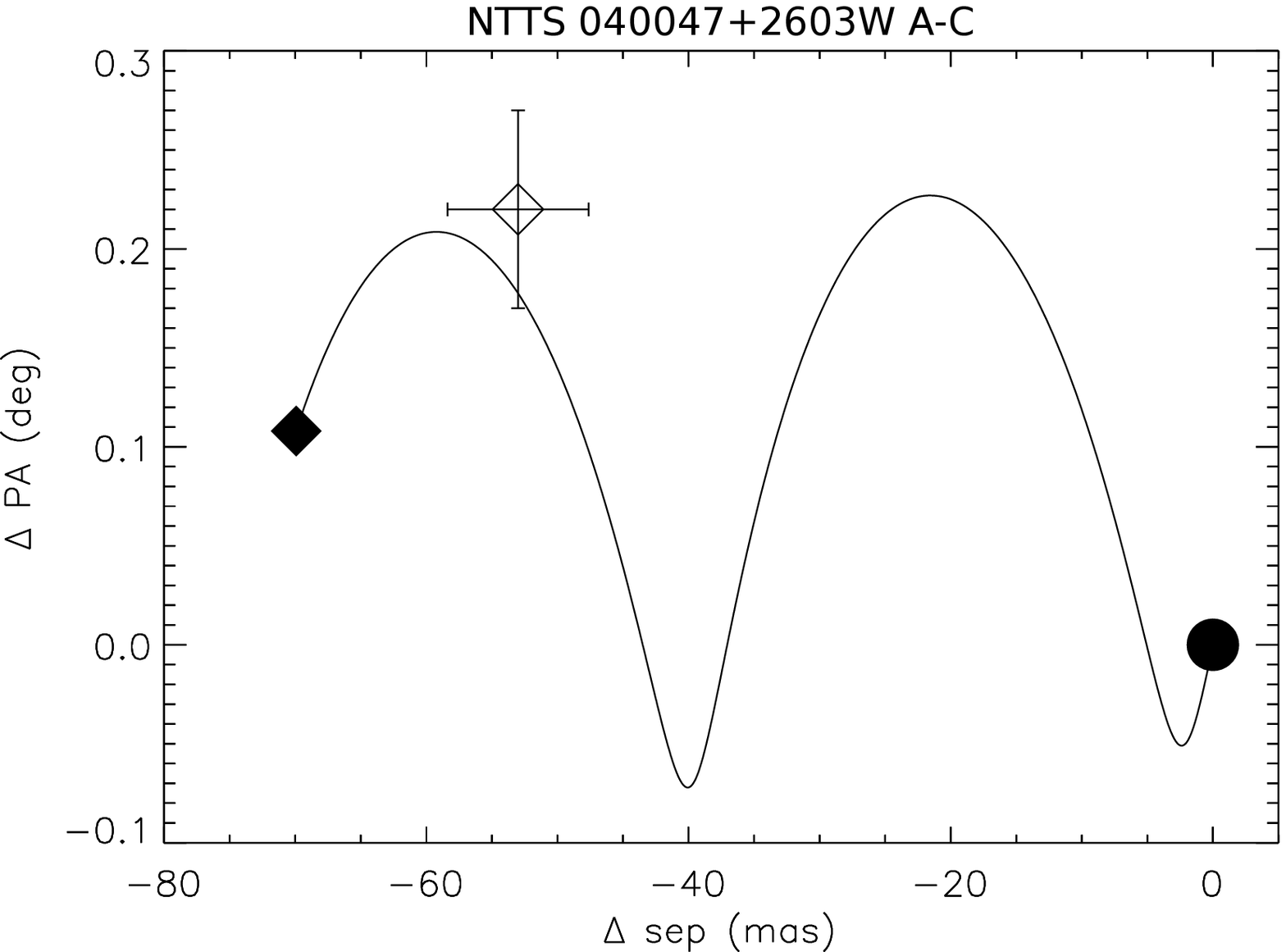}}
\hfill{\includegraphics[angle=0,scale=.32]{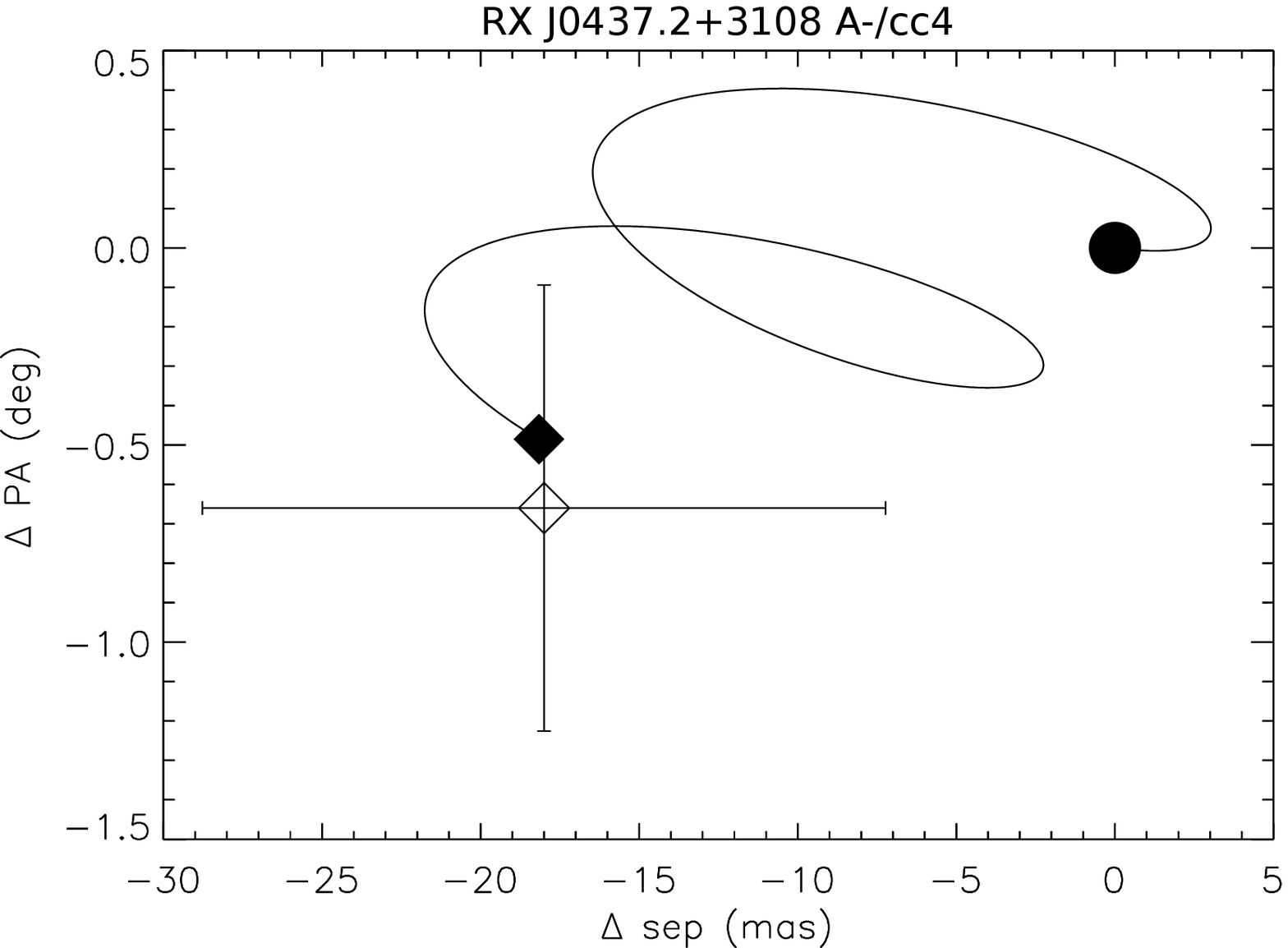}}
\caption{Same as Fig.~\ref{fig:CPMours1} but for likely background sources.\label{fig:CPMours2}}
\end{figure*}

Figs.~\ref{fig:CPMours1}, \ref{fig:CPMours2}, and \ref{fig:CPMours3} show our evaluation of common proper motion between the host star and the companion candidate for all targets with multiple-epoch observations. {We distinguish cases of likely co-moving, background, and inconclusive pairs by square-averaging the dimensionless distances $a=\sum\left[(\mathrm{sep_{measured}}-\mathrm{sep_{expected}})/\delta\mathrm{sep_{measured}}\right]^2+\sum\left[(\mathrm{PA_{measured}}-\mathrm{PA_{expected}})/\delta\mathrm{PA_{measured}}\right]^2$ in $\Delta$sep--$\Delta$PA space between the measured and expected values for a co-moving companion and background star. We assume co-motion if the normalized co-moving distance is 3 times smaller than the background distance. The background case is defined equivalently. If the normalized distances for co-moving and background are comparable (i.e., $1/3<a<3$), then we can make no decision about whether the companion is co-moving or background and future observations with a longer time baseline and/or higher precision must be acquired.}

Targets that have been classified as non-members are included here. Due to the lack of a better parallax estimate, a distance of 140\,pc has been assumed as for the other candidates. This has only a minor effect on the derived classification as co-moving or background since the proper motion dominates the distance parallax estimate in most of the cases and observations were executed during a similar time of year in 2011, 2012, and 2013.
\label{sec:A2}
\addtocounter{figure}{-1}
\begin{figure*}
{\includegraphics[angle=0,scale=.32]{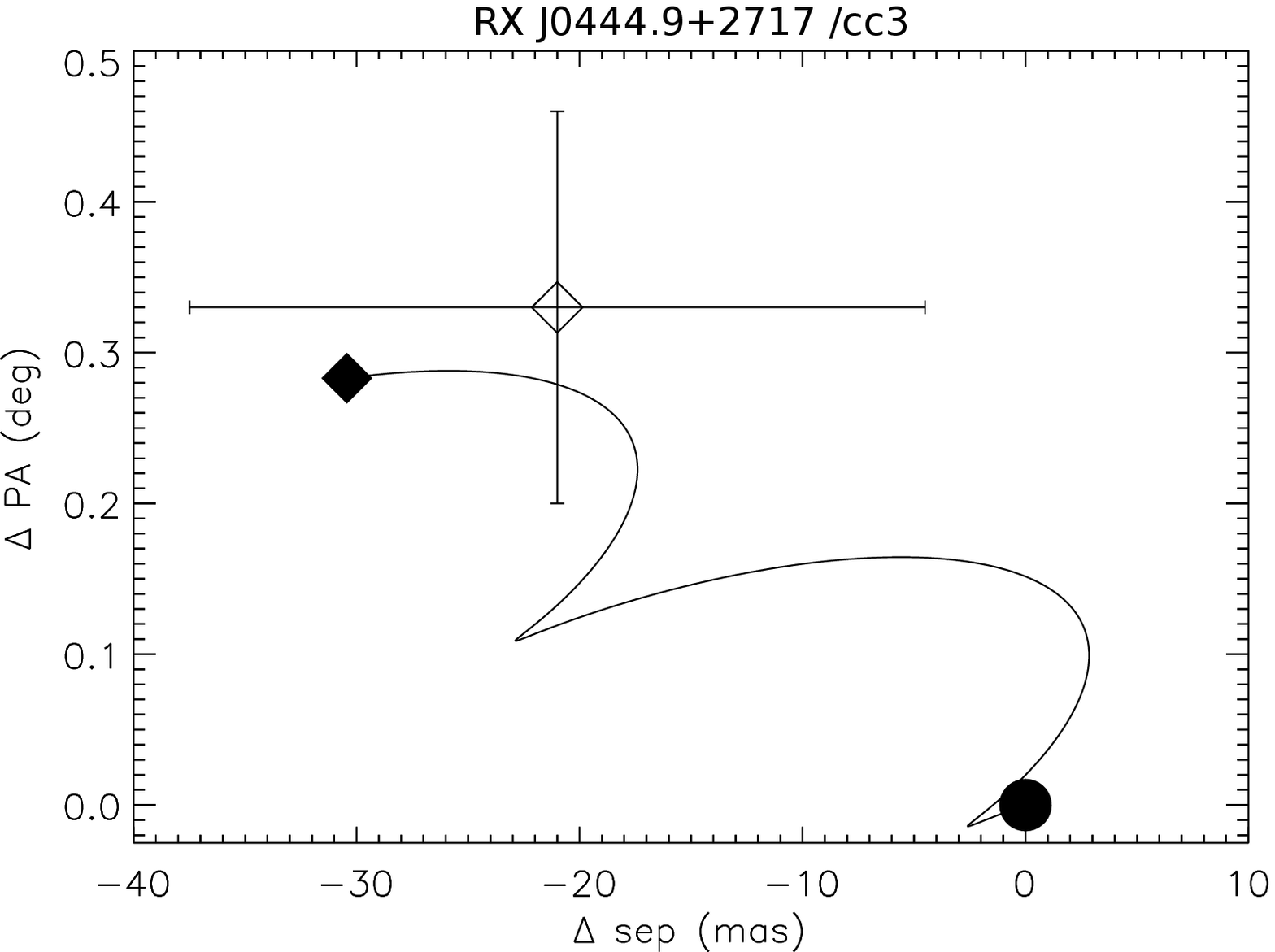}}
\caption{\emph{ctd.}}
\end{figure*}
\begin{figure*}
\includegraphics[angle=0,scale=.32]{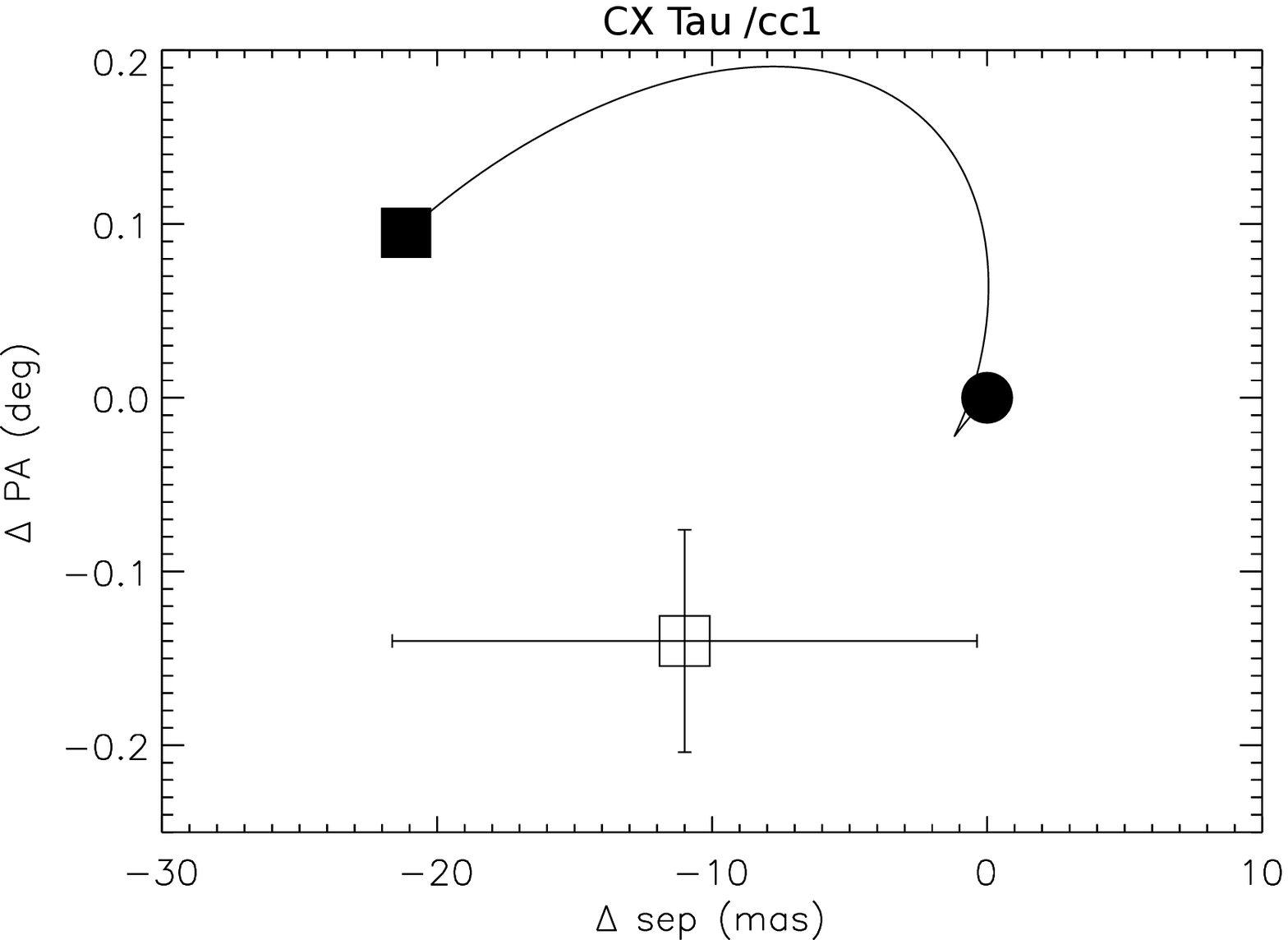}
\hfill\includegraphics[angle=0,scale=.32]{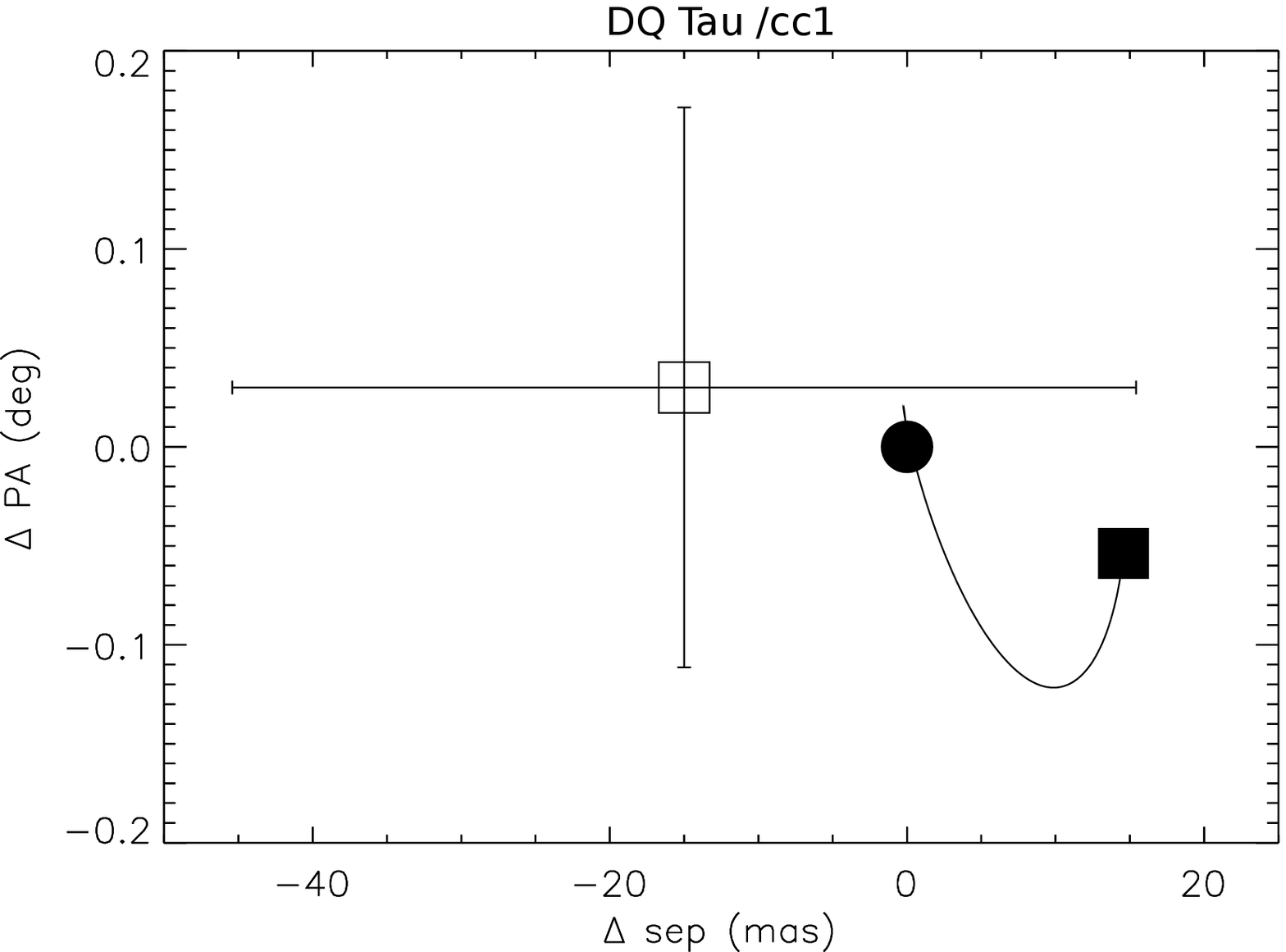}
\hfill\includegraphics[angle=0,scale=.32]{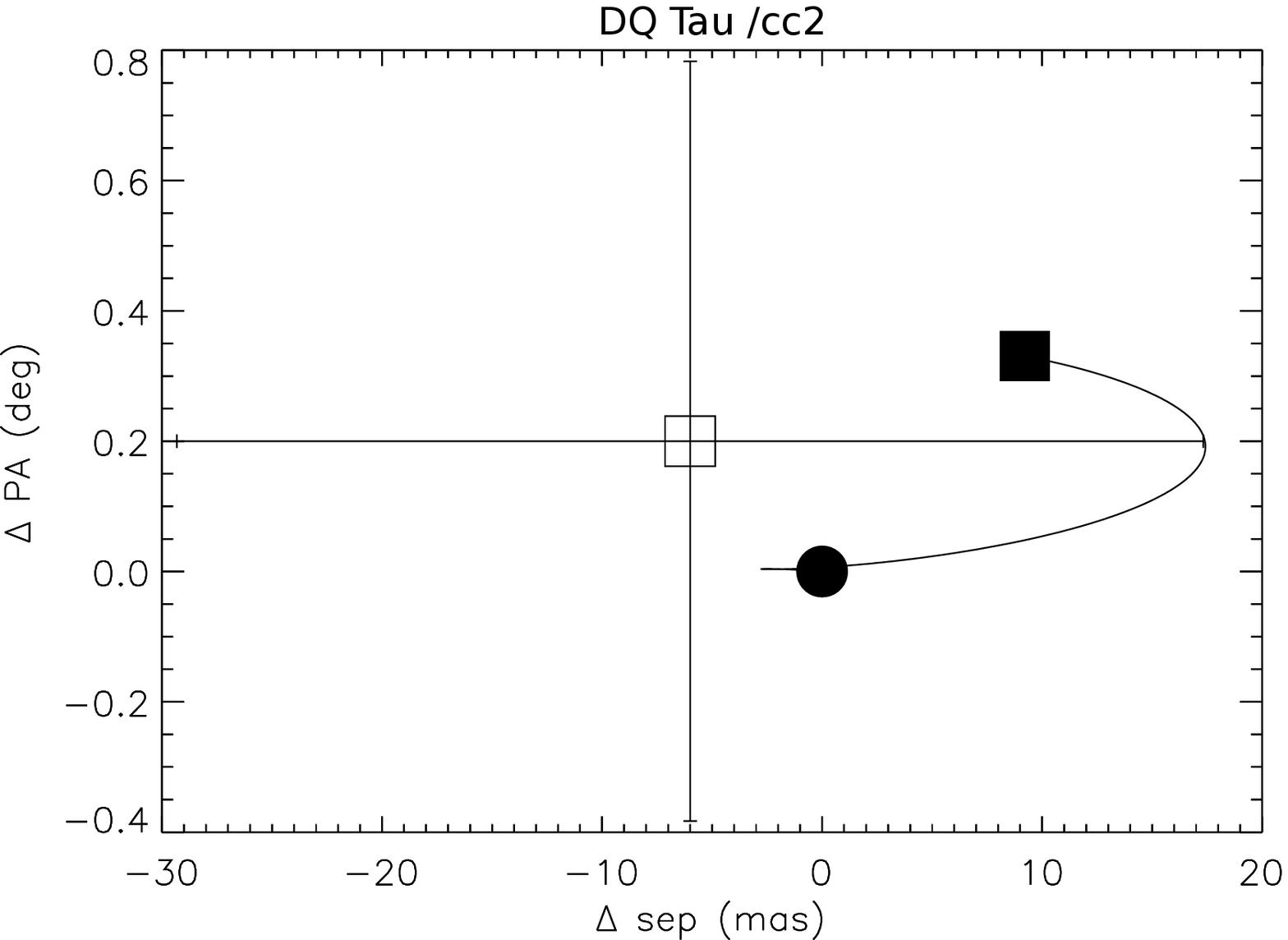}\\[0.4cm]
\hfill\includegraphics[angle=0,scale=.32]{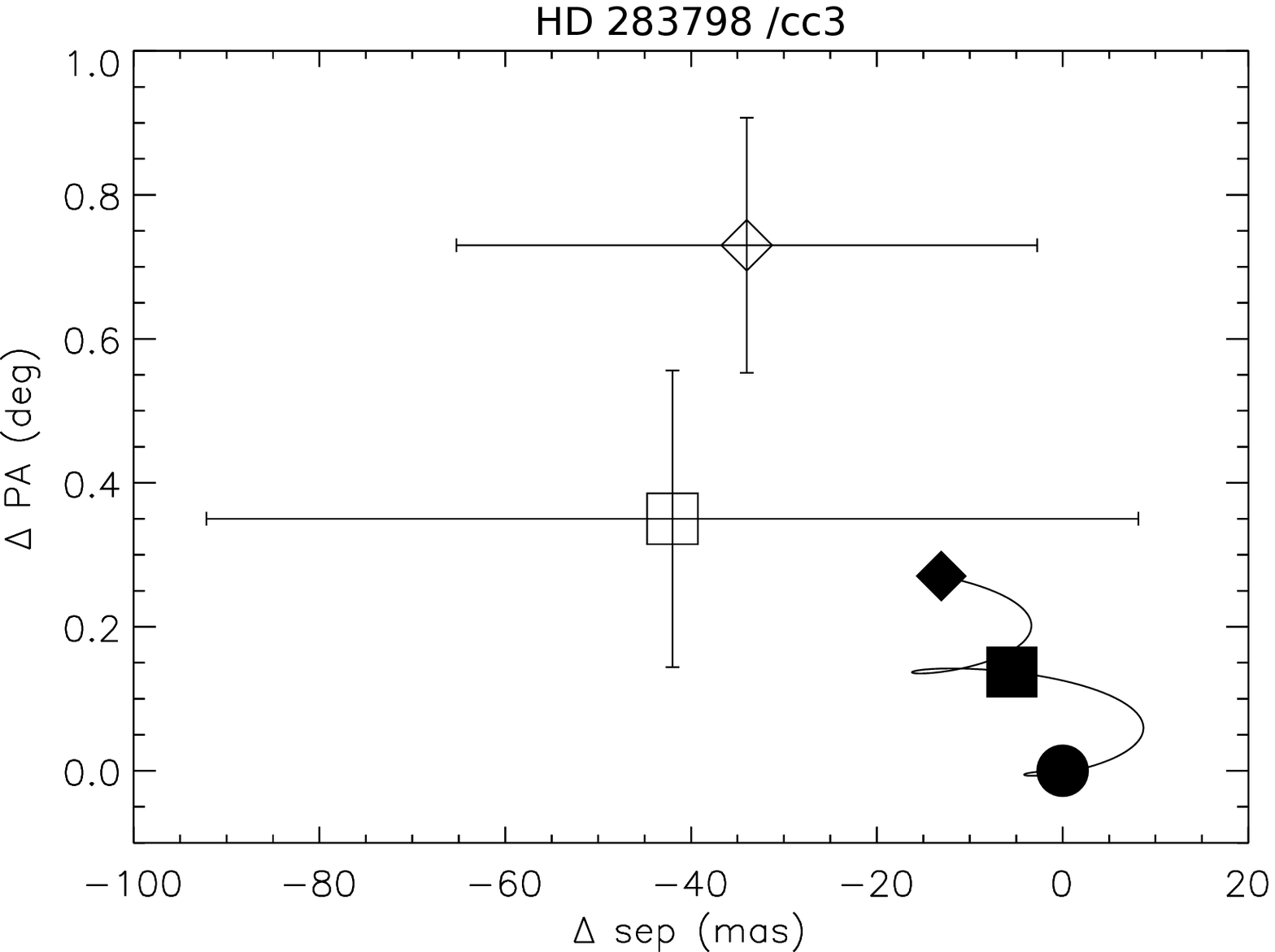}
\hfill\includegraphics[angle=0,scale=.32]{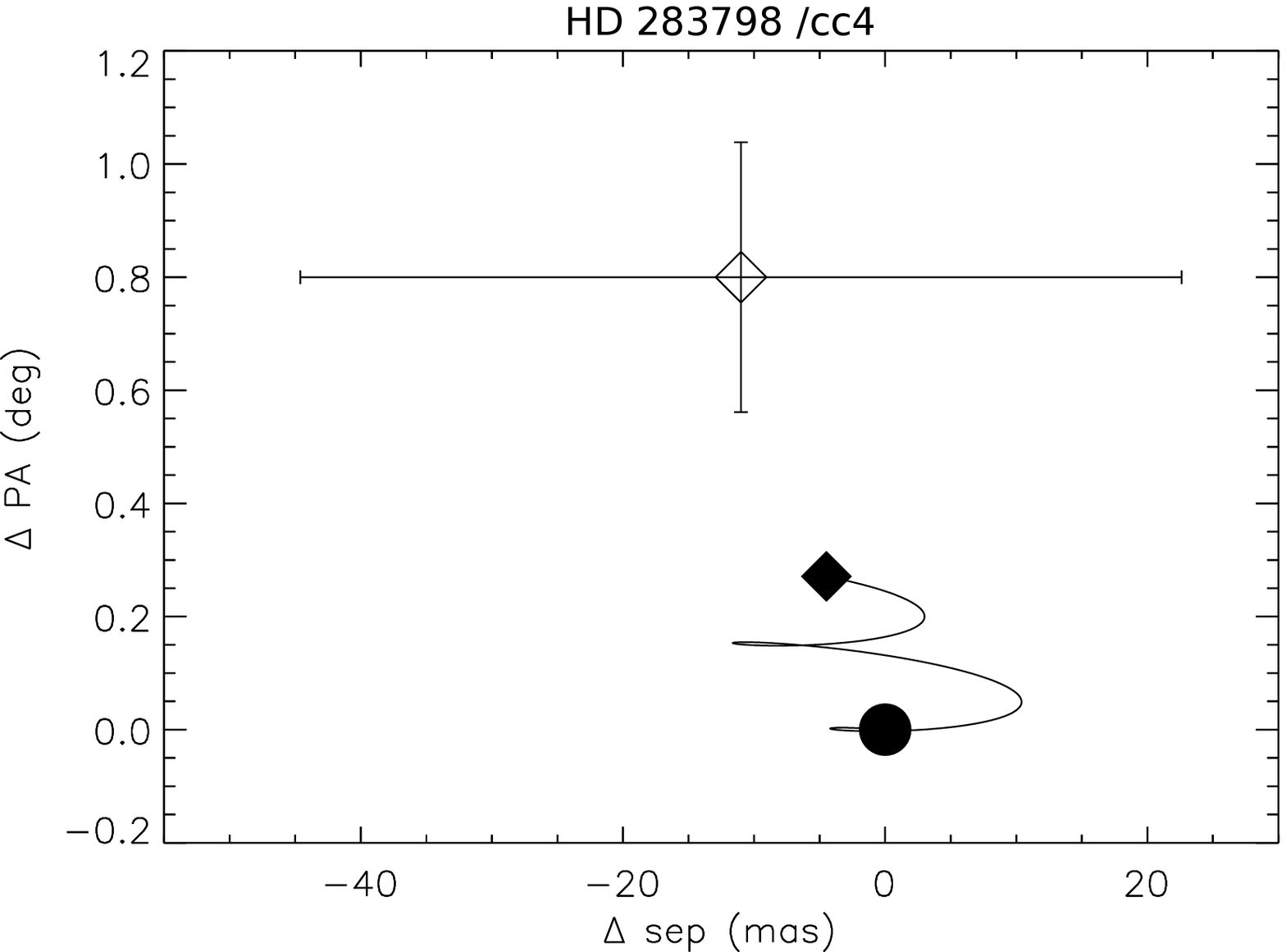} 
\hfill\includegraphics[angle=0,scale=.32]{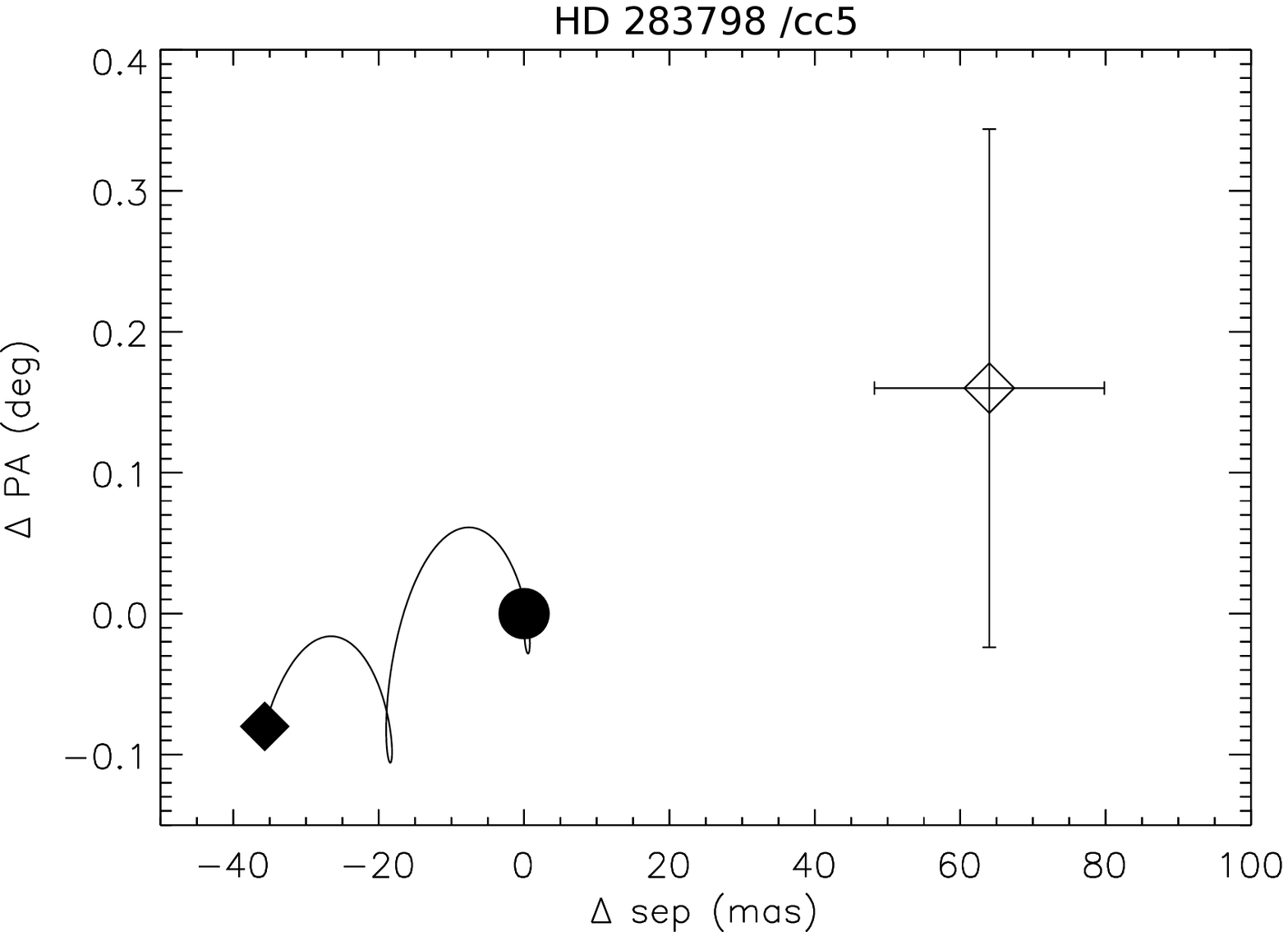}\\[0.4cm]
\hfill\includegraphics[angle=0,scale=.32]{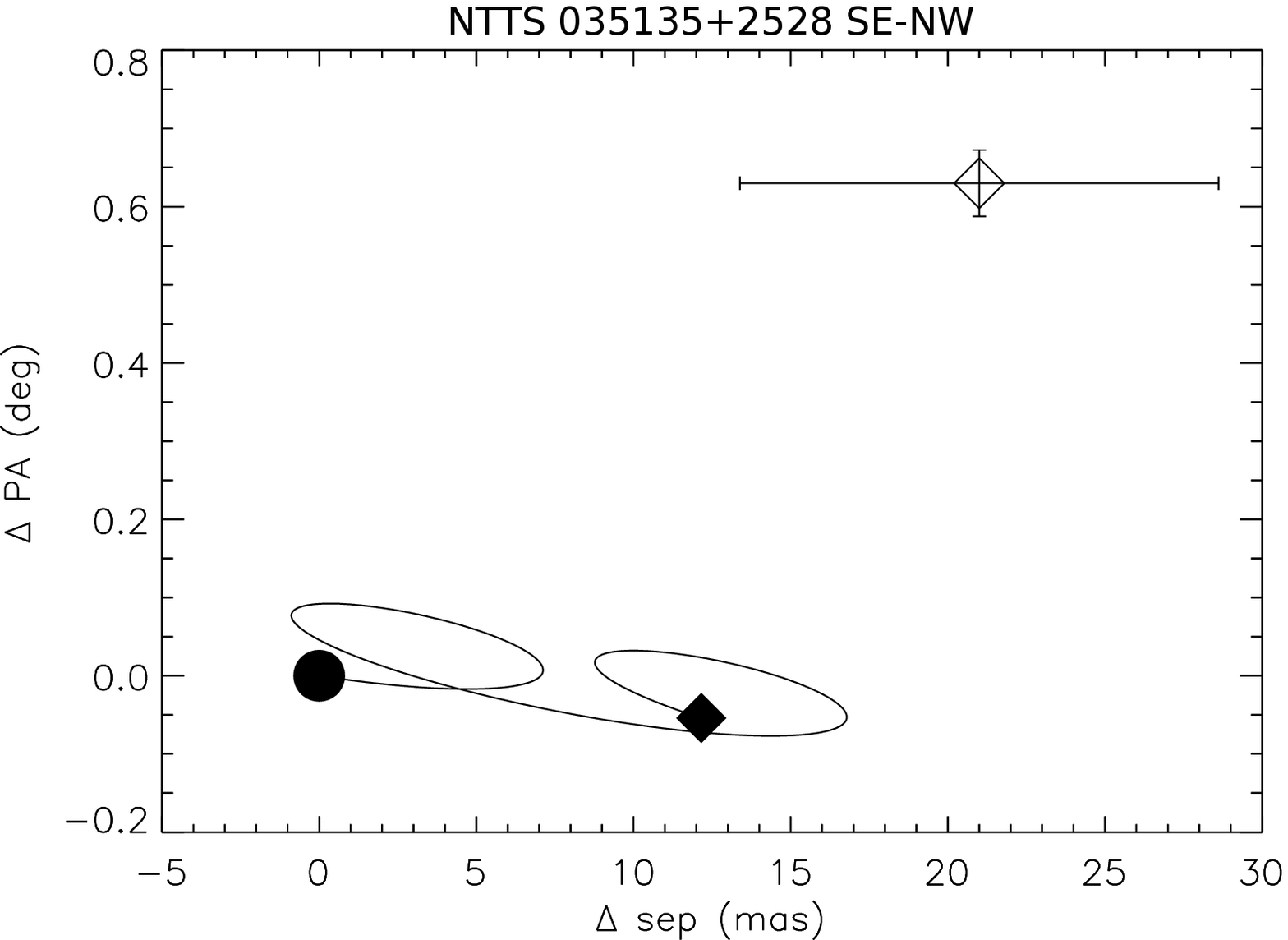} 
\hfill\includegraphics[angle=0,scale=.32]{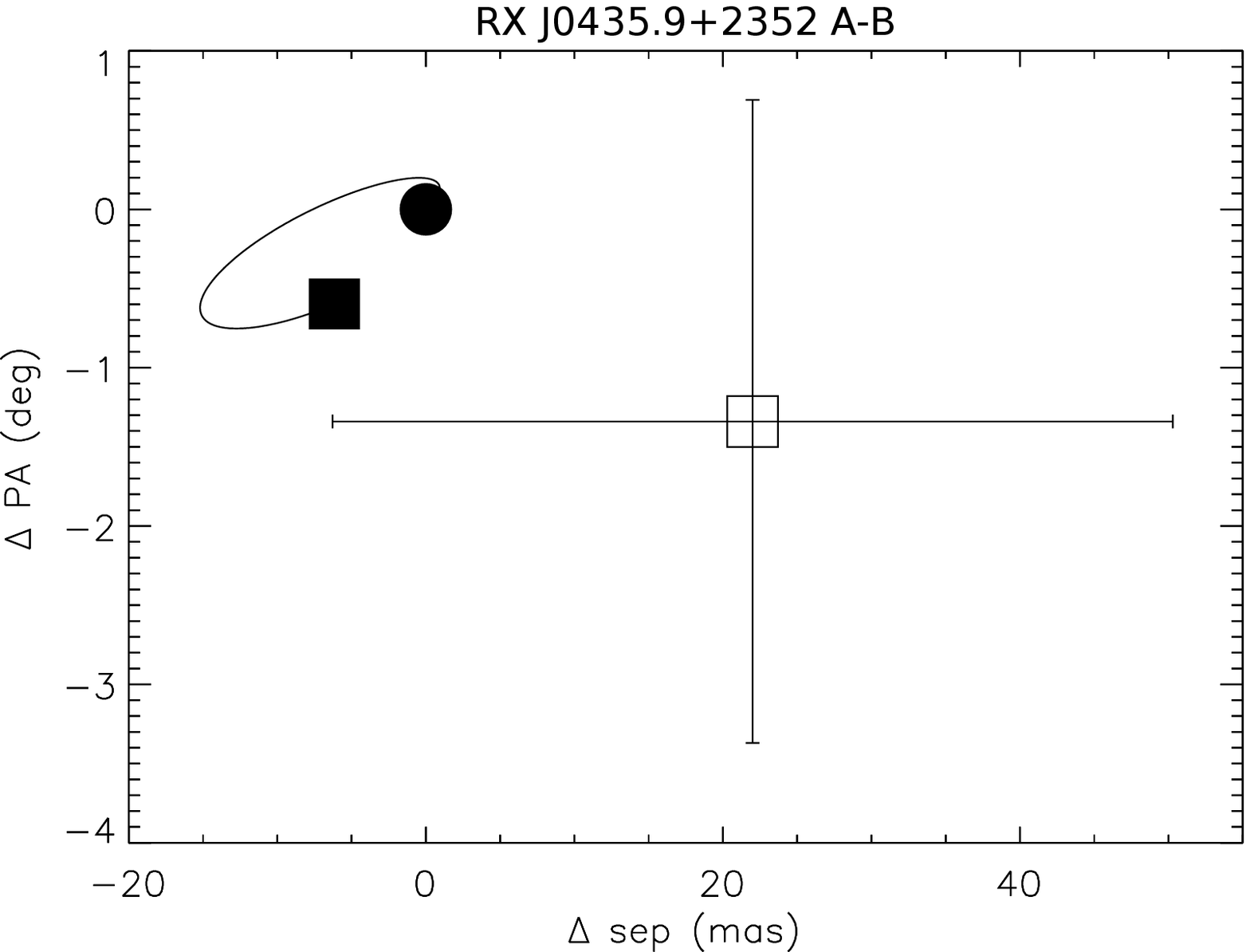}
\hfill\includegraphics[angle=0,scale=.32]{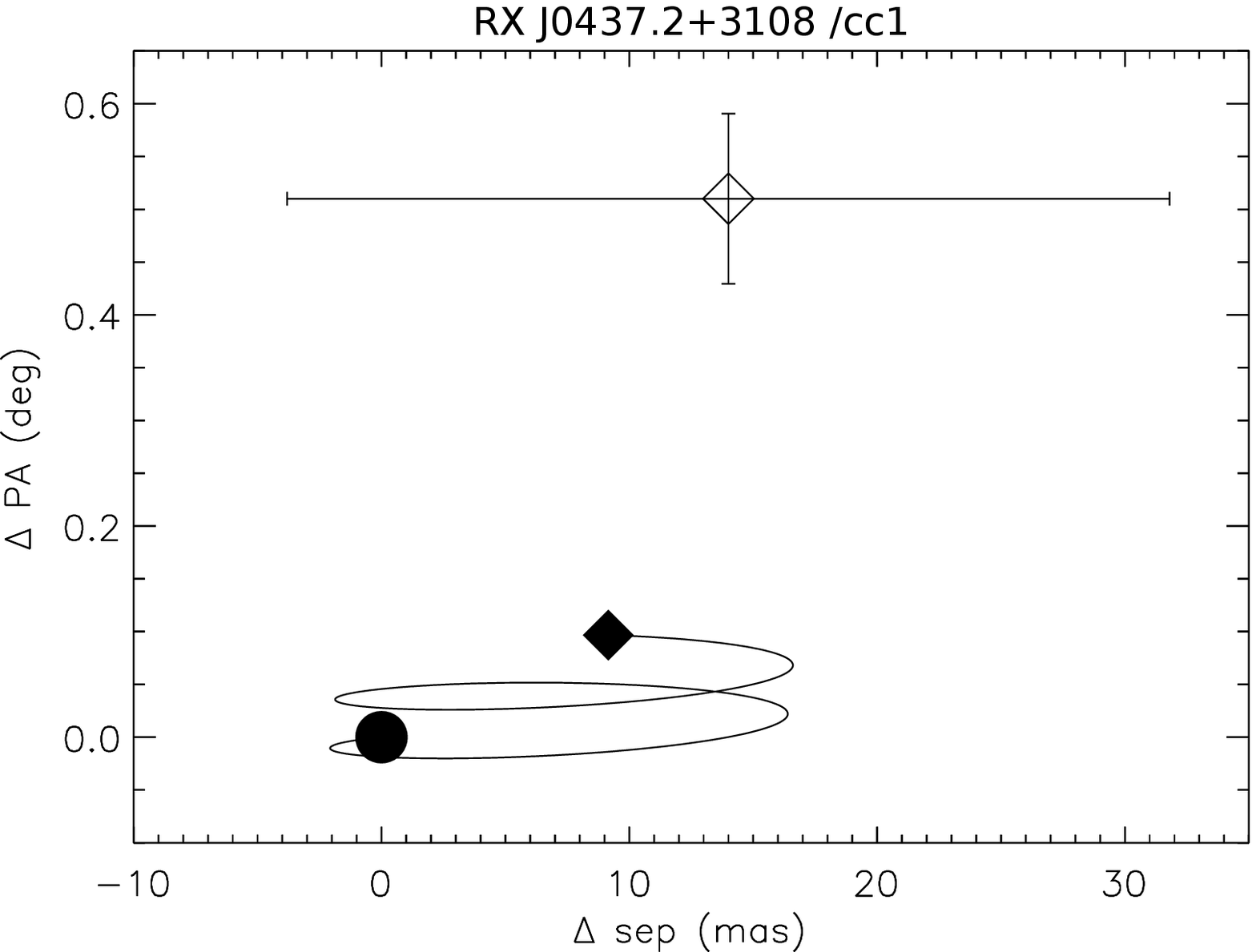}\\[0.4cm]
\hfill\includegraphics[angle=0,scale=.32]{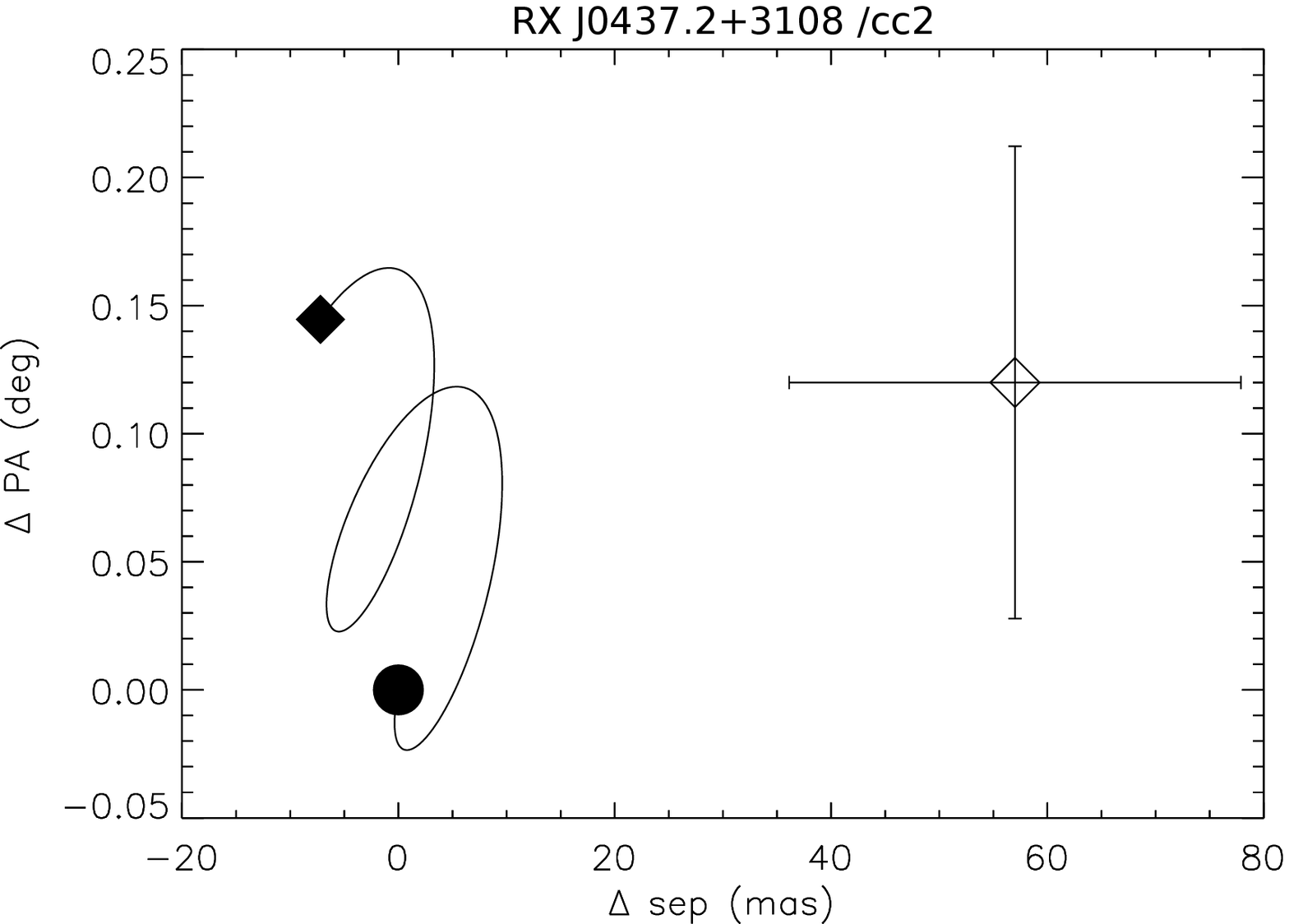}
\hspace{0.8cm}\includegraphics[angle=0,scale=.32]{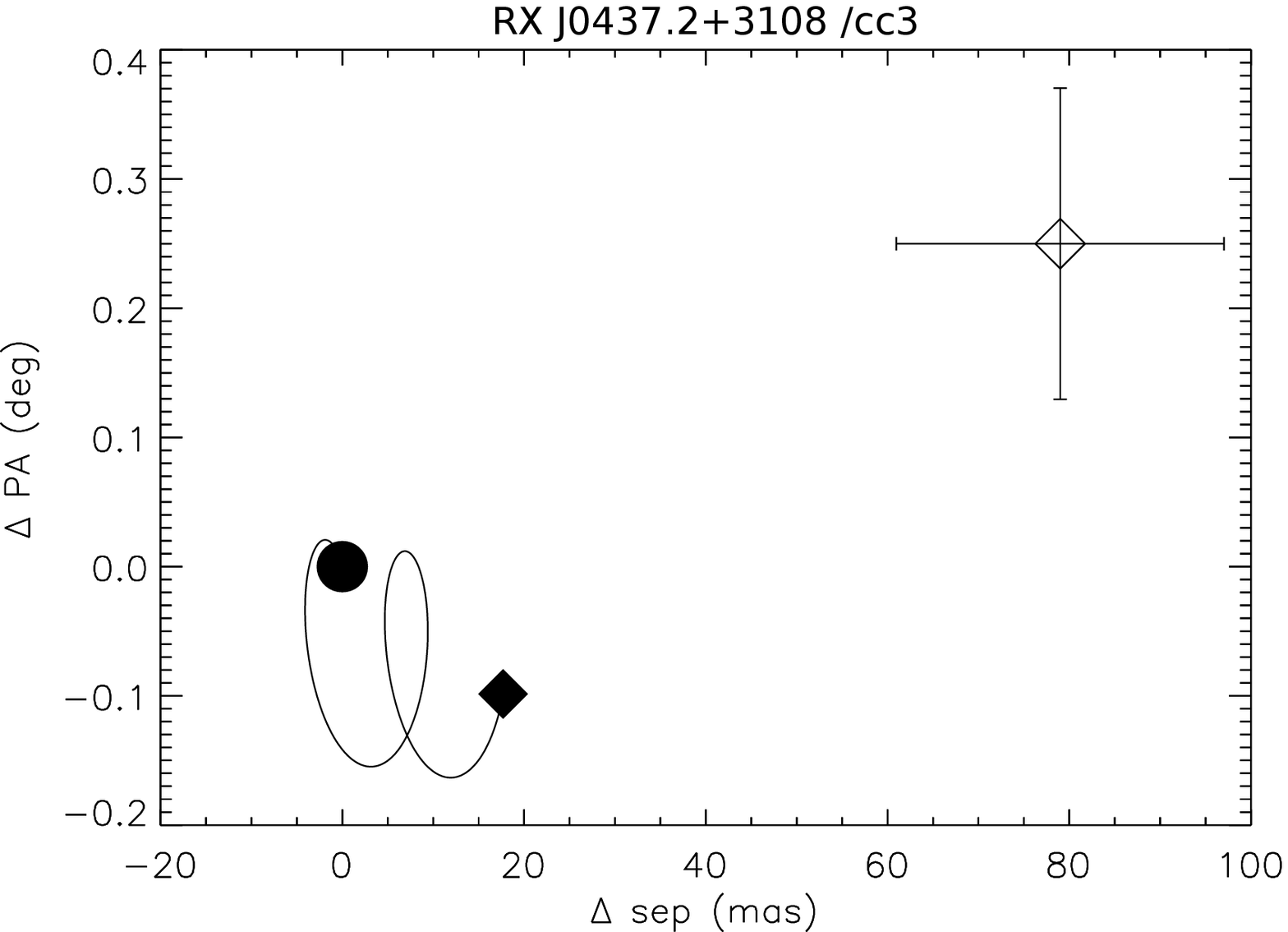}
\caption{Same as Fig.~\ref{fig:CPMours1} for inconclusive proper motion diagrams.\label{fig:CPMours3}}
\end{figure*}

\section{C) Detection limits}
\label{sec:appC}Table~\ref{tab3} lists the measured limiting magnitudes for the detection of companions for all individual targets in the survey at various separations from the star. The limits are measured as described in Sect.~\ref{sec:detection_limits}, each averaged over all five dither positions per star.

\clearpage 
\LongTables 
\begin{deluxetable}{lr@{ }c@{ }lr@{ }c@{ }lccc c}
\tablewidth{0pt}
\tablecaption{The target sample\label{tab1}}
\tablehead{
\colhead{}                        & 
\multicolumn{3}{c}{}              & 
\multicolumn{3}{c}{}              & 
\colhead{$K$\tablenotemark{a}}    & 
\colhead{}                        &
\colhead{SpT}                     &
\colhead{membership}              \\
\colhead{Target name}             & 
\multicolumn{3}{c}{RA}            & 
\multicolumn{3}{c}{DEC}           & 
\colhead{[mag]}                   & 
\colhead{SpT}                     &
\colhead{Ref}                     &
\colhead{evidence\tablenotemark{b}}                
}
\startdata
\multicolumn{11}{c}{Young Members}\\
\hline
\object{CW Tau}                &     04&14&17.00      &     +28&10&57.8      &        7.13      &     K3      &    1    & {\em II,pm         } \\
\object{CX Tau}                &     04&14&47.86      &     +26&48&11.0      &        8.81      &     M2      &    1    & {\em II,pm         } \\
\object{CY Tau}                &     04&17&33.72      &     +28&20&46.8      &        8.6       &     M1.5    &    1    & {\em II,pm         } \\
\object{DE Tau}                &     04&21&55.63      &     +27&55&06.0      &        7.8       &     M1      &    1    & {\em II,Li,pm      } \\
\object{DG Tau}                &     04&27&04.69      &     +26&06&16.3      &        6.99      &     K6      &    1    & {\em II,Li,pm      } \\
\object{DI Tau\tablenotemark{c}}&    04&29&42.47      &     +26&32&49.3      &        8.39      &     M0      &    1    & {\em    Li,pm      } \\
\object{DN Tau}                &     04&35&27.37      &     +24&14&58.9      &        8.02      &     M0      &    1    & {\em II,Li,pm      } \\
\enddata
\tablenotetext{a}{2MASS \citep{cut03}}
\tablenotetext{b}{Evidence for Taurus membership. {\em II:} Class\,II. {\em Li:} Lithium abundance. {\em pm:} proper motion. {\em $\pi$:} Hipparcos parallax. Analysis and references for the individual criteria in Appendix~\ref{sec:A1}.} 
\tablenotetext{c}{DI Tau is in a binary system with classical T Tauri star DH Tau.}
\tablerefs{(1) \citealt{ngu12}; (2) \citealt{wal88}; (3) \citealt{wic00}; (4) \citealt{har03}; (5) \citealt{mar99}; (6) \citealt{luh03}}
\end{deluxetable}

\clearpage
\LongTables 
\begin{landscape}
\begin{deluxetable}{lcr@{$\,\pm\,$}lr@{$\,\pm\,$}lr@{$\,\pm\,$}lccccccc}
\small
\tablewidth{23cm}
\tablecaption{Candidate Companion Parameters\label{tab2}}
\tablehead{
\colhead{}                     & 
\colhead{Obs.}                 & 
\multicolumn{2}{c}{sep}        & 
\multicolumn{2}{c}{PA}         & 
\multicolumn{2}{c}{$\Delta K$} &    
\colhead{$M_1$}                &
\colhead{$M_2$\tablenotemark{b}}&
\colhead{$q$\tablenotemark{b}}  &
\colhead{}                     &
\colhead{}                     &
\colhead{}                     &
\colhead{}                    \\
\colhead{Component\tablenotemark{a}}& 
\colhead{Date}                 & 
\multicolumn{2}{c}{[arcsec]}   & 
\multicolumn{2}{c}{[deg]}      & 
\multicolumn{2}{c}{[mag]}      &          
\colhead{[M$_{\odot}$]}        &
\colhead{[M$_{\odot}$]}        &
\colhead{($M_2/M_1$)}          &
\colhead{$E_\mathrm{det}$}              &
\colhead{$N_\mathrm{det}$}              &
\colhead{Status\tablenotemark{c}}               &
\colhead{ref\tablenotemark{d}}
}
\startdata
\multicolumn{13}{c}{Young Members}\\
\hline
CX Tau                 /cc1                  & 2012-09-02  &  5.783&0.007     &  355.33&0.05  &   8.42&0.05                                  &   0.38   &   0.003  &   0.008  &  $6.6$                &  2  &  B0  &   TP  \\       
                                             & 2013-08-16  &  5.794&0.008     &  355.47&0.04  &   8.40&0.15                                  &          &          &          &                       &     &      &       \\[0.5ex]
DI Tau                 B                     & 2012-10-01  &  5.056&0.010     &   131.8&0.1   &  10.04&0.14                                  &   0.56   &   0.002  &   0.004  &  $11$                 &  4  &  B0  &   1   \\[0.5ex]
DQ Tau                 /cc1                  & 2012-10-01  &  7.573&0.030     &  146.98&0.02  &\multicolumn{2}{c}{\emph{primary sat}$\dots$} &   0.56   &   0.005  &   0.008  &  $3.1$                &  5  &  B0  &   TP  \\       
                                             & 2013-08-19  &  7.588&0.005     &  146.83&0.05  &   7.33&0.13                                  &          &          &          &                       &     &      &       \\[0.5ex]
DQ Tau                 /cc2                  & 2012-10-01  &  2.303&0.030     &   65.46&0.26  &\multicolumn{2}{c}{\emph{primary sat}$\dots$} &   0.56   &   0.002  &   0.004  &  $1.9$                &  2  &   ?  &   TP  \\       
                                             & 2013-08-19  &  2.309&0.009     &   65.17&0.35  &   9.88&0.19                                  &          &          &          &                       &     &      &       \\
\enddata
\tablenotetext{a}{ When available, a commonly used identifier is used to refer to the individual components. Other and newly discovered companion candidates are marked with /cc1, /cc2, ...}
\tablenotetext{b}{Derived secondary masses and mass ratios assume that companions are coeval and physically associated with their primary.}
\tablenotetext{c}{ Companionship status after \citet{laf14}. \emph{C0:} Bound companion based on statistics in this paper, \emph{C1:} Bound companion based on follow-up observations in this paper, \emph{C2}: Bound companion based on other work. \emph{B0:} Unrelated background object based on statistics (this paper), \emph{B1:} Unrelated background object based on follow-up observations in this paper.}
\tablenotetext{d}{ Previous publication of companion or ``TP'' if first reported in the current paper.}
\tablenotetext{e}{ The astrometry and photometry are reported with respect to the astrometric center and combined photometry of the close pair in the same system.}
\tablerefs{
(TP) This paper; 
(1) \citealt{ghe93}; 
(2) \citealt{mon91}; 
(3) \citealt{kra11}; 
(4) \citealt{whi01}; 
(5) \citealt{kra09b}; 
(6) \citealt{ito08}; 
(7) \citealt{koh98}; 
(8) \citealt{har12}; 
(9) \citealt{har94}; 
(10) \citealt{lei93}; 
(11) \citealt{joy44}; 
(12) \citealt{mon91}; 
(13) \citealt{sar98};
(14) \citealt{duc99};
(15) \citealt{bon14}
}
\end{deluxetable}
\clearpage 
\end{landscape}

\clearpage 
\LongTables 
\begin{deluxetable}{lccccccc}
\tabletypesize{\small}
\tablewidth{0pt}
\tablecaption{Detection limits for all observed targets\label{tab3}}
\tablehead{
\colhead{}  &
\multicolumn{7}{c}{$\Delta K_\mathrm{lim}$ (mag) at $\rho$ =}
\\\cline{2-8}
\colhead{Name} &
\colhead{0\farcs10}  &
\colhead{0\farcs25}  &
\colhead{0\farcs50}  &
\colhead{0\farcs75}  &
\colhead{1\arcsec}  &
\colhead{5\arcsec}  &
\colhead{10\arcsec}  
}
\startdata
CW Tau              &  2.3 &  5.5 &  7.2 &  8.6 &  9.3 & 11.0 & 11.5 \\
CX Tau              &  2.6 &  5.5 &  7.2 &  8.6 &  9.2 & 10.6 & 10.8 \\
CY Tau              &  2.1 &  5.8 &  7.4 &  8.9 &  9.4 & 10.9 & 11.2 \\
DE Tau              &  2.3 &  5.3 &  6.6 &  8.1 &  8.8 &  9.9 & 10.1 \\
DG Tau              &  2.3 &  5.9 &  7.7 &  9.1 &  9.7 & 11.5 & 11.9 \\
DI Tau              &  2.4 &  5.7 &  7.3 &  8.8 &  9.4 & 10.9 & 11.1 \\
DN Tau              &  2.7 &  5.2 &  6.6 &  8.0 &  8.6 & 10.2 & 10.6 \\
DQ Tau              &  0.9 &  4.3 &  5.9 &  7.2 &  7.9 &  9.7 & 10.2 \\
\enddata
\end{deluxetable}


\end{document}